\documentclass[onecolumn, 11pt]{article}
\usepackage{amsmath}
\usepackage{amsfonts}
\usepackage{graphicx}
\usepackage{hyperref}
\usepackage{url}
\usepackage{xcolor}
\usepackage{microtype}
\usepackage{booktabs}
\usepackage{fancyhdr}
\usepackage{lipsum}
\usepackage{siunitx}
\usepackage[T1]{fontenc}
\usepackage[a4paper, left=2.1cm, right=2.1cm]{geometry}
\usepackage{caption}
\usepackage{tabularx}
\usepackage{subcaption}

\hypersetup{
    colorlinks=true,
    urlcolor=blue,
    linkcolor=blue,
    citecolor=blue
}
\title{Who burdens the welfare state? Migration and ageing in housing, education, and healthcare demand}
\author{
    Guillermo~Prieto-Viertel$^{1,2}$\thanks{Corresponding author: \texttt{prieto-viertel@csh.ac.at}},
    Carsten~Källner$^1$,
    Elma~Dervic$^1$,\\
    Ola~Ali$^1$,
    Andrea~Vismara$^1$,
    and Rafael~Prieto-Curiel$^1$\\[1em]
    \normalsize\normalfont\itshape
    $^1$Complexity Science Hub, Metternichgasse 8, Vienna, Austria\\
    \normalsize\normalfont\itshape
    $^2$WU Vienna University of Economics and Business, Vienna, Austria
}
\date{}
\begin{document}
\maketitle
\section*{Abstract}
{
Political discourse attributes the pressure on European welfare systems to foreign nationals. Yet projections of service demand rarely disaggregate service demand by citizenship status. We develop a structural demographic model and project healthcare, education, and housing demand in Austria through 2050, disaggregated by citizenship status and regions across migration scenarios. We find that migration, ageing, and fertility shape each sector differently. In healthcare, the ageing of Austrian nationals contributes 4.7 times more to demand growth than immigration, with the most acute pressures in rural, low-migration regions. In housing, migration accounts for the entire net growth in demand, concentrated in metropolitan hubs. In education, aggregate demand contracts regardless of migration assumptions, whereas future needs are driven more by the births of foreigners in Austria than by new arrivals. Foreign nationals consume services in proportion to their demographic weight, with deviations explained by age structure rather than over-utilisation. These results show that the drivers of service demand are sector-specific: migration restrictions could ease housing pressure, but would not address ageing-driven healthcare demand and may accelerate contraction in the education system.
}

\section{Introduction} 

{
The European social model rests on the promise of services of general interest, the essential public services, ranging from healthcare and education to housing and utilities, that society vitally needs to function \cite{gruber2015demographic}. Yet, this foundation faces mounting pressure \cite{europeancommitteeoftheregions2025state}. Ageing populations are intensifying healthcare requirements, urbanisation is clustering housing needs in metropolitan hubs, and declining fertility rates are reducing the needs for educational infrastructure \cite{oecd2024getting, jacobs-crisioni2023big}. These demographic transformations constitute slow-moving yet fundamental shifts that threaten the economic viability of service provision, particularly in depopulating rural areas \cite{kompil2019mapping, gee2002misconceptions}. However, there is a disconnect between these observed realities and the political arena. While the entire demographic structure influences the evolution of service demand \cite{bjornsen2015concept, borges2015europes}, public discourse often reduces these structural transformations to a more tangible factor: migration.
}

{
This concern is deep-rooted. Immigration has consistently ranked as a top concern in Eurobarometer surveys from 2015 through 2024 \cite{europeancommission2015standard, europeancommission2024standard}. The dominant perception is that foreigners are viewed as disproportionate consumers of public goods at the expense of the national populations \cite{banulescu-bogdan2022fear, blasco2025publica}. Such sentiment is widespread; survey data indicate that over 40\% of citizens in major EU states like France, Poland, and Spain believe non-EU migrants enjoy preferential access to services \cite{seiger2025navigating}. This belief fuels welfare chauvinism, the political stance that social rights should be restricted to the national in-group \cite{landini2021exclusion}, and provides the legitimacy for increasingly restrictive policies designed to protect welfare systems from perceived foreign threats \cite{birkebaek2023nordic, europeancommission2025mis}.
}

{
However, migration alone does not threaten the sustainability of welfare systems \cite{hajighasemi2022outcomes, oecd2013international}. Demand for services emerges from the compositional structure of the total population. This is shaped by migration, as well as fertility, mortality, age structure, and geographic distribution \cite{baker2017forecastingd, swanson2012housing}. This raises a critical question: do foreign nationals drive service overburden? Or does attributing the burden to foreigners obscure the broader demographic forces reshaping service demand? Answering this question requires an empirical framework that disentangles the contributions of immigration from broader demographic drivers. We examine this dynamic in Austria, a country where political narratives of service ``overburden'' have recently translated into restrictive migration policies and where public support for conditioning social rights on citizenship is notably high \cite{europeansocialsurveyeuropeanresearchinfrastructureesseric2016ess, parlamentoesterreich2025pause} (see Supplementary Notes C for the Austrian political context and historical patterns of service utilisation by citizenship across all three sectors).
}

{
Previous research has examined migrants' patterns of service utilisation across housing \cite{orok2024osterreichische}, education \cite{choi2023comparison}, and healthcare \cite{dervic2024healthcare}. However, this literature presents two limitations for addressing policy debates about foreign nationals and service overburden. First, utilisation research predominantly examines current or historical patterns, leaving a gap in understanding how the relative contributions of national and foreign populations aggregate and evolve under different demographic futures. Static snapshots cannot reveal whether foreign nationals will drive future overburden or whether structural forces like ageing or fertility decline are the dominant mechanisms. Projecting demand forward, disaggregated by citizenship, is therefore essential for evaluating claims that foreign nationals will increasingly overburden the services of general interest.

Second, social acceptance and perceived entitlement to services are explicitly tied to citizenship in public and political discourse: welfare chauvinist rhetoric targets foreign nationals, irrespective of their actual residence duration or economic contribution \cite{landini2021exclusion}. European Social Survey data reveal that nearly 42\% of Austrians believe social rights should be either conditional on citizenship or withheld entirely, prioritising citizenship over duration of residence \cite{europeansocialsurveyeuropeanresearchinfrastructureesseric2016ess}. While Austria grants foreign residents access to most services based on residence rather than citizenship, citizenship status remains an explicit target of exclusionary rhetoric and policy proposals that condition social rights on legal belonging rather than contribution or residence duration \cite{parlamentoesterreich2025pause, birkebaek2023nordic}. Austria's restrictive \textit{jus sanguinis} naturalisation requirements, which impose lengthy residence periods, language tests, and income thresholds before citizenship can be acquired \cite{stiller2019pathways}, mean that legal foreign status persists long after settlement, creating a large and growing population who remain legally foreign despite lifelong residence. In early 2025, foreign nationals made up 20.2\% of the population. Of these, approximately 15\% were born in Austria and have never migrated \cite{statistikaustria2025statistisches}, a share large enough to constitute a structurally distinct group whose service demands are shaped by integration trajectories rather than arrival patterns. Methodologies that rely on ``country of birth'' classify these individuals as natives, rendering them invisible in migration research, yet they remain the primary target of political rhetoric about foreign service users precisely because of their citizenship status. 
}

{
This study addresses this gap by quantifying future service demand across Austria's housing, education, and healthcare sectors. We restrict our analysis to these three domains as they constitute the core of Social Services of General Interest, the defined ``pillars of welfare'' \cite{fassmann2015services}. This selection is relevant for three reasons. First, housing, education, and healthcare are person-based services whose demand is driven by individual lifecycle stages rather than economic logistics, making them uniquely sensitive to the structural components of demographic change \cite{gruber2015demographic}. Second, political narratives of ``overburden'' specifically target these rivalrous, person-specific goods, implying direct competition between national and foreign users \cite{parlamentoesterreich2025pause}. Third, by analysing these sectors within a unified framework, we move beyond the fragmented approach that characterises most service utilisation research. This enables us to determine whether migration exerts the same pressure across sectors or whether demand drivers are sector-specific, requiring different policy instruments in each domain.
}

{
To analyse these dynamics, we employ a structural demographic model that calculates future service demand by convolving population projections with cohort-specific per capita utilisation rates, disaggregated by region, citizenship, and sex \cite{baker2017forecastingd}. This approach enables us to disaggregate total demand into distinct contributions from Austrian and foreign nationals through 2050. Demand is quantified as the average number of occupied hospital beds per day in healthcare, the number of required dwellings in housing, and the number of full-time equivalent teachers in education. To operationalise migration as a policy lever, we simulate an ensemble of 54 scenarios spanning low to high migration \cite{lempert2003shaping}. This design allows us to distinguish between sectors governed by demographic inertia, where demand trajectories are structurally fixed regardless of border controls, and those in which migration volume shifts the trajectory. Results reveal that the pressure on Austria's healthcare, education, and housing sectors follows three distinct logics, each shaped by a different demographic force. In healthcare, demand is mainly driven by the ageing of the Austrian population. Migration assumptions shift the trajectory by less than 4 percentage points, rendering border policy irrelevant to the sector's future demand. In housing, the opposite holds: migration accounts for the entire net growth in demand. With projections fluctuating by over 40 percentage points across migration scenarios, housing is the sector where migration policy retains substantial leverage. In education, Austria's total fertility rate has fallen to 1.32, among the lowest in Europe \cite{statistikaustria2025demographische}. Aggregate demand, therefore, contracts nationally regardless of migration assumptions, and the growth of foreign national students buffers but cannot reverse this structural decline. Across all three sectors, foreign nationals consume services broadly in proportion to their demographic weight, with marginal deviations explained by age structure rather than over-utilisation. These findings suggest that the drivers of demand for services of general interest are sectorally distinct. Migration restrictions ease housing pressure but cannot address population ageing, which drives healthcare demand, and accelerate the contraction of the education system.
}

\section{Results}\label{sec:results}

\begin{figure}[b!]
\centering
\includegraphics[width=\textwidth]{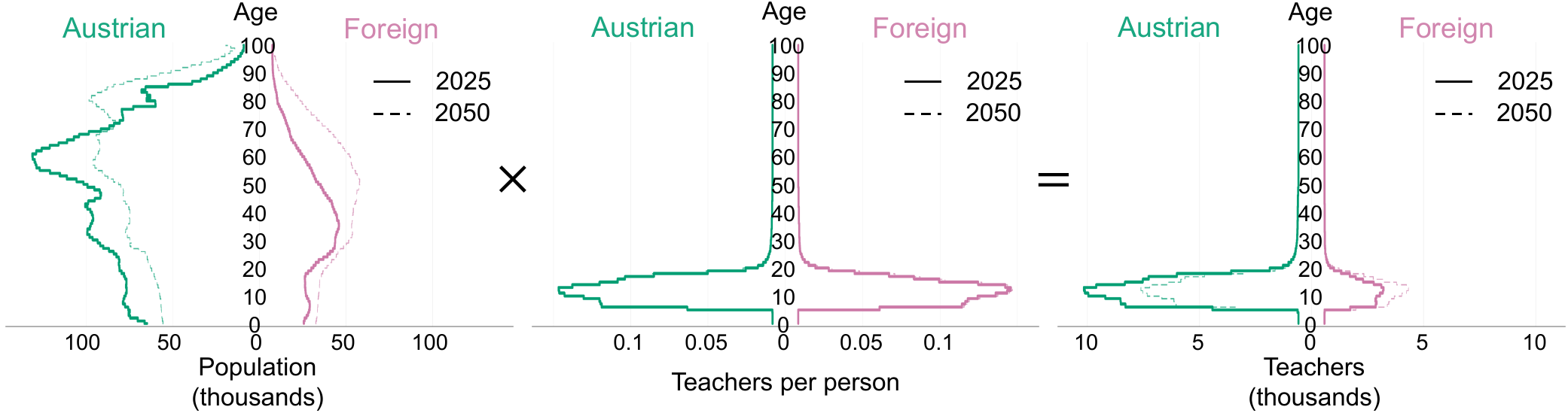}
\caption{Modelling framework applied to the education sector. The figure illustrates how total teacher demand is derived by combining disaggregated population structures (left panel) with cohort-specific per-capita utilisation rates (middle panel), yielding age-specific demand contributions by citizenship (right panel). In each panel, Austrian nationals are shown in green and foreign nationals in pink; solid lines represent 2025, and dashed lines represent 2050. The left panel shows the population pyramid for both groups, reflecting the older age structure of Austrian nationals and the younger, growing foreign-national population. The middle panel shows the per-capita teacher demand rate by age, which is concentrated in school-age cohorts. The student enrollment change from 2025 to 2050 is small. The right panel shows the resulting teacher demand by age group, obtained as the product of the left and middle panels. The shift from solid to dashed lines across all three panels captures the combined effect of demographic change and evolving enrolment behaviour between 2025 and 2050.}
\label{fig:method_results}
\end{figure}

{
To project service demand across Austria's healthcare, education, and housing sectors through 2050, we employ a structural demographic model combining population projections \cite{eurostat2019europop2019,eurostat2023europop2023} (EUROPOP), adapted to distinguish citizenship status, with cohort-specific utilisation rates through 2050 (Figure~\ref{fig:method_results}). To isolate migration's impact, we generate a scenario ensemble spanning low to high migration assumptions, enabling us to distinguish sectors governed by demographic inertia from those where migration is the primary driver. While the housing and education models incorporate behavioural trends, healthcare projections assume fixed utilisation rates due to data limitations. All results are reported as median values, with the full migration scenario range [Min–Max] in brackets, and all percentage changes are measured relative to 2025.
}

{
The demographic trend in Austria is an ageing domestic population living alongside a younger migrant population. This is because migrants migrate at a young age \cite{kallner2025arriving}. Future service demand in Austria follows three distinct paths, determined by the interaction between migration flows and the age sensitivity of each sector. Housing demand follows migration volume, education faces structural contraction regardless of migration scenario, and healthcare expands country-wide, driven predominantly by the ageing of the Austrian population (Figure~\ref{fig:demand_components}. Tables with scenario ranges in the Supplementary Materials B). 
}

\begin{figure}[h!]
\centering
\includegraphics[width=\textwidth]{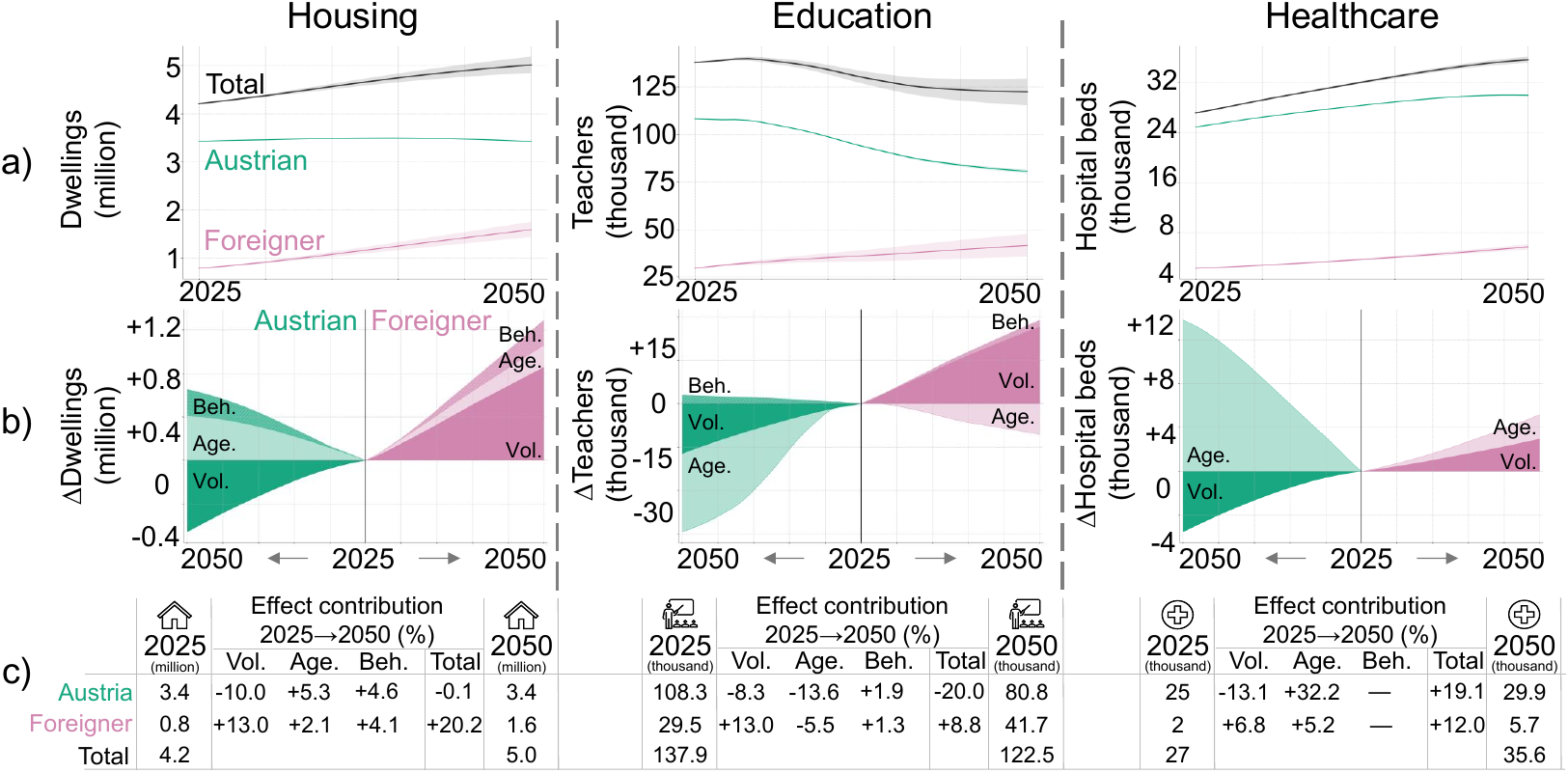}
\caption{Sectoral demand projections and driver decomposition (2025--2050). Columns show results for housing (left), education (middle), and healthcare (right). (a) Projected evolution of total service demand (black line), disaggregated into demand from Austrian nationals (green) and foreign nationals (pink). Shaded regions denote the sensitivity of the results to the migration scenario ensemble. (b) Cumulative absolute contribution of specific drivers to changes in demand ($\Delta$) relative to the 2025 baseline for the median migration scenario, shown separately for Austrian nationals (left, green) and foreign nationals (right, pink). Drivers are classified as volume (Vol., changes in total population size), ageing (Age., shifts in population age structure), and behaviour (Beh., changes in per-capita utilisation intensity). The healthcare model assumes constant utilisation rates, thereby excluding behavioural effects. (c) Summary tables reporting baseline demand in 2025, the percentage contribution of each driver decomposed by citizenship, and total projected demand for 2050.}\label{fig:demand_components}
\end{figure}

\subsection{Aggregate Demand Dynamics}

{
Demand in the housing sector is characterised by migration-driven expansion. Demand from Austrian nationals stagnates ($-0.1\%$ $[-0.5, +0.3]$), reflecting a population that has reached peak household formation, whereas foreign national demand doubles ($+102.3\%$ $[+82.7, +122.3]$). Migration thus accounts for the sector's entire net growth. For context, under the baseline scenario, foreign nationals will constitute roughly $32\%$ of Austria's population by 2050, up from $20.2\%$ in 2025, meaning that the volume of new arrivals drives the sector's trajectory.
}

{
The education system will face a contraction in teacher demand, combined with a compositional shift in the student body. Total teacher demand declines by $-11.1\%$ ($[-16.2, -6.1]$), driven by a $-25.4\%$ decrease in demand from Austrian nationals. This contraction holds across all scenarios; even under the highest migration scenario, teachers' demand shrinks by at least $6.1\%$. Indeed, immigration can buffer but not reverse demographic decline. The defining feature is the transformation of the user base: the foreign share of demand rises from $21.5\%$ to $34.0\%$, while per-capita teacher demand declines for both groups, indicating that the ageing of the overall population drives the contraction.
}

{
The healthcare sector exhibits a distinct pattern in which demand from Austrian nationals increases ($+20.8\%$) even as their population declines ($-11.7\%$). This inverse relationship reflects the steep age gradient in healthcare consumption: as the Austrian population ages, per-capita intensity surges. Demand from foreign nationals more than doubles, but due to the older age profile of the national population, Austrian nationals still account for $84.5\%$ of total demand in 2050. This divergence represents a quantitative shift in the country's functional burden: by 2050, the system requires $14.8\%$ less capacity per capita in education but $26.1\%$ more in healthcare.
}

\subsection{Decomposing the Drivers: Volume vs. Ageing}

{ 
To isolate the specific contribution of population size versus ageing, we decompose the total change in demand into three drivers: a \emph{volume effect} (changes in total population size), an \emph{ageing effect} (shifts in the population's age composition), and a \emph{behavioural effect} (changes in per-capita service intensity) (Figure~\ref{fig:demand_components}).
}

{
The decomposition reveals that the primary contributor to system pressure varies by sector. In housing, the volume effect of foreigners is the primary driver ($+13.0$ pp from 2025 to 2050). The second driver is the combined behavioural effect of Austrians and foreign nationals ($+8.7$ pp), reflecting the structural trend toward smaller household sizes as populations age, partnerships form later, and single-person occupancy becomes more prevalent \cite{bergin2024population} In education, both Austrian volume ($-8.3$ pp) and ageing ($-13.6$ pp) effects contribute to a contraction in teacher demand, whereas the volume effect of foreign nationals ($+13.1$ pp) is the primary positive component. In healthcare, the Austrian ageing effect accounts for $+32.2$ percentage points of national demand growth, compared to only $+6.9$ pp from the volume effect of foreigners, a ratio of nearly 4:1 even under the highest migration scenario. The hierarchy of drivers is thus sectorally distinct: foreign migration expands housing demand, ageing drives healthcare demand, and Austria's own fertility collapse contracts education demand.
}

{
Foreigners are not a static young cohort. As this population ages, their impact on service demand reverses across sectors (contracting $-5.4$ pp in education to expanding $+5.2$ pp in healthcare). This means the same settled population who moderates school contraction is simultaneously ageing into high-morbidity age groups. Among foreign citizens, nearly $46\%$ of healthcare demand growth stems from ageing rather than from new arrivals.
}

{
Yet immigration is only part of the story of foreign population growth. Over one-third of the growth of the population of foreign nationals in Austria through 2050 originates from births within the country rather than from new arrivals. By 2050, the model projects an additional 1.2 million foreign-born immigrants settling in Austria alongside 0.6 million births to foreign-national parents, roughly two new arrivals for every birth to foreign nationals. This ratio holds even under the highest migration scenario (1.7 to 2.3 arrivals per foreign birth). Most will remain legally as foreign nationals. Austria's naturalisation rate stood at 0.72\% of the foreign-national resident stock in 2024 \cite{statistikaustria2025naturalisations}, one of the lowest in Western Europe, reflecting stringent eligibility requirements \cite{stiller2019pathways}. In education and housing, where demand is more sensitive to young cohorts and household formation, this reproductive momentum of the settled foreign population cannot be addressed by border policy.
}

\subsection{Do Foreign Nationals Consume Services Above Their Demographic Weight?}

\begin{figure}[htbp]
\centering
\includegraphics[width=\textwidth]{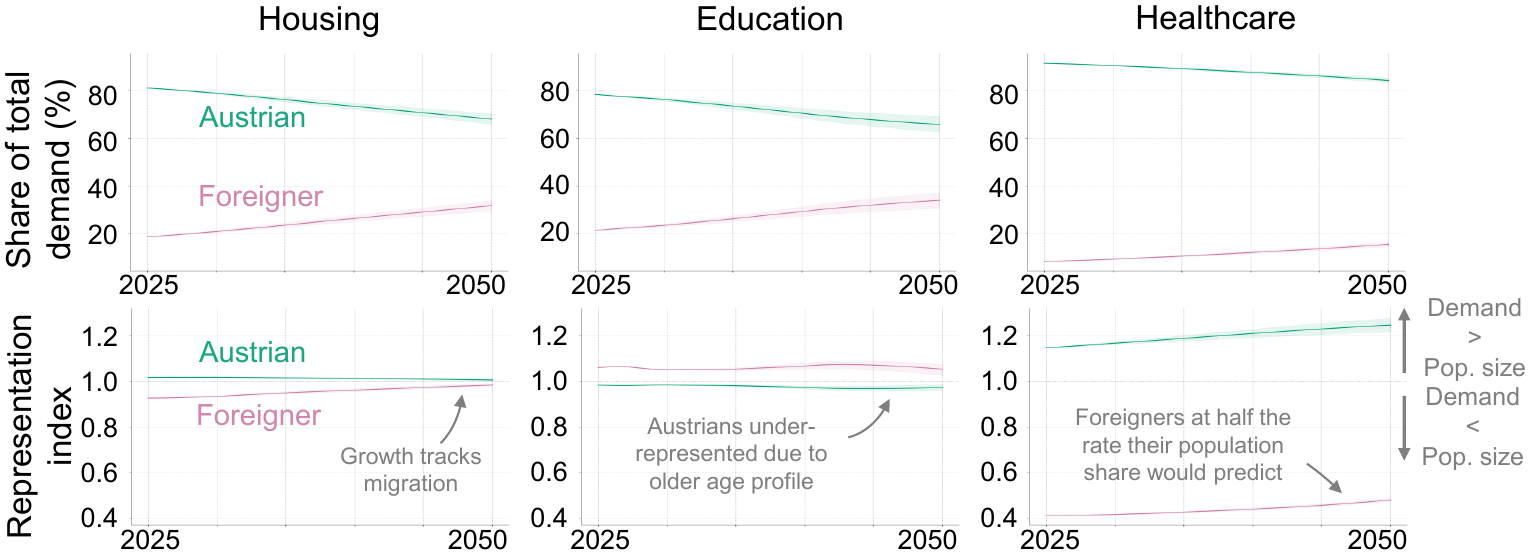}
\caption{Service demand shares and representation index by sector and citizenship (2025--2050). Columns show results for housing (left), education (middle), and healthcare (right). (a) Share of total demand generated by Austrian nationals (green) and foreign nationals (pink). (b) Representation index, defined as each group's demand share divided by its population share; a value of $1.0$ indicates that a group consumes services in exact proportion to its demographic weight, values above $1.0$ indicate over-representation, and values below $1.0$ indicate under-representation. Shaded regions represent the sensitivity to the migration scenario ensemble.}
\label{fig:metrics}
\end{figure}

{

A central claim in the overburden debate is that foreigners are disproportionate consumers of public goods at the expense of the national populations \cite{blasco2025publica}. To test whether this holds in Austria, we introduce the \emph{representation index}, which measures each group's share of total demand relative to its share of the population, with a value of 1.0 indicating exact proportionality (Figure~\ref{fig:metrics}).
}

{
In housing, both groups remain near parity throughout the projection period. Foreign demand doubles while Austrian demand stagnates, and the foreign share of dwellings rises from $18.7\%$ to $31.8\%$, but this growth is proportionate to population growth, not above it. The expansion is driven by the growth of the foreign population, not by disproportionate demand per person.
}

{

In education, the foreign share of teacher demand rises from $21.5\%$ to $34.0\%$, not because foreign nationals consume more education per person, but because the Austrian student population contracts more rapidly than the foreign one expands due to low Austrian fertility. The representation index remains stable, only marginally above parity, attributable to the younger age profile of foreign nationals rather than to per-person over-utilisation. The sector contracts in volume while shifting in composition: classrooms that were $78.5\%$ Austrian in 2025 become two-thirds Austrian by 2050.
}

{
Healthcare exhibits the starkest asymmetry and is most directly relevant to the overburden narrative. Foreign demand grows fastest in relative terms ($+151.9\%$), yet Austrians still account for $84.0\%$ of total demand in 2050. The representation index explains why: foreign nationals consume services at roughly half the rate predicted by their population share, while Austrian over-representation increases as national cohorts age into high-morbidity groups. Across all three sectors, the representation index tells a consistent story: where foreign nationals exceed parity, as marginally in education, this reflects their younger age profile rather than disproportionate per-person consumption.
}

\subsection{Sectoral Decoupling and Regional Divergence}

{
Service overburden is rarely felt at the national level, but rather it materialises in specific regions where demographic pressures concentrate. Whether the appropriate response is border policy or long-term infrastructure planning depends on a prior question: is local demand driven by ongoing migration flows, or by the demographic momentum of populations already settled? To answer this, we track two spatial relationships over the projection horizon: how demand growth correlates with the regional foreign population share ($\beta^{\text{conc}}$, the \emph{spatial concentration}), and how sensitive demand growth is to regional foreign population growth ($\beta^{\text{growth}}$, the \emph{migration elasticity}). The results reveal a clear divergence across sectors. Housing demand is strongly coupled to migration inflows. Education demand decouples from new arrivals while concentrating spatially in high-foreign-share regions because future teacher demand is increasingly driven by children born within Austria to settled foreign-national parents rather than by newly arrived students. Lastly, healthcare demand remains spatially independent of migration altogether until the settled foreign population begins to age.
}

\begin{figure}[htbp]
\centering
\includegraphics[width=\textwidth]{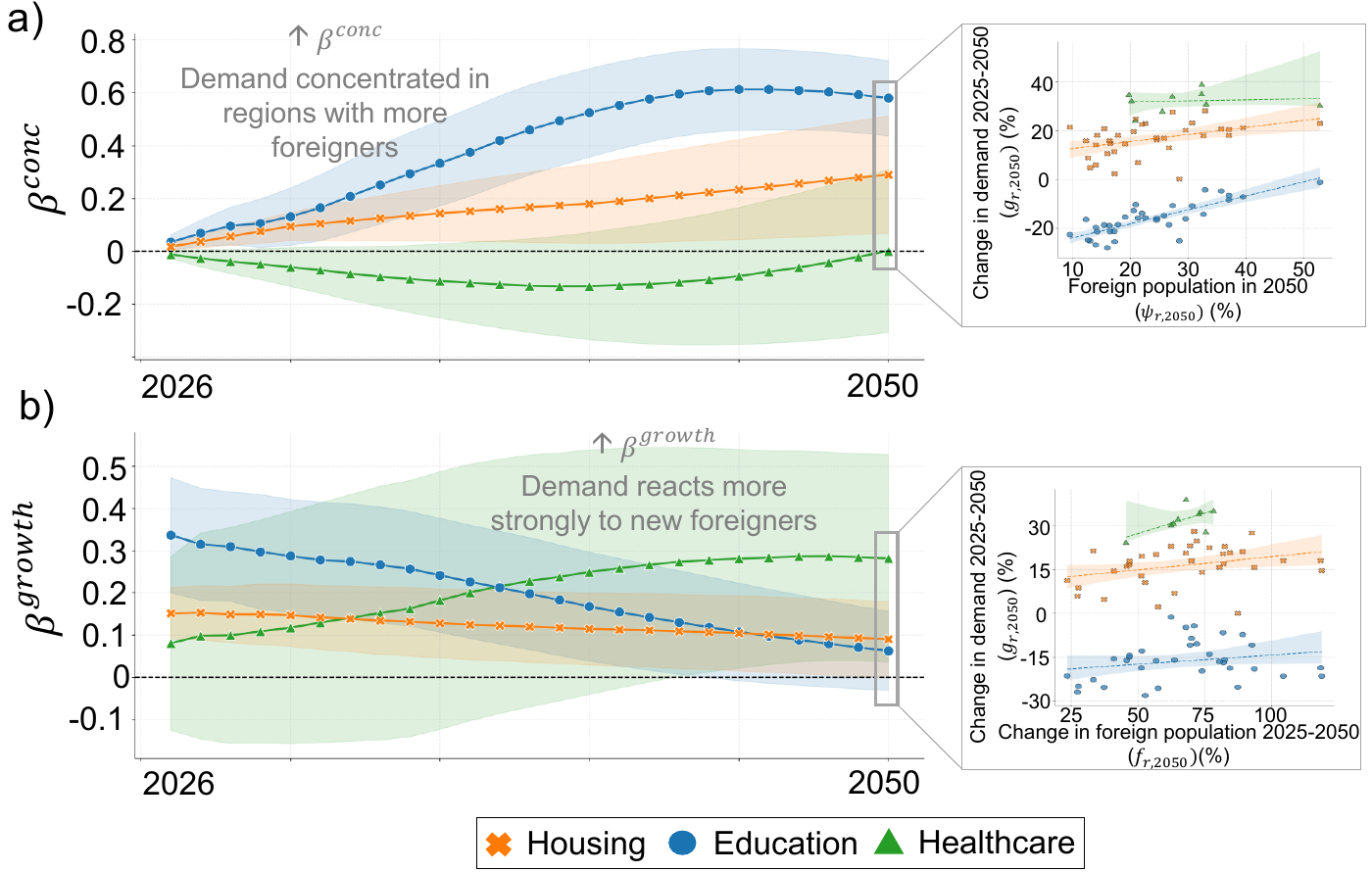}
\caption{Spatial concentration and growth elasticity of service demand to foreign population (2026--2050). (a) Temporal evolution of the spatial concentration coefficient ($\beta^{\text{conc}}$), which measures whether cumulative demand growth localises in regions with higher foreign population shares. A positive and rising coefficient indicates increasing spatial concentration; a coefficient indistinguishable from zero indicates that demand growth is spatially independent of foreign settlement, pointing instead to universally operating drivers such as population ageing. The inset scatter plot shows the underlying regional regression at 2050, illustrating the cross-sectional relationship between foreign population share ($\psi_{r,2050}$) and cumulative demand growth ($g_{r,2050}$) across regions. (b) Temporal evolution of the foreign population growth elasticity ($\beta^{\text{growth}}$), which measures the sensitivity of demand growth to the growth of the foreign population. A declining coefficient indicates decoupling, in which demand is increasingly governed by the demographic momentum of the settled population rather than by new arrivals. The inset scatter plot shows the underlying regional regression at 2050, illustrating the cross-sectional relationship between cumulative foreign population growth ($f_{r,2050}$) and cumulative demand growth ($g_{r,2050}$). Shaded areas represent 95\% confidence intervals; coefficients are statistically significant where the interval does not cross zero. Values are based on the median across the migration scenario ensemble. Housing and education use $N=35$ NUTS\,3 regions; healthcare uses $N=9$ NUTS\,2 regions.}
\label{fig:regression_results}
\end{figure}

\begin{table}[htbp]
\centering
\small
\caption{Regional variation in projected service demand growth across federal states (2025--2050).}
\label{tab:regional_services}
\resizebox{\textwidth}{!}{
\begin{tabular}{lrrrr}
\toprule
\textbf{Region} & \textbf{Foreign Share 2050 (\%)} & \textbf{Housing Growth (\%)} & \textbf{Education Growth (\%)} & \textbf{Healthcare Growth (\%)} \\
\midrule
Vienna      & 52.9 [50.6, 54.9] & +23.0 [+15.4, +30.6] & -1.0 [-8.8, +6.9]   & +29.4 [+23.8, +34.9] \\
Salzburg    & 33.2 [31.2, 35.2] & +17.3 [+12.5, +22.2] & -10.8 [-17.3, -4.1] & +30.6 [+28.5, +32.8] \\
Vorarlberg  & 32.5 [30.4, 34.6] & +26.3 [+21.8, +30.9] & -6.8 [-13.3, -0.1]  & +38.8 [+36.8, +40.7] \\
Tyrol       & 32.4 [30.3, 34.5] & +21.8 [+16.7, +27.0] & -9.8 [-16.8, -2.7]  & +34.9 [+32.8, +36.9] \\
Upper Austria & 27.3 [25.0, 29.5] & +20.1 [+15.1, +25.1] & -12.8 [-18.4, -7.2] & +34.1 [+32.7, +35.5] \\
Styria      & 25.6 [23.2, 27.9] & +14.4 [+9.8, +19.0]  & -14.9 [-20.0, -9.6] & +28.4 [+26.9, +29.9] \\
Carinthia   & 20.9 [18.5, 23.3] & +9.4 [+5.6, +13.3]   & -21.7 [-26.3, -17.0] & +24.5 [+22.5, +26.5] \\
Lower Austria & 20.2 [16.8, 23.3] & +18.1 [+12.1, +24.0] & -16.6 [-24.9, -8.4] & +32.6 [+29.3, +36.1] \\
Burgenland  & 19.9 [16.1, 23.1] & +17.8 [+10.5, +24.9] & -20.7 [-27.6, -13.7] & +35.1 [+30.3, +40.1] \\
\bottomrule
\end{tabular}%
}
\end{table}

\begin{figure}[htbp]
\centering
\includegraphics[width=\textwidth]{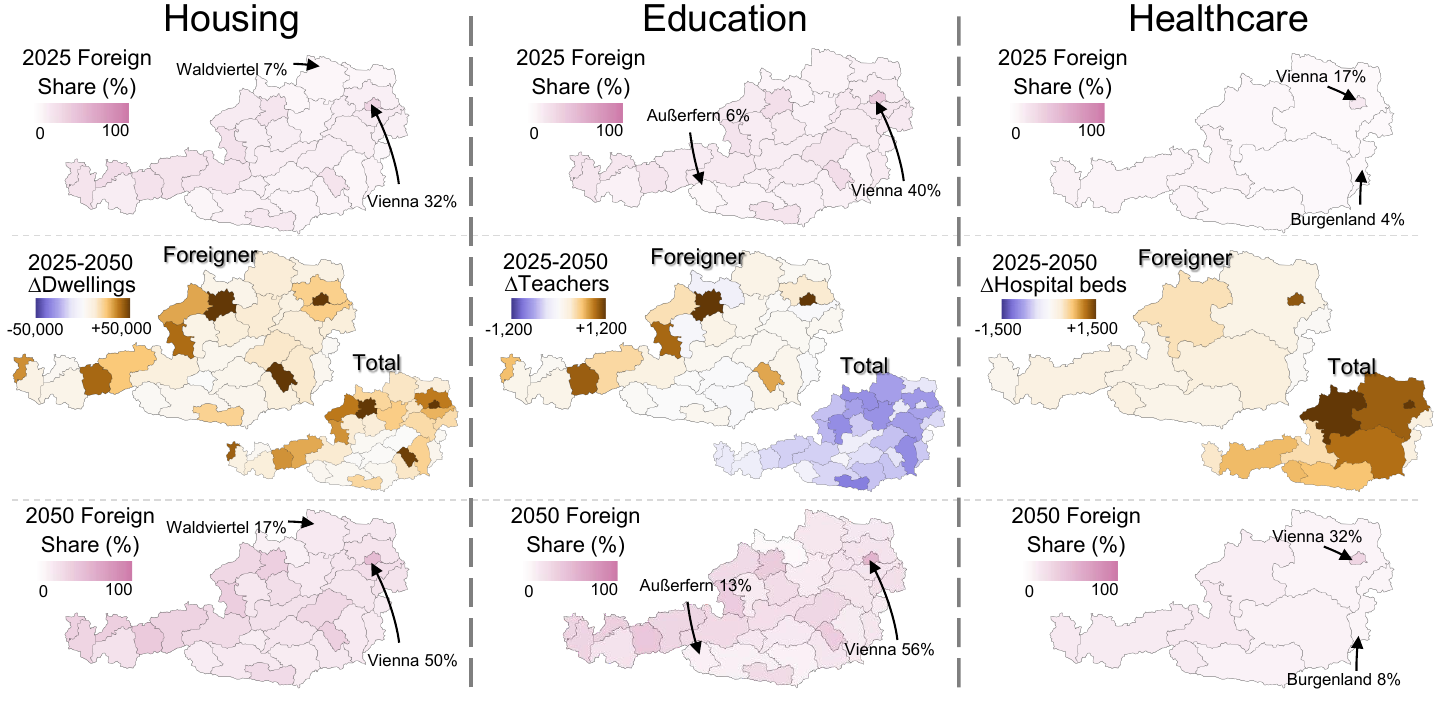}
\caption{Spatial distribution and projected change in service demand by sector and citizenship (2025--2050). Columns show results for housing (left), education (middle), and healthcare (right). The top row shows the baseline foreign national share of total sectoral demand in 2025. The middle row shows the absolute change in demand ($\Delta$) from 2025 to 2050, decomposed into the foreign national contribution (left larger map) and the total change (right smaller map); orange indicates growth and blue indicates contraction. The bottom row shows the projected foreign national share in 2050, enabling direct comparison with the 2025 baseline. Selected regions are annotated with their foreign share values, highlighting the regions with the highest and lowest foreign national shares in 2025. Spatial units correspond to NUTS\,2 regions for healthcare and NUTS\,3 regions for housing and education. Absolute change values are capped at the 95th percentile to mitigate the influence of outliers. All projections are reported as median values across the migration scenario ensemble.}
\label{fig:map}
\end{figure}

{
Housing demand remains closely coupled with foreign population growth. Demand continues to localise in urban hubs ($\beta^{\text{conc}} = 0.29$) and retains sensitivity to new arrivals ($\beta^{\text{growth}}_{2050} = 0.09$), confirming that the volume of foreign nationals has genuine leverage over the sector's spatial distribution. Outside Vienna, however, the dominant growth mechanism shifts: volume effects from new arrivals contribute minimally, and demand is instead driven by household dilution as ageing populations consume more space per capita. The maps illustrate this duality, with absolute housing demand growth concentrated in Vienna and western urban corridors while behavioural effects sustain moderate growth even in regions with modest foreign population shares (Figure~\ref{fig:map}). At the NUTS\,3 level, a ``dormitory'' typology emerges: in Mostviertel-Eisenwurzen and Oststeiermark, foreign housing demand exceeds foreign education demand by ratios of $91:1$ and $75:1$, indicating concentrations of single workers without accompanying families. These dormitory regions decouple housing and education planning, requiring differentiated infrastructure responses that aggregate national projections cannot capture.

}

{
In education, the spatial pattern undergoes a regime shift over the projection period. Spatial concentration rises ($\beta^{\text{conc}}$ from 0.03 to 0.57, CI: 0.43--0.72), meaning that future teacher demand will increasingly localise in regions with large foreign populations. Yet simultaneously, the elasticity of demand to foreign population growth falls from $0.34$ in 2026 to becoming statistically insignificant in 2050. Rising concentration alongside collapsing elasticity indicates that demand is no longer driven by new arrivals but by the age structure and reproductive momentum of the existing foreign population. The maps confirm this spatial polarisation, with demand contraction spreading across rural regions while urban centres hold or grow (Figure~\ref{fig:map}). Vienna exemplifies both dynamics simultaneously: the foreign share of education demand crosses the majority threshold, rising from $39.6\%$ to $53.5\%$, while the city is the only region where the direction of demand growth depends on migration assumptions, contracting by $-8.8\%$ under low migration but expanding by $+6.9\%$ under high migration. Everywhere else, teacher demand falls regardless of border policy.
}

{
Healthcare presents the clearest case of spatial independence from migration. The spatial concentration coefficient is effectively zero or slightly negative throughout ($\beta^{\text{conc}}_{2026} = -0.01$, CI: -0.03--0.01), and the migration elasticity is negligible and statistically indistinguishable from zero after 2040. 
This does not mean regional variation is absent. Demand growth shows differences across federal states, ranging from $+24.5\%$ in Carinthia to $+38.8\%$ in Vorarlberg (Figure~\ref{fig:map} and Table~\ref{tab:regional_services}). This variation follows the geography of Austrian ageing, not of foreign settlement: Vienna, despite having the highest foreign share of any region, ranks only 7th of 9 in healthcare demand growth. After 2040, however, the migration elasticity turns positive and statistically significant, reflecting the delayed demographic maturation documented earlier: foreign cohorts begin ageing into high-morbidity age groups by the 2040s. Migration thus leaves a deferred footprint in healthcare, spatially insignificant in the near term but emerging as the settled population ages, making the geography of foreign settlement a secondary yet non-negligible factor for future planning.
}

\section{Discussion}

{
Our analysis identifies a sectoral decoupling: the drivers of demand for services of general interest follow divergent logics across welfare domains. Housing demand tracks foreign migration volume; education demand contracts nationally while shifting compositionally toward the settled foreign population; and healthcare demand is driven predominantly by the ageing of Austrian nationals. This divergence implies that the reductionist narrative of service ``overburden'' due to the migration of foreign nationals obscures the actual mechanisms at work: migration restrictions could ease housing pressure but could not address healthcare needs driven by ageing, while potentially accelerating the shrinkage of the education system.
}


{
Migration of foreign nationals is the dominant contributor to net growth in housing demand, and, unlike the other sectors, migration policy retains genuine leverage here. Housing demand varies by over 40 percentage points between the lowest and highest migration scenarios. Austria is already showing signs of acute housing pressure: building permits reached a record low in 2024, with approved apartments falling from 84 thousand units in 2019 to 51 thousand, while projected household growth continues to rise \cite{statistikaustria2025wohnen, orok2024osterreichische}. This insight extends the Austrian regional planning agency (\textit{ÖROK}) household projections by attributing demand growth to its source \cite{orok2024osterreichische} and aligns with European studies on housing shortages in high-immigration regions \cite{jacobs-crisioni2023big}. Yet the impact of this aggregate growth is exacerbated by a mismatch between household size and new housing supply. Foreign households are, on average, larger than Austrian ones (2.28 persons compared to 1.92) \cite{statistikaustria2025housing}. However, residential construction in primary settlement hubs like Vienna yields small units ($<60\text{m}^2$) \cite{statistikaustria2025wohnen}. This misalignment contributes to the overcrowding conditions already prevalent for non-nationals \cite{statistikaustria2025wohnen}. Supply-side planning failures, not migration alone, therefore bear substantial responsibility for the housing shortage. At the regional scale, dormitory regions identified in the results (foreign housing-to-education demand ratios exceeding 90:1 in industrial corridors such as Mostviertel-Eisenwurzen) further decouple housing from education planning. Regional characteristics require differentiated infrastructure responses that aggregate national projections cannot capture.
}

{
In education, our results directly contradict the narrative of overburden: even under the highest migration scenario, aggregate demand still contracts. Migration functions as a demographic buffer, moderating but not reversing the structural decline attributable to Austrian fertility collapse. More significantly, future demand decouples from immigration and is rather governed by the age structure and reproductive momentum of the foreign settled population. The challenge thus shifts from the reception capacity for newly arrived migrants to the integration capacity for second- and third-generation migrants who, despite holding foreign citizenship, have never migrated. Language acquisition is central to this change in classroom composition, as currently 7.7\% of compulsory school pupils require intensive German support, rising to 23.2\% in first-grade primary school \cite{statistikaustria2025bildung}. This challenge is spatially concentrated in Vienna, the only region where the direction of demand depends on migration scenarios. Teacher demand in all other regions, including Burgenland, which has already seen 29 school closures between 2000 and 2015 \cite{gruber2015demographic}, contracts regardless of border policy. Our projections assume constant student-to-teacher ratios; pedagogical adaptations, such as language support and teacher diversification, would increase actual requirements beyond the projected baseline.
}

{
The ageing of Austrian nationals dominates healthcare demand growth and expenditure \cite{oecd2021access}. Foreign nationals consume services at half the rate predicted by their population share, consistent with the ``healthy migrant effect'' \cite{graetz2017utilization, dervic2024healthcare}. Yet, Austrian research also implicates access barriers that channel migrants toward emergency care rather than preventive, primary care with general practitioners \cite{perchinig2016care, crede2018international}. While this creates highly visible patient volume in emergency rooms, these outpatient emergency visits rarely translate into inpatient admissions that drive hospital bed demand \cite{crede2018international}. Even if foreign utilisation increased to match Austrian levels, this would not overturn the structural hierarchy: the ageing of Austrian nationals contributes 4.7 times more to demand growth than the volume effect of foreigners. Migration assumptions shift total demand by less than 4 percentage points, rendering border policy effectively irrelevant to the sector's trajectory. This independence is not permanent. The foreign cohorts that arrived (mostly young) during the 2010s and 2020s will age into high-morbidity groups by the 2040s, at which point migration leaves a deferred but measurable footprint on regional healthcare demand. In the near term, however, the most acute pressure will materialise not in migrant-dense urban centres but in the rural regions where foreign nationals are scarcest, and the Austrian population is ageing fastest.
}

{
The divergent regional elasticity of each sector to the growth of the foreign population reflects the interaction between the age profile of migration and the age sensitivity of each service. Migration flows are concentrated among young adults \cite{kallner2025arriving}, a demographic group with low healthcare consumption, moderate housing demand, and limited direct education demand. Housing is migration-elastic because household formation is concentrated precisely in the age groups that migrants occupy on arrival. Healthcare is migration-independent in the near term because its demand is driven by the steep age gradient of morbidity, which foreign cohorts have not yet reached—-though they will. Education occupies an intermediate position: direct demand from newly arrived students is modest, but the reproductive momentum of the settled foreign population generates a lagged wave of child cohorts that sustains demand in urban regions while Austrian fertility collapse dominates elsewhere. The sectoral decoupling is therefore a structural consequence of the lifecycle mismatch between when migrants arrive, when they form households, when they have children, and when they age into high-morbidity groups.
}


{
These findings engage directly with welfare-chauvinist arguments that portray foreign nationals as illegitimate claimants threatening fiscal sustainability \cite{landini2021exclusion}. The political economy literature suggests this perception has structural roots: ethnic fragmentation has been shown to reduce public willingness to fund shared goods, as polarised groups are less willing to pool resources for services benefiting outgroups \cite{alesina1999public}. Yet comparative research shows that welfare chauvinist sentiment often responds to perceived service degradation rather than actual utilisation patterns \cite{bello2022prejudice}. The representation index tells that foreign nationals consume services broadly in proportion to their demographic weight, with marginal deviations explained by age structure rather than over-utilisation in education. The policy implications are therefore more complex than the overburden narrative suggests. Foreign nationals account for a minority of healthcare consumption but represent 44.3\% of beginning medical students \cite{statistikaustria2025jahrbuch}; in education, an ageing teaching workforce (38.5\% over 50 years old) \cite{statistikaustria2025bildung} must adapt to a student body undergoing rapid compositional change. Restrictionist migration policies thus risk reducing the very supply of skilled workers on which an ageing welfare state depends \cite{falkenbach2019austria, cangiano2014elder}, a trade-off that cannot be resolved at the level of border policy and that requires sector-specific workforce planning.
}

{
Citizenship status is central to our analysis, and it highlights a structural limitation of migration policy. Over one-third of foreign demographic expansion through 2050 originates from births within Austria rather than from immigration, a population that border restrictions cannot reach. Austria's restrictive \textit{jus sanguinis} regime maintains a growing category of \emph{native-born foreigners} (foreign nationals with no direct migration experience) \cite{stiller2019pathways}, whose service demands are shaped by integration trajectories rather than admission policies. The population often targeted by welfare-chauvinist rhetoric is increasingly one that has never crossed a border. Naturalisation policy may therefore be a more relevant long-term lever than migration restriction, not because it would reduce service use, but because it would dissolve the legal category through which that use is politically framed.
}

{
Our analysis is subject to several limitations. We treat ``foreign nationals'' as a single aggregate group, obscuring heterogeneity between, for instance, a recent refugee and a long-settled professional migrant. Our healthcare model captures inpatient hospital demand only, quantified as occupied hospital beds per day; outpatient utilisation (including general practitioner visits and emergency ambulatories) is excluded. To the extent that outpatient services are less age-graded than inpatient care, our results may overstate the dominance of ageing as the primary demand driver across the healthcare system as a whole. Our healthcare model also assumes constant age-specific utilisation rates, excluding potential shifts from technological advances or care delivery reforms. While this isolates the demographic component, actual future demand may diverge, likely upward given trends toward more intensive care. Static naturalisation rates likely underestimate the long-term integration of demand into the ``Austrian'' category, thereby narrowing the citizenship gap across all three sectors. Moreover, we model demand as a direct function of demographics, ignoring supply-side feedback loops where capacity constraints may suppress household formation or healthcare utilisation. Our scenario ensemble varies migration rates while holding fertility and mortality at baseline assumptions, likely understating total projection uncertainty, particularly in the tails. Finally, as a single-country case study, our findings are bounded by Austria's specific institutional context. Further limitations and assumptions are described in Supplementary Methods A.
}

{
Despite these limitations, the mechanisms apply to other states with universalist welfare aspirations. The dominance of ageing over migration in healthcare is consistent with the broader European demographic trajectory, suggesting that the tension between demand and supply is a structural feature of modern European demography rather than a unique Austrian outcome. Our findings are most directly applicable to other \textit{jus sanguinis} countries such as Germany and Switzerland, where second-generation residents may remain legally foreign despite lifelong residence \cite{stiller2019pathways}, creating a comparable native-born foreigner phenomenon. In \textit{jus soli} regimes (such as Brazil or Canada), the citizenship-residency wedge we document would be substantially narrower, and the decoupling between migration flows and service demand would emerge earlier. The tripartite sectoral framework we developed, treating healthcare, housing, and education together, distinguishes migration-elastic, migration-buffered, and migration-independent demand trajectories. Such a framework provides a transferable analytical template for evaluating overburden claims in other countries.
}

\section{Methods}\label{sec:methods}

\subsection{General Framework}

{
We employ a structural demographic demand model that projects service requirements by combining disaggregated population projections with cohort-specific per-capita utilisation rates \cite{baker2017forecastingd, swanson2012housing}. The model operates through two components: a demographic projection module that evolves the population annually by age, sex, region, and citizenship; and three sector-specific demand modules that translate these projections into required dwellings, teachers, and hospital beds per day. A complete technical description of all model components, estimation procedures, validation tests, and data processing steps is provided in Supplementary Methods A.

Formally, the demand $D_{n}(t)$ for a service by nationality $n$ at time $t$ is:
\begin{equation}
D_{n}(t) = \sum_{r} \sum_{a} \sum_{s} P_{r,a,s,n}(t)\,d_{r,a,s,n}(t),
\label{eq:demand}
\end{equation}
where $P_{r,a,s,n}(t)$ is the population of region $r$, age $a$, sex $s$, and nationality $n$ at time $t$, and $d_{r,a,s,n}(t)$ is the corresponding per-capita utilisation rate. Separating these two terms allows us to isolate the contribution of demographic change from that of behavioural change in utilisation patterns (Figure~\ref{fig:method_sketch}).

\begin{figure}[htbp]
\centering
\includegraphics[width=\textwidth]{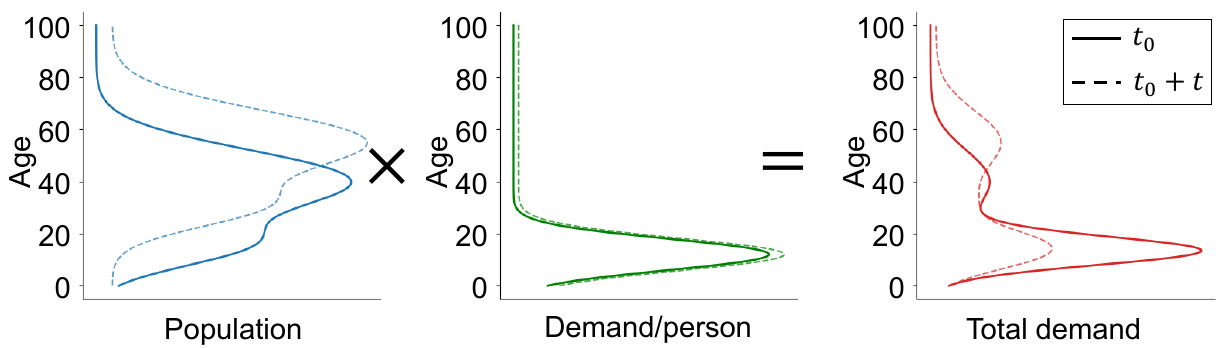}
\caption{Conceptual framework of the structural demographic demand model. The total demand (Right panel) is derived as the product of the projected population structure (Left panel) and the age-specific per-capita demand intensity (Middle panel). Solid lines represent the base year ($t_0$) conditions, while dashed lines illustrate the evolution toward a future projection year ($t_0+t$). The concentration of demand at younger ages shown here is representative of educational services, where the size and composition of student-age cohorts shape needs. The shifts illustrate the three core drivers of demand: the volume effect (overall growth/shrinkage in population), the ageing effect (shifts in the population age composition), and the behavioural effect (changes in per capita service utilisation rates).}
\label{fig:method_sketch}
\end{figure}
}

{
All three sectors are calibrated to a common base year of 2025, which represents the first year of the full projection pipeline. Because the sectoral utilisation data predate this year, the 2025 baseline demand values are model-derived rather than directly observed; full details of this extrapolation and its implications are provided in Supplementary Methods A. All percentage changes and decomposition effects reported in the main text are measured relative to this modelled baseline.
}

\subsection{Demographic Projections}

\subsubsection{Population framework and nationality decomposition}

The demographic foundation is the EUROPOP2019 projection framework produced by Eurostat \cite{eurostat2019europop2019}, which provides age-, sex-, and region-specific components of demographic change at the NUTS\,3 level via a standard cohort-component method. EUROPOP2019, however, does not distinguish between Austrian and foreign nationals. We therefore develop a decomposition that models the foreign-national population explicitly by sex, single-year age, and NUTS\,3 region, and derives the Austrian-national population as the residual difference from the EUROPOP total. This residual construction ensures exact consistency with official projection totals while enabling flexible modelling of citizenship-specific dynamics: differential fertility, naturalisation, and the formation of mixed-nationality partnerships.

Each projection year proceeds through seven sequential steps: application of mortality rates, addition of net migration by nationality, transfer of foreign nationals to Austrian status through naturalisation, calculation of births with nationality assignment, application of infant mortality, ageing of all cohorts by one year, and assembly of the updated population.

\subsubsection{Migration calibration}

The central challenge is that EUROPOP2019 provides the NUTS\,3 regional detail required for the model, but was calibrated before the Ukrainian refugee crisis and post-pandemic migration surge. EUROPOP2023 incorporates these shocks but is published only at the national level \cite{eurostat2023europop2023}. We bridge the two vintages by decomposing projected net migration into a structural baseline derived from EUROPOP2019 regional patterns and a national-level correction calculated as the divergence between EUROPOP2023 and the EUROPOP2019 national aggregate. This correction is allocated to regions using a time-varying blending of two spatial keys: a crisis-period key estimated from observed 2022--2024 settlement patterns, and a long-run structural key reflecting historical regional attractiveness. The weight assigned to the crisis key declines linearly from 1.0 in 2026 to 0.0 by 2035, after which allocation follows the structural key exclusively. This design captures the gradual normalisation of migration patterns following the refugee shock while preserving the long-run spatial logic of economic migration.

The Austrian component of migration is modelled as a historical average with a region-specific responsiveness to changes in total migration, estimated from pre-crisis, crisis-era, and post-COVID observation windows \cite{statistikaustria2025internal, statistikaustria2025international}. The foreign component is obtained as a residual, ensuring that regional totals remain consistent with EUROPOP2023 national control totals at all times.

\subsubsection{Additional demographic processes}

Naturalisation rates are held constant at their observed 2022--2024 averages (approximately 0.7\% of the foreign-national resident stock per year), supported by the historical stability of this rate over the preceding 15 years \cite{statistikaustria2025naturalisations}. Fertility is differentiated by nationality using scaling ratios derived from historical data \cite{statistikaustria2025demographische}, with foreign-national fertility exceeding Austrian fertility but converging gradually toward parity by 2100. Nationality assignment to newborns follows Austrian \textit{jus sanguinis} law: a child acquires Austrian citizenship if at least one parent is Austrian. The probability that a child born to a foreign-national mother has an Austrian father is governed by a homophily parameter ($h = 0.61$), estimated from birth registry data via grid search and validated against observed nationality-specific birth counts for 2010--2024 \cite{statistikaustria2025births}.

\subsubsection{Scenario structure}

To bound projection uncertainty, we generate 54 demographic scenarios by combining the three EUROPOP2023 migration variants (low, main, high) with alternative specifications for migration allocation, Austrian responsiveness, and calibration-period weighting. All results are reported as medians across this ensemble, with the full scenario range [Min--Max] indicating sensitivity to migration assumptions.

\subsection{Housing Demand}

\subsubsection{Outcome and data}

Housing demand is quantified as the number of occupied dwellings required to accommodate the projected population. Three administrative datasets from Statistik Austria are combined: population registers  \cite{statistikaustria2025population}, the register-based housing census \cite{statistikaustria2025housing}, and institutional household records, covering reference years 2011, 2021 \cite{statistikaustria2025registera}, and 2022 \cite{statistikaustria2025registerb}. The analysis is conducted at NUTS\,3 level, with regions grouped into five housing market clusters based on the \textit{ÖROK} classification to enable robust trend estimation through regional pooling \cite{orok2024osterreichische}.

\subsubsection{Modelling approach}

Housing demand is driven by the household formation rate ($\mathrm{HHFR}$), the per-capita probability that a person heads a private household. This rate integrates two empirically estimated components: the group quarters rate ($\mathrm{GQR}$, the share of the population in institutional settings) and the headship rate ($\mathrm{HR}$, the probability of heading a private household):

\begin{equation}
\mathrm{HHFR}_{r,a,s,n,t} = \bigl(1 - \mathrm{GQR}_{r,a,s,n,t}\bigr)\,\mathrm{HR}_{r,a,s,n,t}.
\end{equation}

To project future housing behaviour while preventing extrapolation into infeasible values, trends in household formation are estimated in logit space. We model the evolution of the $\mathrm{HHFR}$ for each demographic cell as a linear trend applied to the base-year logit:

\begin{equation}
\mathrm{logit}\left(\frac{\mathrm{HHFR}_{r,a,s,n,t}}{100}\right) = \alpha_{r,a,s,n} + \beta_{v(r),g(a),s,n} \; t,
\end{equation}
where $\alpha$ is the cohort's baseline intercept and $\beta$ captures the specific behavioural trend estimated across regional housing market clusters ($v$) and age groups ($g$). Total dwelling demand ($H$) for a future year $t$ is then derived by applying these forecasted rates to the projected population:

\begin{equation}
\hat{H}_{r,a,s,n,t} = P_{r,a,s,n,t} \; \frac{\mathrm{HHFR}_{r,a,s,n,t}}{100}.
\end{equation}

This preserves the overarching $D = P \times d$ structure, where the per-capita demand $d = \mathrm{HHFR}/100$. Finally, the projected $\mathrm{HHFR}$ values are scaled to remain consistent with external regional household-size targets from the Austrian Conference on Spatial Planning (\textit{ÖROK}). Model validation demonstrated national-level prediction errors below 0.5\% in both backtesting and forward testing, with 91.2\% of the 14,140 demographic cohorts exhibiting errors within $\pm$10

\subsection{Education Demand}

\subsubsection{Outcome and data}

Education demand is quantified as the number of full-time equivalent teachers required to serve the projected student population. Administrative data from Statistik Austria cover school enrolment \cite{statistikaustria2025school} and teaching staff at the political district level across six school categories (primary, lower secondary, special education, polytechnic, academic secondary, and vocational) \cite{statistikaustria2025teaching}, for the academic years 2015--2023. District-level data are mapped to 35 NUTS\,3 regions using population-proportional allocation for districts that cross regional boundaries.

\subsubsection{Modelling approach}

The projection employs a two-stage model. In the first stage, a binomial GLM with a logit link estimates the probability $\hat{\pi}_{r,a,n,t}$ that a person in demographic cell $(r, a, n)$ is enrolled in any school:

\begin{equation}
\mathrm{logit}(\hat{\pi}_{r,a,n,t}) = \beta_0 + \beta_{\mathrm{nat}(i)} + \beta_{\mathrm{age}(i)} + \beta_{\mathrm{NUTS3}(i)} + \beta_{\mathrm{age}(i)}\,\beta_{\mathrm{nat}(i)},
\end{equation}
where the age-nationality interaction captures observed heterogeneity in enrolment patterns across citizenship groups. A trend variant augments this by including a standardised year covariate, allowing enrolment probabilities to evolve over time. In the second stage, total projected enrolment is allocated across school types using historically pooled empirical proportions ($\chi_{k \mid r,a,n}$, for school type $k$). Projected student counts are then:

\begin{equation}
\hat{S}_{r,a,n,k,t} = P_{r,a,n,t} \; \hat{\pi}_{r,a,n,t} \; \chi_{k \mid r,a,n},
\end{equation}
which preserves the $D = P \times d$ structure, with per-capita demand $d = \hat{\pi} \; \chi$. Teacher requirements are derived by dividing projected student counts by the base-year student-to-teacher ratio:

\begin{equation}
\hat{T}_{r,n,k,t} = \frac{\hat{S}_{r,n,k,t}}{\eta_{r,k,t_0}},
\end{equation}
where $\eta_{r,k,t_0}$ is the observed ratio in the base year. This assumption isolates the pure demographic contribution to staffing requirements, holding class-size policy constant. Model validation on the 2023 holdout data showed that more than 91\% of demographic cohorts exhibit prediction errors within $\pm$10\%.

\subsection{Healthcare Demand}

\subsubsection{Outcome and data}

Healthcare demand is quantified as the average number of hospital beds occupied per day, enabling direct comparison with inpatient capacity. Patient-level records of hospital admissions and length of stay were obtained from the Austrian Federal Ministry of Health. The analysis uses 2019 as the base year, selected as the most recent complete observation period before the COVID-19 pandemic. The model operates at NUTS\,2 level, comprising nine federal states.

\subsubsection{Modelling approach}

Healthcare utilisation is modelled using a Poisson GLM that estimates the per-capita bed-day rate ($\hat{\lambda}_{r,a,s,n}$) for each demographic cell:

\begin{equation}
\log(\hat{\lambda}_{r,a,s,n}) = \beta_0 + \beta_{\mathrm{age}(i)} + \beta_{\mathrm{sex}(i)} + \beta_{\mathrm{nat}(i)} + \beta_{\mathrm{NUTS2}(i)} + \beta_{\mathrm{age}(i)}\,\beta_{\mathrm{nat}(i)} + \beta_{\mathrm{age}(i)}\,\beta_{\mathrm{sex}(i)},
\end{equation}
where the age--nationality interaction captures the differential utilisation gap between nationals and foreign nationals at advanced ages. The expected annual bed-days for a cohort of size $P_i$ follow directly as $\nu_i = P_i \; \hat{\lambda}_{r,a,s,n}$, preserving the $D = P \times d$ structure with $d = \hat{\lambda}$. Projected average daily beds for a future year $t$ are then:

\begin{equation}
\bar{BD}_{r,a,s,n,t} = P_{r,a,s,n,t} \; \hat{\lambda}_{r,a,s,n} \; \frac{1}{365}.
\end{equation}

Heteroskedasticity-consistent (HC3) robust standard errors are used to account for the overdispersion typical of healthcare utilisation data. The model assumes constant utilisation rates, so all projected demand changes reflect demographic composition effects. Uncertainty in estimated rates was quantified via Monte Carlo simulation with 1,000 draws from the asymptotic parameter distribution.

\subsection{Analytical Framework}

\subsubsection{Decomposition of demand drivers}

To disentangle the contributions of population size, age structure, and behavioural change to future demand, we apply a three-component symmetric decomposition to the total change in demand relative to the 2025 baseline.

First, we isolate the \emph{behavioural effect} ($E_{\text{Beh}}$). This measures the change in demand attributable exclusively to evolving per-capita utilisation habits (e.g., shrinking household sizes or shifting school enrolment rates). We calculate this by taking the difference between the full trend projection ($D^{\text{Trend}}_{t}$) and a status quo counterfactual ($D^{\text{SQ}}_{t}$) that holds behaviour frozen at 2025 levels. Because both scenarios use the identical projected population, their difference is purely behavioural:

\begin{equation}
E_{\text{Beh}} = D^{\text{Trend}}_{t} - D^{\text{SQ}}_{t}.
\end{equation}

Second, we decompose the remaining status quo demand into demographic components. Under status quo conditions, per-capita utilisation rates are fixed; therefore, any change in the aggregate demand intensity ($I^{\text{SQ}}_t = D^{\text{SQ}}_t / P_t$) relative to the base year ($I_{t_0}$) is driven entirely by shifts in the population's age composition.

To separate the impact of a growing population (\emph{volume effect}) from an ageing one (\emph{ageing effect}) without generating an unassigned interaction residual, we employ a two-component symmetric (Fisher-type) decomposition.

The \emph{volume effect} ($E_{\text{Vol}}$) isolates the change due to population growth by multiplying the absolute change in population by the inter-period average intensity:
\begin{equation}
E_{\text{Vol}} = \left(P_{t} - P_{t_0}\right) \frac{I_{t_0} + I^{\text{SQ}}_{t}}{2}.
\end{equation}

Conversely, the \emph{ageing effect} ($E_{\text{Age}}$) isolates the impact of the shifting age structure by multiplying the change in intensity by the inter-period average population:

\begin{equation}
E_{\text{Age}} = \left(I^{\text{SQ}}_{t} - I_{t_0}\right) \frac{P_{t_0} + P_{t}}{2}.
\end{equation}

The symmetric weighting ensures that neither the base-year nor target-year structure is arbitrarily privileged, allowing the three computed effects to sum perfectly to the total demand change ($\Delta D = E_{\text{Beh}} + E_{\text{Vol}} + E_{\text{Age}}$). Each effect is computed separately for the Austrian and foreign sub-populations and expressed as a percentage contribution relative to the base-year demand, enabling direct cross-sectoral comparison of the forces that demand essential services.

\subsubsection{Comparative indices}

Three standardised indices characterise the evolution of demand across sectors and citizenship groups. Together, they distinguish absolute system pressure from compositional shifts and utilisation parity.

The \emph{cumulative demand growth} ($CDG$) tracks the percentage change in total sectoral demand relative to the 2025 base year ($t_0$), quantifying absolute system expansion or contraction:

\begin{equation}
CDG_{t} = \left(\frac{D_{t}}{D_{t_0}} - 1\right) \times 100.
\end{equation}

The \emph{demand share} ($DS$) quantifies each group's proportion of total sectoral demand at a given point in time, revealing shifts in the composition of the user base:

\begin{equation}
DS_{n,t} = \frac{D_{n,t}}{D^{\text{tot}}_{t}}.
\end{equation}

The \emph{representation index} ($RI$) normalises this demand share by each group's population weight ($P_{n,t} / P^{\text{tot}}_{t}$), yielding a measure of proportionality between demographic size and service consumption:

\begin{equation}
RI_{n,t} = \frac{DS_{n,t}}{P_{n,t} / P^{\text{tot}}_{t}},
\end{equation}
where a value of $RI = 1.0$ indicates parity: a group consumes services in exact proportion to its demographic weight. Values above 1.0 indicate that a group consumes services disproportionately more than its population share, whereas values below 1.0 indicate under-representation. This index provides an empirical test for evaluating political narratives that foreign nationals disproportionately overburden essential services.

\subsubsection{Spatial analysis}

To formalise the distinction between the spatial footprint of the settled foreign population and the marginal impact of new arrivals, we estimate two repeated cross-sectional regression models for each sector across every projection year from 2026 to 2050.

The \emph{spatial concentration} model regresses cumulative demand growth in each region on that region's foreign population share:

\begin{equation}
g_{r,t} = \alpha^{\text{conc}}_t + \beta^{\text{conc}}_t \, \psi_{r,t} + \epsilon_{r,t},
\end{equation}
where $g_{r,t} = (D_{r,t} - D_{r,t_0}) / D_{r,t_0}$ is the cumulative demand growth and $\psi_{r,t} = P_{r,t}^{\text{For}} / P_{r,t}^{\text{tot}}$ is the regional foreign population share. A positive and significant $\beta^{\text{conc}}_t$ indicates that regions with a higher share of foreign population experience systematically larger demand growth. A coefficient indistinguishable from zero indicates spatial independence, pointing instead to population ageing rather than the localised presence of foreign nationals.

The \emph{foreign population growth elasticity} model regresses cumulative demand growth on the cumulative growth rate of the foreign population:

\begin{equation}
g_{r,t} = \alpha^{\text{growth}}_t + \beta^{\text{growth}}_t \, f_{r,t} + \nu_{r,t},
\end{equation}
where $f_{r,t} = (P_{r,t}^{\text{For}} - P_{r,t_0}^{\text{For}}) / P_{r,t_0}^{\text{For}}$ is the cumulative foreign population growth rate. The slope $\beta^{\text{growth}}_t$ is the elasticity of demand with respect to foreign population growth: a value near 1.0 indicates a direct relationship, while a value statistically indistinguishable from zero indicates full decoupling, where demand is governed by the demographic momentum of the settled population rather than by new arrivals. The intercept $\alpha^{\text{growth}}_t$ captures the baseline demand trajectory in the absence of foreign population growth; a negative intercept indicates structural contraction driven by the Austrian population alone, as observed in education.

The combination of the two models enables a diagnosis of each sector's relationship to migration. A sector with a rising $\beta^{\text{conc}}$ but a declining $\beta^{\text{growth}}$, as observed in education, exhibits increasing spatial concentration of demand in regions with high foreign population shares, yet this concentration is no longer sensitive to how fast the foreign population grows. This pattern arises because the spatial footprint of the foreign population is already established; the demand it generates derives from the age structure and reproductive momentum of that settled population rather than from continued inflows. Restricting new arrivals would not neutralise the spatial redistribution of demand, because the driver has already shifted from border flows to domestic demographic processes.

Both models are estimated by ordinary least squares independently for each projection year, with 95\% confidence intervals used to assess statistical significance. Housing and education regressions use $N = 35$ NUTS\,3 regions, while healthcare uses $N = 9$ NUTS\,2 regions.

\subsection{Data Sources}

The model integrates administrative data from three principal sources. Population registers, school enrolment records, teaching staff statistics, housing census data, and naturalisation records were obtained from Statistik Austria via the STATcube database system. Population projections derive from the EUROPOP2019 and EUROPOP2023 frameworks produced by Eurostat, with the nationality decomposition developed as part of this study. Healthcare utilisation data were obtained from the Austrian Federal Ministry of Health. Table~\ref{tab:data_sources} summarises the primary inputs; full details of data cleaning, harmonisation, and variable construction are provided in Supplementary Methods A.

\begin{table}[htbp]
\centering
\caption{Primary data sources.}
\label{tab:data_sources}
\small
\begin{tabular}{llll}
\toprule
\textbf{Domain} & \textbf{Source} & \textbf{Years} & \textbf{Spatial Level} \\
\midrule
Population and migration & Statistik Austria \cite{statistikaustria2025population, statistikaustria2025internal, statistikaustria2025international} & 2002--2025 & NUTS\,3 \\
Population projections & Eurostat EUROPOP2019/2023 \cite{eurostat2019europop2019, eurostat2023europop2023} & 2025--2050 & NUTS\,3 / National \\
School enrolment \& staff & Statistik Austria \cite{statistikaustria2025school, statistikaustria2025teaching} & 2015--2023 & District \\
Housing census & Statistik Austria \cite{statistikaustria2025housing, statistikaustria2025registera, statistikaustria2025registerb} & 2011, 2021--2022 & NUTS\,3 \\
Healthcare utilisation & Austrian Federal Ministry of Health & 2019 & NUTS\,2 \\
Naturalisation & Statistik Austria \cite{statistikaustria2025naturalisations} & 2022--2024 & NUTS\,2 \\
Fertility by nationality & Statistik Austria \cite{statistikaustria2025demographische} & 2010--2023 & National \\
\bottomrule
\end{tabular}
\end{table}

\renewcommand{\figurename}{Supplementary Figure}
\renewcommand{\tablename}{Supplementary Table}

\setcounter{section}{1} 

\renewcommand{\thefigure}{A.\arabic{figure}}
\renewcommand{\thetable}{A.\arabic{table}}
\renewcommand{\thesection}{A} 

\section{Supplementary Methods}\label{sec:supp_methods}

This section details the modelling framework used to project future demand for services of general interest in Austria. The methodology, which links demography to service demand through a mathematical structure, is designed to isolate the specific contributions of migration, population ageing, and behavioural change to sectoral demand growth. It is based on demographic forecast, with particular focus on the calibration of migration flows and the decomposition of the population by nationality.

The framework operates on a modular logic. At its core is a high-resolution demographic projection that disaggregates the population by nationality (Austrian vs. foreign national), single-year age, sex, and NUTS\,3 region. This demographic backbone is coupled with sector-specific demand modules (i.e., housing, education, and healthcare) that translate population structure into service requirements using empirically estimated utilisation rates.

\subsection{General Framework for Sectoral Projections}
\label{sec:general_framework}

\subsubsection{Structural Demand Equation}

{
The sectoral demand projections are governed by a unified structural equation linking the demographic output to service requirements. We employ a cohort-component demand model where the national demand $D_{n}(t)$ for a service by nationality $n$ at time $t$ is defined as:

\begin{equation} \label{EqnService}
D_{n}(t) = \sum_{r} \sum_{a} \sum_{s} P_{r,a,s,n}(t)\,d_{r,a,s,n}(t),
\end{equation}
where $P_{r,a,s,n}(t)$ represents the projected population of region $r$, age $a$, sex $s$, and nationality $n$ at time $t$ (derived from the demographic forecast in Section Demographic Forecast: Technical details), and $d_{r,a,s,n}(t)$ is the corresponding per-capita utilisation rate (derived from the sectoral models described below).
}

{
This unified architecture comprises two linked modules: a demographic projection module that evolves $P$ annually, and sector-specific demand modules that estimate $d$ to produce teachers, dwellings, or hospital bed-days. The separation of these terms allows us to isolate the contribution of population change ($P$) from behavioural change ($d$).
}

\subsubsection{Temporal Alignment and Base Year}

{
The sectoral demand models are calibrated on administrative data whose most recent available years differ across sectors: 2021--2022 for housing (census data), 2015--2023 for education (school enrolment records), and 2019 for healthcare (inpatient hospital records). However, the demographic projection model produces its first complete population structure for 2025, the jump-off year of the adapted EUROPOP framework. Because all three demand models treat demand as a direct function of demographic composition convolved with per-capita utilisation rates, the 2025 demand estimates are obtained by applying sector-specific utilisation parameters, estimated from the historical data described in each sectoral section, to the 2025 projected population.

The year 2025 thus represents an extrapolation rather than an observation: it is the first year in which the full modelling pipeline (demography $\times$ utilisation) operates jointly, and it is adopted as the base year for all results reported in the main paper. This convention ensures a common temporal anchor across sectors, enabling cross-sectoral comparisons of demand growth driven by the same underlying demographic trajectory. All percentage changes, decomposition effects, and index values reported in the main text are measured relative to this 2025 baseline.
}

\subsubsection{Behavioural Scenario Definitions}

{
Each sectoral demand model produces projections under two behavioural scenarios that bracket the range of plausible futures for per-capita utilisation intensity. The \emph{status quo} scenario holds all per-capita utilisation rates constant at their most recently observed levels. Under this assumption, changes in projected demand arise exclusively from demographic shifts, population growth, ageing, and changing nationality composition, while individual service consumption behaviour remains frozen. The status quo scenario thereby isolates the pure demographic effect on demand and provides a counterfactual against which behavioural change can be measured.

The \emph{trend} scenario allows per-capita utilisation rates to evolve according to trends estimated from the historical calibration data. In housing, this captures the observed tendency toward smaller household sizes (rising headship rates extrapolated in logit space). In education, it captures gradual shifts in the probability of school enrolment, as estimated by the year covariate in the binomial model. The trend scenario thus represents a trajectory in which both demographic structure and individual behaviour change simultaneously.

The healthcare model constitutes an exception: because it is fitted on a single cross-sectional year (2019) and does not include a temporal covariate, it produces only a status quo projection. All demand variation in the healthcare sector therefore reflects compositional demographic change under fixed age-specific utilisation rates. Unless otherwise noted, the results presented in the main paper correspond to the trend scenario for housing and education and to the status quo scenario for healthcare.
}

\subsection{Demographic Forecast: Technical Details}

\subsubsection*{Overview and data}
{
Our projection builds on the EUROPOP2019 framework produced by Eurostat \cite{eurostat2019europop2019}, which provides age-, sex-, and region-specific components of demographic change (fertility, mortality, and migration) at the NUTS\,3 level using a standard cohort-component method. EUROPOP2019, however, does not distinguish between Austrian and foreign nationals. We therefore develop a decomposition that models the foreign-national population explicitly by sex, single-year age, and region, and obtains the Austrian-national population as the residual difference from the EUROPOP total. This residual construction guarantees exact consistency with the official projection totals while allowing flexible treatment of migrant-specific dynamics such as differential fertility, naturalisation, and mixed-nationality partnership formation.

The main technical challenge is migration. EUROPOP2019 provides the regional detail we need, but it was calibrated before the Ukrainian refugee crisis and the post-pandemic migration surge. EUROPOP2023 incorporates these shocks but is published only at the national level \cite{eurostat2023europop2023}. We bridge the two by decomposing projected migration into a structural baseline (from EUROPOP2019's regional patterns) and a national-level correction (the gap between EUROPOP2023 and EUROPOP2019 at the national level), which is then distributed to regions using empirical settlement patterns that transition gradually from crisis-era to long-run spatial structure \cite{statistikaustria2025internal, statistikaustria2025international}. The Austrian component of migration is estimated from its historical relationship to total migration, and the foreign component is obtained as the residual, ensuring that regional totals are always consistent with EUROPOP2023 national control totals. Supplementary table \ref{tab:data_sources_demo} summarises the principal data inputs used in the projection model.

\begin{table}[htbp]
\centering
\caption{Data sources for the components of the demographic model.}
\label{tab:data_sources_demo}
\small
\setlength{\tabcolsep}{4pt}
\begin{tabular}{
>{\raggedright\arraybackslash}p{4.2cm}
>{\raggedright\arraybackslash}p{3.8cm}
>{\raggedright\arraybackslash}p{2.0cm}
>{\raggedright\arraybackslash}p{4.0cm}
}
\toprule
\textbf{Data} & \textbf{Source} & \textbf{Period} & \textbf{Resolution} \\
\midrule

Population by nationality 
& Statistik Austria (STATcube) 
& 2002--2025 
& NUTS\,3 $\times$ sex $\times$ single age \\

Internal migration by nationality 
& Statistik Austria (STATcube) 
& 2002--2024 
& NUTS\,3 $\times$ sex $\times$ single age \\

International migration by nationality 
& Statistik Austria (STATcube) 
& 2002--2024 
& NUTS\,3 $\times$ sex $\times$ single age \\

Ukrainian refugee flows 
& Statistik Austria (STATcube) 
& 2022--2024 
& NUTS\,3 $\times$ sex $\times$ 5-year age \\

Naturalisations (resident) 
& Statistik Austria (STATcube) 
& 2022--2024 
& NUTS\,2 $\times$ sex $\times$ age group \\

Live births by parental nationality 
& Statistik Austria 
& 2010--2024 
& National \\

TFR by nationality 
& Statistik Austria 
& 2010--2023 
& National \\

Mortality rates 
& Eurostat (EUROPOP2019) 
& 2019--2100 
& NUTS\,3 $\times$ sex $\times$ single age \\

Fertility rates (ASFR) 
& Eurostat (EUROPOP2019) 
& 2019--2100 
& NUTS\,3 $\times$ single age \\

Population projection (regional) 
& Eurostat (EUROPOP2019) 
& 2019--2100 
& NUTS\,3 $\times$ sex $\times$ single age \\

Net migration projection (regional) 
& Eurostat (EUROPOP2019) 
& 2019--2100 
& NUTS\,3 $\times$ sex $\times$ single age \\

Net migration projection (national) 
& Eurostat (EUROPOP2023) 
& 2024--2100 
& National $\times$ sex $\times$ single age \\

Population projection (national) 
& Eurostat (EUROPOP2023) 
& 2024--2100 
& National $\times$ sex $\times$ single age \\

\bottomrule
\end{tabular}
\end{table}

The projection updates the population annually using a standard cohort-component method. Each year proceeds through seven sequential steps:

\begin{enumerate}
    \item \textbf{Mortality.} Age-, sex-, and geography-specific survival rates are applied to all ages except age~0 (newborns are handled separately in step~5). The same rates are used for Austrian and foreign nationals (see justification below).
    \item \textbf{Migration.} Net migration is added by region, age, sex, and by nationality (Austrian or foreign national).
    \item \textbf{Naturalisation.} A fraction of the foreign-national stock acquires Austrian citizenship based on age-, sex-, and geography-specific naturalisation rates, reducing the foreign population accordingly.
    \item \textbf{Fertility.} Births are calculated from women of reproductive age using age-specific fertility rates. Each newborn is assigned a nationality based on the parental-nationality model.
    \item \textbf{Age-0 mortality.} Survival rates are applied to the age-0 cohort (both existing infants and newborns), and survivors are aged to age~1.
    \item \textbf{Ageing.} All remaining cohorts advance by one year.
    \item \textbf{Assembly.} The updated population is compiled for the next projection year.
\end{enumerate}

Because the foreign-national population is tracked at fine resolution (35~NUTS\,3 regions $\times$ 101~single-year ages $\times$ 2~sexes), some cells can turn negative when large net outflows or naturalisations exceed the cell count. Rather than simply setting these cells to zero, which would create population from nothing at the national level, we redistribute each deficit to nearby age groups within the same region and sex, using Gaussian weights with dispersion $\sigma=5$~years. Concretely, for a cell at age $a$ with a deficit $d>0$, each positive-population cell at age $a'$ in the same group absorbs a share proportional to $\exp\!\bigl(-(a'-a)^2 / 2\sigma^2\bigr)$ times its current population. Any residual negative values after this step (from floating-point rounding) are set to zero. Across all scenarios, the total amount corrected in this way never exceeds 800 people over 25 years of projections.
}

{
\subsubsection*{Mortality}

We apply age-specific mortality rates from EUROPOP2019 uniformly across nationality groups. This simplification is supported by Austrian mortality statistics provided by Statistik Austria \cite{statistikaustria2025demographische}. At the national level, life expectancy at birth differs by only $+0.47$~years for males and $+0.10$~years for females between foreign and Austrian nationals, with foreign nationals living marginally longer. Across working ages (15--64), foreign nationals exhibit approximately 15--18\% lower age-specific mortality than Austrian nationals (mortality rate ratios of 0.82 for males and 0.86 for females, averaged across single-year ages), consistent with the ``healthy-migrant effect'' in European populations. Infant mortality runs in the opposite direction, roughly 80\% higher among foreign nationals, but the absolute rates are very low (2--5~per~1,000 live births), so the impact on projected population totals is negligible (Supplementary Figure \ref{fig:mortality_comparison}).

\begin{figure}[htbp]
    \centering
    \includegraphics[width=\textwidth]{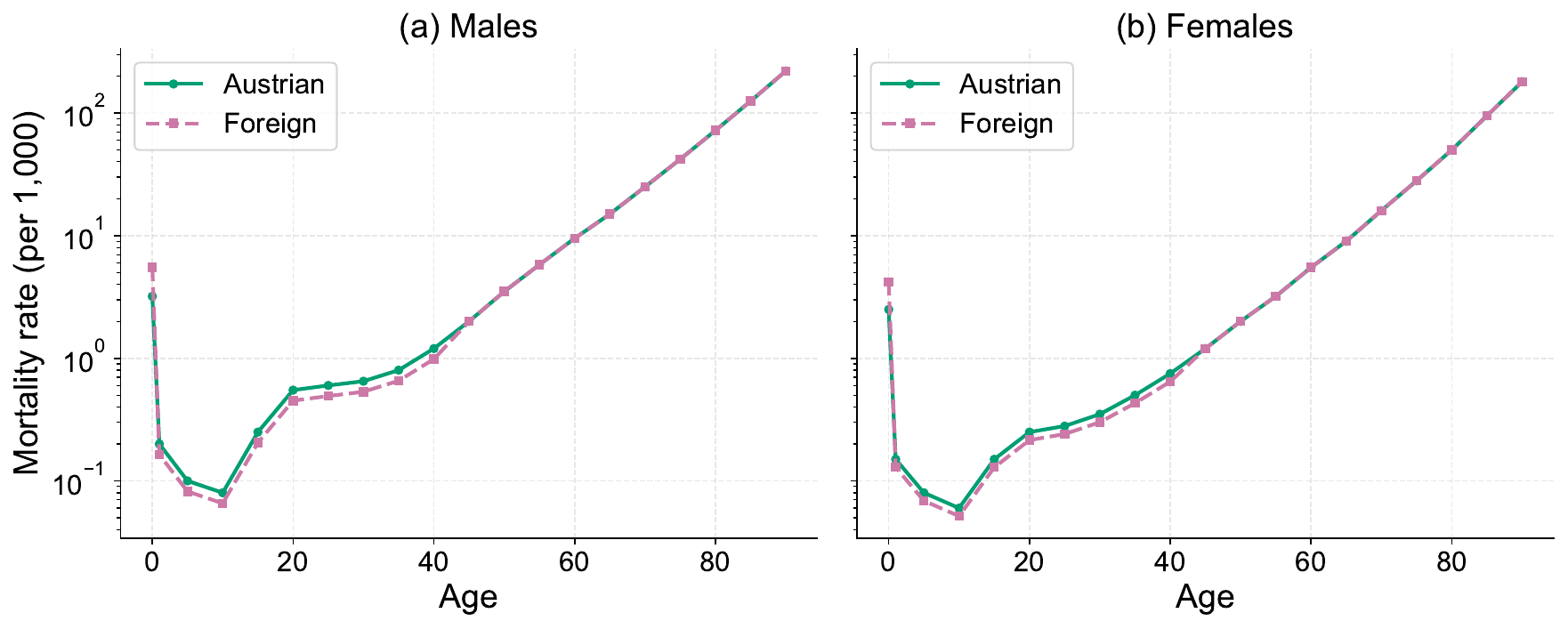}
    \caption{Age-specific mortality rates for Austrian and foreign nationals (2022). 
    The near-overlap across most of the age range supports the use of uniform rates.}
    \label{fig:mortality_comparison}
\end{figure}

}

\subsubsection{Migration Flow Calibration}\label{sec:migration}

{
The central challenge is to produce NUTS\,3 migration projections disaggregated by nationality, given that the two available Eurostat projection vintages each lack a key dimension. EUROPOP2019 provides regional (NUTS\,3) detail but was calibrated before the Ukrainian refugee crisis and the post-pandemic migration surge. EUROPOP2023 incorporates these shocks but is published only at the national level, with no regional breakdown. A simple proportional downscaling of EUROPOP2023 using pre-crisis regional shares would misallocate migrants: regions that received large refugee inflows in 2022--2023 (such as Vienna) would not be distinguished from regions with historically stable migration patterns. Conversely, using EUROPOP2019 regional structure without updating the national totals would ignore the scale of recent shocks.

To address this, we decompose projected net migration into two additive components: a \emph{structural baseline} that captures long-run regional attractiveness, and a \emph{national-level correction} that captures the divergence between the two projection vintages.
}

{
\paragraph{Additive decomposition.}
Let $M_{r,c,t}$ denote projected net migration for region $r$, demographic cohort $c$ (defined by single-year age and sex), and year $t$. We write:
\begin{equation}\label{eq:decomposition}
    M_{r,c,t} = B_{r,c,t} + E_{r,c,t},
\end{equation}
where $B_{r,c,t}$ is the structural baseline taken directly from the EUROPOP2019 NUTS\,3 projection, and $E_{r,c,t}$ is the regionalised correction derived from the gap between EUROPOP2023 and EUROPOP2019 at the national level. Specifically, we first compute the national-level divergence:
\begin{equation}\label{eq:national_delta}
    \Delta_{c,t} = M^{\text{EP2023}}_{c,t} - \sum_{r} B_{r,c,t},
\end{equation}
where $M^{\text{EP2023}}_{c,t}$ is the EUROPOP2023 national projection for cohort $c$ in year $t$, and $\sum_r B_{r,c,t}$ is the national total implied by summing the EUROPOP2019 regional baselines. The regionalised correction is then:
\begin{equation}\label{eq:regional_correction}
    E_{r,c,t} = \Delta_{c,t} \, \lambda^{\text{fin}}_{r,c,t},
\end{equation}
where $\lambda^{\text{fin}}_{r,c,t}$ is a spatial distribution key described below. By construction $\sum_r \lambda^{\text{fin}}_{r,c,t} = 1$, so the combined regional corrections is the national divergence.
}

{
\paragraph{Structural distribution key.}
To allocate the EUROPOP2019 baseline flows across regions, we compute an intensity-based share $\lambda^{\text{str}}_{r,c,t}$ from the absolute value of net migration rather than from signed flows. The reason is practical: when a cohort experiences net emigration nationally, dividing regional flows by a negative national total produces negative shares and sign reversals. Using absolute values avoids this problem.

We define the migration intensity of region $r$ in cohort $c$ at time $t$ as:
\begin{equation}
    I_{r,c,t} = |B_{r,c,t}|.
\end{equation}
To reduce year-to-year noise, we smooth with a centred 3-year rolling average:
\begin{equation}
    \tilde{I}_{r,c,t} = \frac{1}{3}\sum_{\tau = t-1}^{t+1} I_{r,c,\tau}.
\end{equation}

The structural share is then:
\begin{equation}
    \lambda^{\text{str}}_{r,c,t} = \frac{\tilde{I}_{r,c,t}}{\sum_{r'} \tilde{I}_{r',c,t}}.
\end{equation}

For cohorts where the national total intensity is very small ($\sum_r \tilde{I}_{r,c,t} < 10$~persons), we fall back to a uniform distribution ($1/R$, where $R=35$ is the number of NUTS\,3 regions) to avoid numerically unstable shares.
}

{
\paragraph{Crisis-period distribution key.}
The structural key reflects long-run regional attractiveness but cannot capture the distinct spatial pattern of the 2022--2023 migration surge, which was dominated by the settlement of Ukrainian refugees. We therefore construct a separate distribution key $\lambda^{\text{ref}}_{r,c}$ from observed migration during the calibration period $\mathcal{T} = \{2022, 2023, 2024\}$.

This key is a weighted combination of two empirical spatial distributions: the settlement pattern of Ukrainian nationals and that of all other foreign nationals. The idea is that different origin groups settle differently: Ukrainian refugees were initially concentrated in larger cities and reception centres, whereas other foreign migrants follow longer-established labour-market and network channels.

We compute the spatial distribution for each group using only positive net migration (i.e., inflow regions), weighted by calibration-period temporal weights $w_t$:
\begin{equation}
    \psi^{\text{Ukr}}_{r,c} = \frac{\sum_{t \in \mathcal{T}} w_t \, M^{+,\text{Ukr}}_{r,c,t}}{\sum_{r'}\sum_{t \in \mathcal{T}} w_t \, M^{+,\text{Ukr}}_{r',c,t}}, \qquad
    \psi^{\text{Oth}}_{r,c} = \frac{\sum_{t \in \mathcal{T}} w_t \, M^{+,\text{Oth}}_{r,c,t}}{\sum_{r'}\sum_{t \in \mathcal{T}} w_t \, M^{+,\text{Oth}}_{r',c,t}},
\end{equation}
where $M^{+,g}_{r,c,t} = \max(M^{g}_{r,c,t}, 0)$ denotes the positive part of net migration for group $g$. The restriction to positive flows isolates regions experiencing net inflow pressure, which is the relevant signal for allocating future inflows.

The two distributions are blended using a cohort-specific weight $\alpha_c$ that reflects the Ukrainian share of total foreign inflows for that age--sex group:
\begin{equation}
    \alpha_c = \frac{\sum_r \sum_{t \in \mathcal{T}} w_t \, M^{+,\text{Ukr}}_{r,c,t}}{\sum_r \sum_{t \in \mathcal{T}} w_t \bigl(M^{+,\text{Ukr}}_{r,c,t} + M^{+,\text{Oth}}_{r,c,t}\bigr)}.
\end{equation}

The composite crisis key is then:
\begin{equation}\label{eq:refugee_key}
    \lambda^{\text{ref}}_{r,c} = \alpha_c \, \psi^{\text{Ukr}}_{r,c} + (1 - \alpha_c)\,\psi^{\text{Oth}}_{r,c}.
\end{equation}

Two implementation details are worth noting. First, the Ukrainian migration data from Statistik Austria are published in 5-year age groups \cite{statistikaustria2025internal, statistikaustria2025international}. To produce single-year-of-age distributions consistent with the rest of the model, we expand these to single ages using monotone piecewise cubic Hermite interpolation (PCHIP), applied to the cumulative distribution to guarantee non-negative counts that sum to the original group total. Second, the crisis key is computed at the 5-year age-group level and then applied uniformly to all single ages within each group, since the finer Ukrainian data do not support single-year estimation. The structural key, by contrast, is computed directly at the single-year level from EUROPOP2019.
}

{
\paragraph{Time-varying blending of keys.}
The final distribution key $\lambda^{\text{fin}}_{r,c,t}$ blends the crisis key and the structural key using a time-varying weight $\Omega(t)$ that governs how quickly the spatial allocation of the correction term transitions from crisis-driven to structural:
\begin{equation}\label{eq:final_key}
    \lambda^{\text{fin}}_{r,c,t} = \Omega(t)\,\lambda^{\text{ref}}_{r,c} + \bigl(1 - \Omega(t)\bigr)\,\lambda^{\text{str}}_{r,c,t},
\end{equation}
with:
\begin{equation}\label{eq:omega}
    \Omega(t) = \begin{cases}
        1 & \text{if } t \le 2026, \\[4pt]
        \displaystyle\frac{2035 - t}{2035 - 2026} & \text{if } 2026 < t < 2035, \\[6pt]
        0 & \text{if } t \ge 2035.
    \end{cases}
\end{equation}

This creates three regimes. Up to 2026, the correction is allocated entirely by the crisis key, so that any reversal or continuation of crisis-era flows is concentrated in the most affected regions. Between 2027 and 2034, the weight shifts linearly from the crisis pattern to the structural pattern, reflecting the assumption that the spatial logic of migration gradually returns to long-run economic fundamentals. From 2035 onward, allocation follows only the structural key (Supplementary Figure \ref{fig:omega_decay}). After blending, the final key is renormalised to sum to~1 across regions for each cohort and year.

\begin{figure}[htbp]
    \centering
    \includegraphics[width=\textwidth]{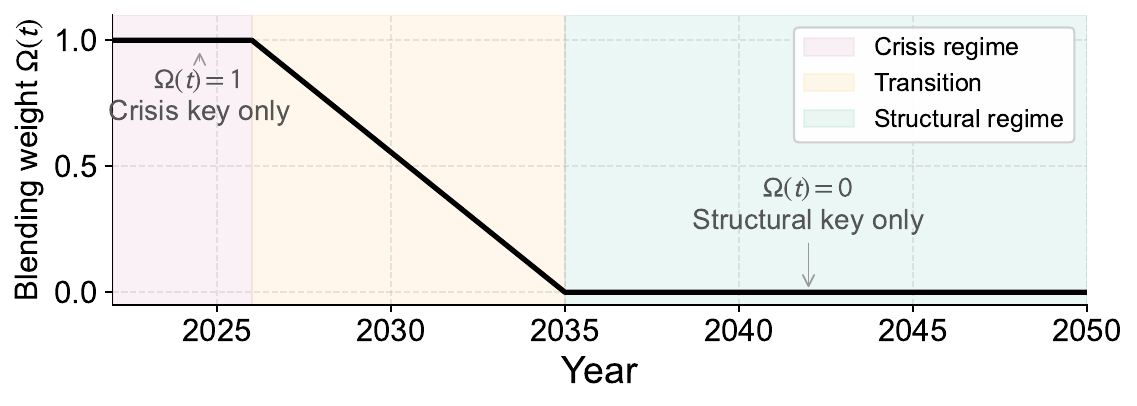}
    \caption{Illustration of the blending weight $\Omega(t)$. Before 2026, the national correction is allocated entirely using the crisis-period settlement pattern. After 2035, allocation follows the long-run structural key from EUROPOP2019.}
    \label{fig:omega_decay}
\end{figure}
}

{
\paragraph{Decomposition into Austrian and foreign components.}
Having distributed the total correction to regions, we next decompose the baseline $B_{r,c,t}$ into Austrian ($M^{\text{Aut}}_{r,c,t}$) and foreign ($M^{\text{For}}_{r,c,t}$) components. A proportional split based on historical nationality shares would fail in regions where Austrian and foreign migration move in opposite directions. For example, Vienna during 2022--2024 experienced simultaneous Austrian net outflows (approximately $-6{,}000$ per year) and foreign net inflows ($+20{,}000$ per year). Applying a fixed share to the positive total would force the model to predict Austrian inflows, which would contradict the data.

Instead, we model the Austrian component as a weighted historical average that responds to future changes in total migration with region-specific elasticity:
\begin{equation}\label{eq:austrian_component}
    M^{\text{Aut}}_{r,c,t} = \kappa \, \bar{M}^{\text{Aut}}_{r,c} + \beta_r \, \bigl[B_{r,c,t} - B_{r,c,2024}\bigr].
\end{equation}

The first term sets the baseline level of Austrian migration. $\bar{M}^{\text{Aut}}_{r,c}$ is the weighted average of observed Austrian net migration during 2022--2024, computed as:
\begin{equation}
    \bar{M}^{\text{Aut}}_{r,c} = \frac{\sum_{t \in \mathcal{T}} w_t \, M^{\text{Aut,obs}}_{r,c,t}}{\sum_{t \in \mathcal{T}} w_t},
\end{equation}
where $w_t$ are temporal weights (see Section~\ref{sec:sensitivity}). This average is smoothed over age using a 7-year centred rolling mean to reduce noise in thin cells. The multiplier $\kappa \in \{0.8, 1.0\}$ scales this historical average to account for the possibility that the calibration period (which includes crisis-driven peaks) may overstate the long-run structural level of Austrian migration.

The second term introduces a dynamic response: $\beta_r$ measures the change in Austrian net migration in region $r$ per unit change in total net migration. We estimate $\beta_r$ from historical data using Theil-Sen regression (regressing Austrian net migration on total net migration across years), which is robust to outliers from crisis periods (for example, the European refugee crisis in 2015 or the Ukrainian refugee crisis in 2022). $B_{r,c,2024}$ serves as the reference point, so the correction term is zero at the start of the projection and grows with the projected deviation of total migration from its 2024 level.

We estimate $\beta_r$ using three alternative time windows, reflecting different assumptions about which historical period best represents future behaviour:
\begin{itemize}
    \item \emph{Pre-crisis} (2010--2019): excludes all crisis periods; mean $\beta = 0.28$.
    \item \emph{Crisis era} (2015--2024): spans both the European 2015 and the Ukrainian 2022 refugee crises; mean $\beta = 0.20$.
    \item \emph{Post-COVID} (2018--2024): captures recent structural shifts; mean $\beta = 0.07$.
\end{itemize}
The decline in mean $\beta$ across windows reflects a growing decoupling of Austrian migration from foreign-dominated surges: as refugee-driven flows increased their share of total migration, Austrian migration became less responsive to overall fluctuations. We retain all three windows as separate scenarios rather than selecting a single estimate.
}

{
\paragraph{Foreign component as residual.}
The foreign component is calculated as the remainder:
\begin{equation}\label{eq:foreign_residual}
    M^{\text{For}}_{r,c,t} = \bigl[B_{r,c,t} + E_{r,c,t}\bigr] - M^{\text{Aut}}_{r,c,t}.
\end{equation}
This residual construction ensures that the sum of Austrian and foreign migration in each cell exactly equals the total projected migration (which itself is consistent with EUROPOP2023 national totals by construction). If the projected total exceeds the Austrian component's estimated responsiveness, the excess is attributed to foreign nationals.
}

{
\subsubsection{Scenario Structure}\label{sec:sensitivity}

We generate projections under a factorial combination of parameter choices to capture uncertainty from sources that cannot be resolved empirically. The dimensions are:

\begin{enumerate}
    \item \textbf{EUROPOP2023 variant} (3~levels): baseline, high-migration, and low-migration, as published by Eurostat. Each implies a different national trajectory for total net migration.

    \item \textbf{Calibration-period weights} $w_t$ (3~schemes): these govern the temporal weighting of observed 2022--2024 data when computing the crisis distribution key and the Austrian historical average.
    \begin{itemize}
        \item Equal: $w_{2022} = w_{2023} = w_{2024} = 1/3$.
        \item Recent-biased: $w_{2022} = 0.2$, $w_{2023} = 0.3$, $w_{2024} = 0.5$.
        \item Early-biased: $w_{2022} = 0.5$, $w_{2023} = 0.3$, $w_{2024} = 0.2$.
    \end{itemize}

    \item \textbf{Sensitivity estimation window} (3~windows): pre-crisis, crisis-era, and post-COVID, as described above.

    \item \textbf{Structural level multiplier} $\kappa$ (2~levels): 0.8 and 1.0.
\end{enumerate}

For each of the 3 EUROPOP2023 variants, the full cross of dimensions~2--4 yields $3 \times 3 \times 2 = 18$ calibrations, giving $3 \times 18 = 54$ scenarios. In the results, we report medians and full ranges (min--max) across all scenarios, unless otherwise noted (Supplementary Figures \ref{fig:national_fan} and \ref{fig:regional_fans}).

\begin{figure}[htbp]
    \centering
    \includegraphics[width=\textwidth]{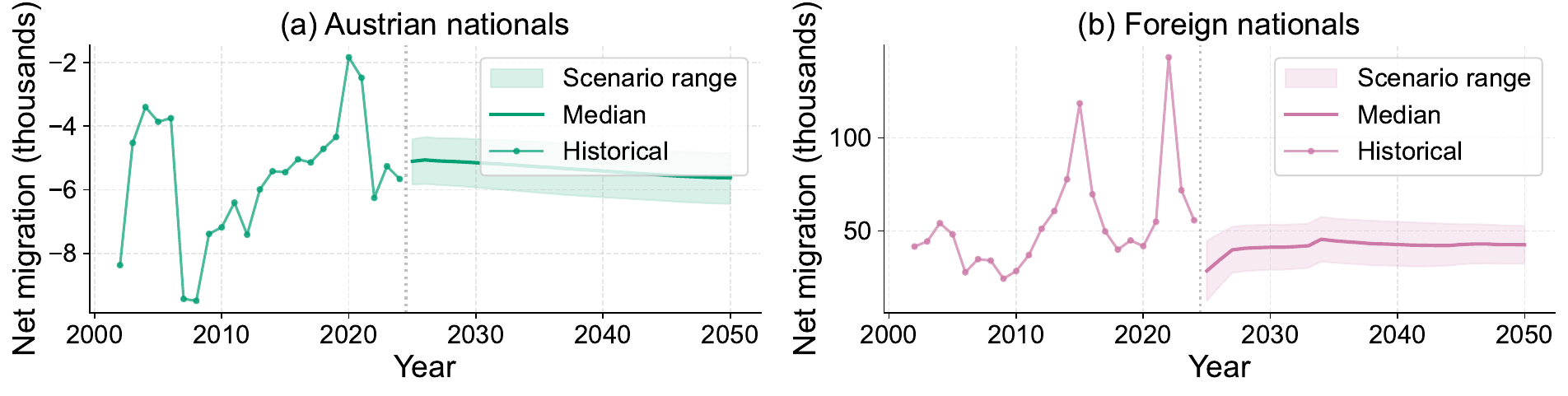}
    \caption{National net migration projections by nationality (2010--2050). Historical data (solid markers, 2010--2024) and projected trajectories (2025--2050) for Austrian nationals (left) and foreign nationals (right). Solid lines: median across all 54~scenarios. Shaded areas: full min--max envelope. The correction in 2025 reflects the modelled transition from calibration-period levels to the EUROPOP2023 trajectory.}
    \label{fig:national_fan}

    \includegraphics[width=\textwidth]{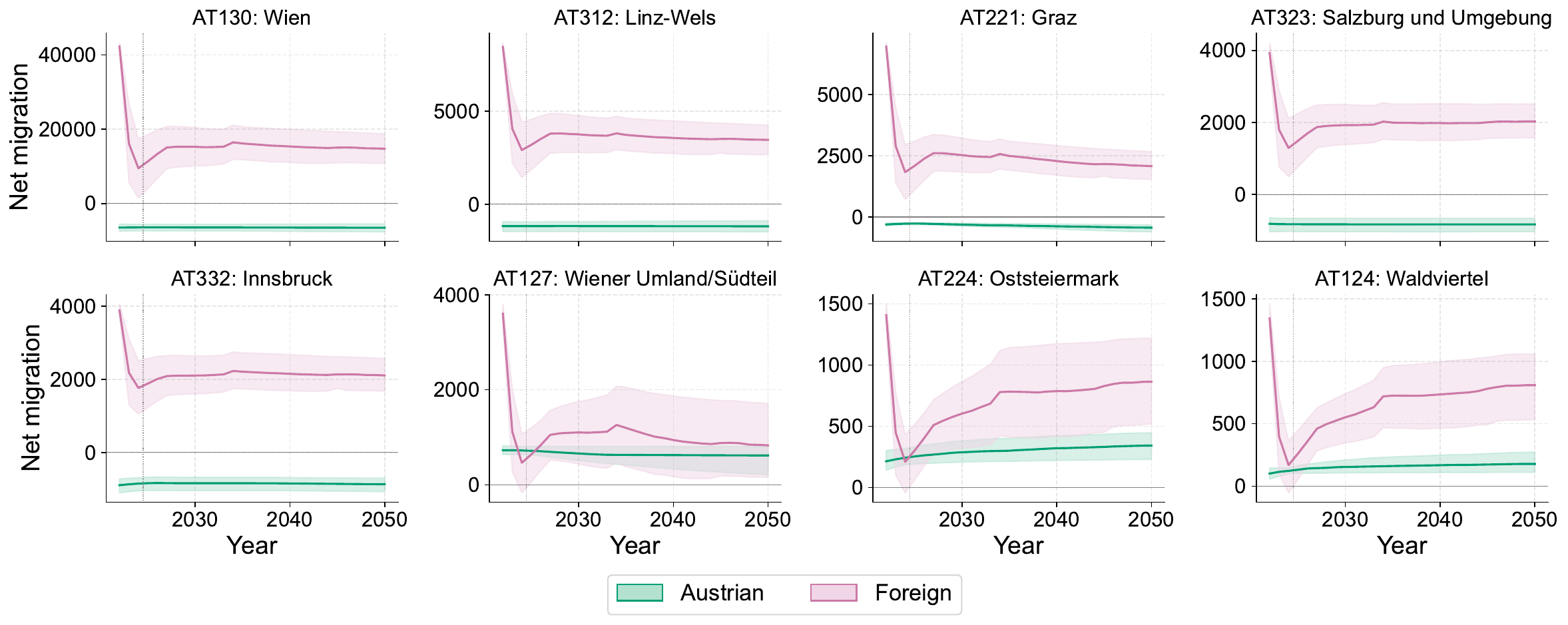}
    \caption{Regional net migration projections (selected NUTS\,3 regions, 2010--2050). Green: Austrian nationals; pink: foreign nationals. Shaded areas: full scenario range. The width of the fan reflects region-specific sensitivity to parameter choices, with regions that experienced large refugee inflows exhibiting greater short-term uncertainty.}
    \label{fig:regional_fans}
\end{figure}
}

{
\subsubsection{Naturalisation}\label{sec:naturalisation}

Naturalisation transfers individuals from the foreign to the Austrian population stock. We model this using administrative data from Statistik Austria on the number of naturalisations granted to residents of Austria during 2022--2024, disaggregated by NUTS\,2 region ($j$), sex ($s$), and coarse age group ($g$) \cite{statistikaustria2025naturalisations}. We exclude naturalisations granted to persons residing abroad (primarily under \S58c~StbG, which restores citizenship to descendants of victims of National Socialism) to isolate the demographic dynamics of the resident population. Over 2022--2024, resident naturalisations averaged approximately 11,900 per year.

We compute the naturalisation rate for each NUTS\,2--sex--age-group cell as:
\begin{equation}
    \hat{\rho}_{j,s,g} = \frac{\bar{N}_{j,s,g}}{\sum_{a \in g} P_{j,s,a}},
\end{equation}
where $\bar{N}_{j,s,g}$ is the average annual number of naturalisations in region $j$, sex $s$, age group $g$ during 2022--2024, and $P_{j,s,a}$ is the 2024 foreign-national population at single-year age $a$. To obtain the fine-grained rates required by the projection model (single ages within NUTS\,3 regions), we apply the NUTS\,2-level rate uniformly to all single ages within the corresponding age group and to all NUTS\,3 sub-regions within the parent NUTS\,2 region:
\begin{equation}
    \rho_{r,a,s} = \hat{\rho}_{j(r),\, s,\, g(a)},
\end{equation}
where $j(r)$ denotes the NUTS\,2 parent of region $r$ and $g(a)$ denotes the age group containing age $a$.

We hold naturalisation rates constant over the projection horizon. This assumption is supported by the historical record: the resident naturalisation rate has fluctuated within a narrow band of 0.6--0.75\% over the past 15~years \cite{statistikaustria2025naturalisations}. The temporary dip in 2020--2021 is attributable to pandemic-related administrative delays; by 2024, the rate had recovered to approximately 0.72\% (Supplementary Figure \ref{fig:naturalisation}). 

\begin{figure}[htbp]
    \centering
    \includegraphics[width=0.85\linewidth]{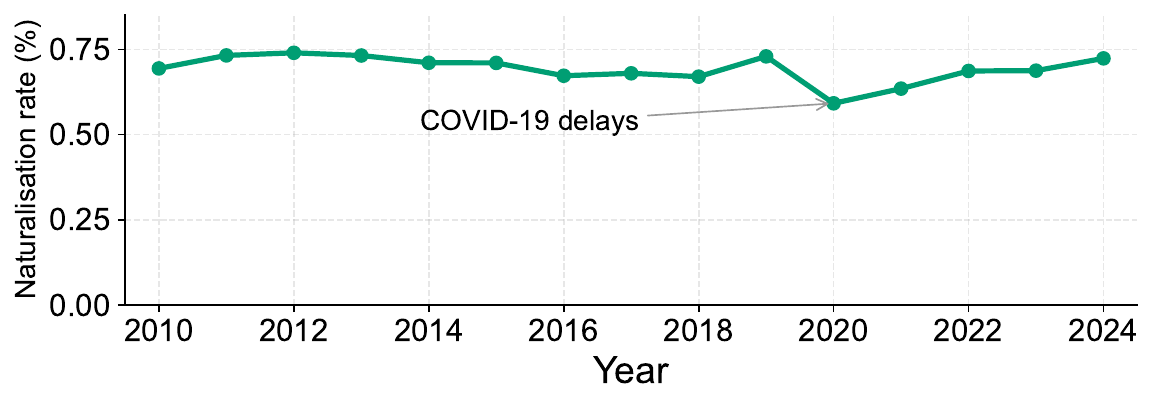}
    \caption{Historical naturalisation rate of residents (2010--2024). Excluding non-resident cases reveals a stable long-term rate of approximately 0.7\%, supporting the assumption of constant rates over the projection horizon.}
    \label{fig:naturalisation}
\end{figure}
}

\subsubsection{Fertility by Nationality}\label{sec:fertility}

{
\paragraph{Calibration to observed Total Fertility Rate.}
The EUROPOP2019 projection baseline implies a total fertility rate (TFR) substantially higher than the value observed in 2023 (1.32 children per woman). To correct this, we scale all baseline age-specific fertility rates (ASFRs) by a constant factor chosen to match the 2023 observation. The scaling factor is applied uniformly across all projection years, preserving the long-term trend shape from EUROPOP2019 while shifting the level downward.
}
{
\paragraph{Nationality-specific fertility rates.}
Austrian and foreign nationals have persistently different fertility levels. To capture this, we define scaling ratios $\phi_{\text{Aut}}(t)$ and $\phi_{\text{For}}(t)$ that express each group's TFR as a fraction of the overall TFR (Supplementary Figure \ref{fig:fertility}):
\begin{equation}
    \phi_{n}(t) = \frac{\text{TFR}_{n}(t)}{\text{TFR}_{\text{tot}}(t)}, \qquad n \in \{\text{Aut}, \text{For}\}.
\end{equation}

Supplementary table \ref{tab:fertility_ratios} reports these ratios for the recent historical period. In 2023, $\phi_{\text{Aut}} \approx 0.932$ (Austrian women have about 7\% lower fertility than average) and $\phi_{\text{For}} \approx 1.182$ (foreign women have about 18\% higher fertility).

\begin{table}[htbp]
    \centering
    \caption{Historical Total Fertility Rates and derived scaling ratios, 2019--2023. Source: Statistik Austria.}
    \label{tab:fertility_ratios}
    \begin{tabular}{lccccc}
        \toprule
        \textbf{Year} & \textbf{TFR Total} & \textbf{TFR Aut.} & \textbf{TFR For.} & $\phi_{\text{Aut}}$ & $\phi_{\text{For}}$ \\
        \midrule
        2019 & 1.46 & 1.35 & 1.85 & 0.925 & 1.267 \\
        2020 & 1.44 & 1.35 & 1.76 & 0.938 & 1.222 \\
        2021 & 1.48 & 1.40 & 1.75 & 0.946 & 1.182 \\
        2022 & 1.41 & 1.33 & 1.64 & 0.943 & 1.163 \\
        2023 & 1.32 & 1.23 & 1.56 & 0.932 & 1.182 \\
        \bottomrule
    \end{tabular}
\end{table}
}

{
\paragraph{Long-run convergence.}
For the projection horizon, we model a gradual convergence of fertility behaviour between groups, reflecting assimilation and compositional changes in the foreign-national population over time. The scaling ratios evolve linearly from their 2023 values to $\phi = 1.0$ (parity) by the year 2100:
\begin{equation}
    \phi_{n}(t) = \phi_{n,2023} + \frac{1 - \phi_{n,2023}}{2100 - 2023}\,(t - 2023).
\end{equation}

Nationality-specific ASFRs for each NUTS\,3 region $r$ and age $a$ are then obtained by applying the national ratio to the calibrated regional baseline (Supplementary Figure \ref{fig:asfr_curves}):
\begin{equation}
    \text{ASFR}_{r,a,n}(t) = \text{ASFR}_{\text{adj},r,a}(t) \, \phi_{n}(t).
\end{equation}

The nationality scaling ratios are estimated at the national level and applied uniformly across NUTS\,3 regions. If the fertility gap between Austrian and foreign women varies across regions, this introduces spatial misspecification.

\begin{figure}[htbp]
    \centering
    \includegraphics[width=\linewidth]{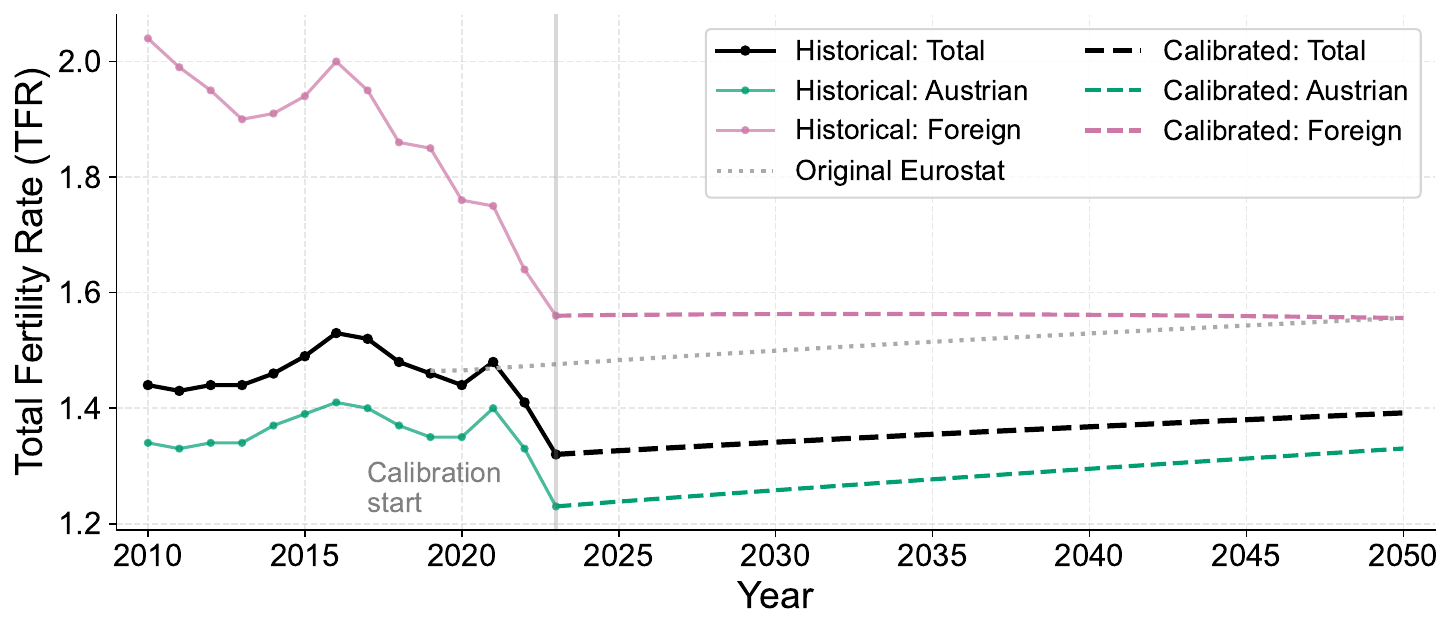}
    \caption{Historical and projected TFR by nationality (2010--2050). Solid lines: historical data. Dotted lines: original (uncorrected) EUROPOP2019 baseline. Dashed lines: calibrated projection. The baseline level is adjusted to match the 2023 observed TFR of 1.32, while the original projection slope is preserved. Nationality-specific rates converge linearly toward parity by 2100.}
    \label{fig:fertility}

    \includegraphics[width=0.7\linewidth]{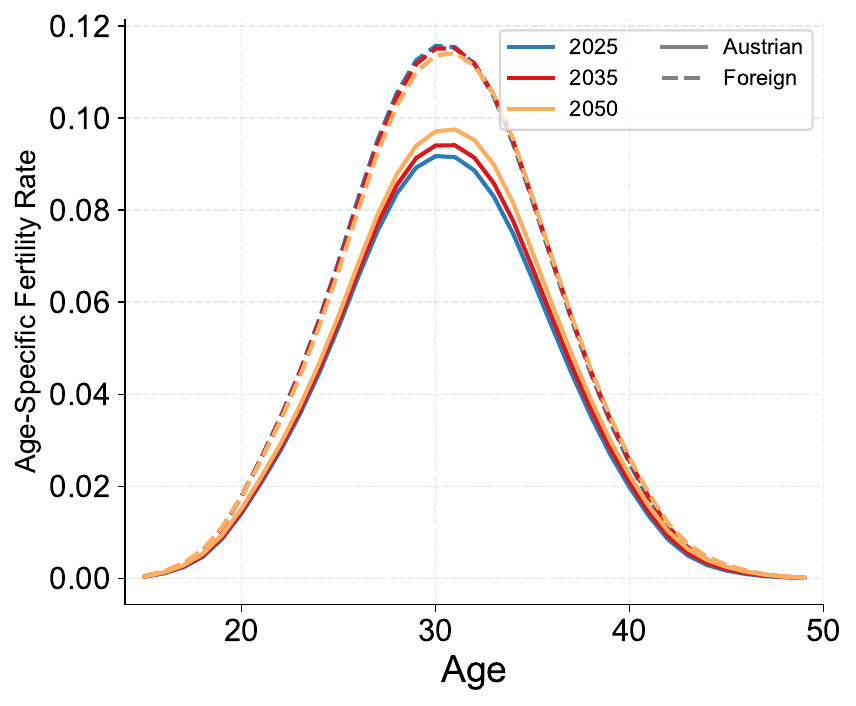}
    \caption{Age-specific fertility rate profiles for Austrian and foreign nationals at selected projection years (e.g., 2025, 2035, 2050), illustrating the gradual convergence of the curves over time.}
    \label{fig:asfr_curves}
\end{figure}
}

\subsubsection{Nationality Assignment to Births}\label{sec:homophily}

{
The projection model must assign a nationality to each newborn. Under Austrian \textit{jus sanginis} citizenship law, a child acquires Austrian citizenship if at least one parent is an Austrian national. We model this by estimating the probability that a child born to a foreign-national mother has an Austrian-national father, which we call the \emph{mix rate}. This probability depends on two quantities: the availability of Austrian-national men in the local partnership market, and the tendency of individuals to form partnerships within their own nationality group.

We define the mix rate for a given region $r$ as:
\begin{equation}\label{eq:mix_rate}
    \text{MixRate}_{r} = \theta^{\text{Aut}}_r \, (1 - h),
\end{equation}
where $\theta^{\text{Aut}}_r$ is the share of Austrian-national men within the ``marriageable'' age range in region $r$, calculated from the total population model, and $h \in [0,1]$ is the \emph{homophily parameter}, the probability that a foreign-national woman forms a partnership with a man of the same (foreign) nationality group. When $h=1$, all partnerships are within-group, and no mixed births occur; when $h=0$, partner choice is random with respect to nationality.

The share $\theta^{\text{Aut}}_r$ is computed each year as the number of Austrian-national men aged 20--65 in region $r$ divided by the total number of men aged 20--65 in the same region. Austrian male counts are obtained by subtracting the simulated foreign male population from the total male population (from the total-population model). Births assigned to Austrian fathers are removed from the foreign-national birth count and added to the Austrian population.
}

{
\paragraph{Parameter estimation.}
We estimated the homophily $h$ and the male age range jointly by grid search, testing all combinations of five age ranges (20--50, 20--55, 20--60, 20--65, 20--70) and homophily values from 0.50 to 0.69 (in steps of 0.01). For each parameter combination, we ran the full projection model and compared simulated regional birth counts at age~0 (by nationality) against observed data from Statistik Austria for 2022--2025. We evaluated the fit using a combined Root Mean Squared Error ($\text{RMSE}_{\text{comb}}$) that equally weights the errors for Austrian and foreign nationalities, while applying heavier temporal weights to recent observations:
\begin{equation}
    \text{RMSE}_{\text{comb}} = 0.5 \, \text{RMSE}_{\text{Aut}} + 0.5 \, \text{RMSE}_{\text{For}},
\end{equation}
with year-specific weights $w_{2022}=0$, $w_{2023}=0.2$, $w_{2024}=0.35$, $w_{2025}=0.45$ (giving more weight to recent observations).

The optimal configuration is an age range of 20--65 with $h=0.61$ (Supplementary Figure \ref{fig:homophily_validation}). This result can be cross-checked against aggregate data by rearranging Equation~\ref{eq:mix_rate} to isolate $h$ and deducing the mix rate through Austrian \textit{jus sanguinis} citizenship laws: 

\begin{equation}\label{eq:homphily_approx}
h = 1 - \frac{\left( \frac{\text{Foreign Mothers} - \text{Foreign Children}}{\text{Foreign Mothers}} \right)}{\left( \frac{\text{Austrian Men aged 20--65}}{\text{Total Men aged 20--65}} \right)}.
\end{equation}

Using aggregate data from 2023 \cite{statistikaustria2025births}, the deduced share of mixed births in the numerator was 30.1\%, and the Austrian male share in the denominator was approximately 78\%. Substituting these values yields $h = 1 - 0.301/0.78 \approx 0.614$, consistent with the grid-search optimum.
}

{
\paragraph{Temporal stability.}
A key question is whether $h$ can be treated as a constant or varies over time as the population composition shifts. We reconstructed the implied homophily index from 2010 to 2024 using Eq. \ref{eq:homphily_approx} with annual birth registry \cite{statistikaustria2025births} and population data \cite{statistikaustria2025population}. Despite the Austrian male share declining from 87.4\% to 77.0\% over this period, and two major migration shocks (Syria 2015, Ukraine 2022), the implied $h$ has remained between 0.57 and 0.63 throughout (Supplementary Figure \ref{fig:homophily_historical}). This stability over 14~years of substantial demographic change supports treating $h$ as a structural parameter rather than a time-varying coefficient.

\begin{figure}[htbp]
    \centering
    \includegraphics[
        width=\linewidth
    ]{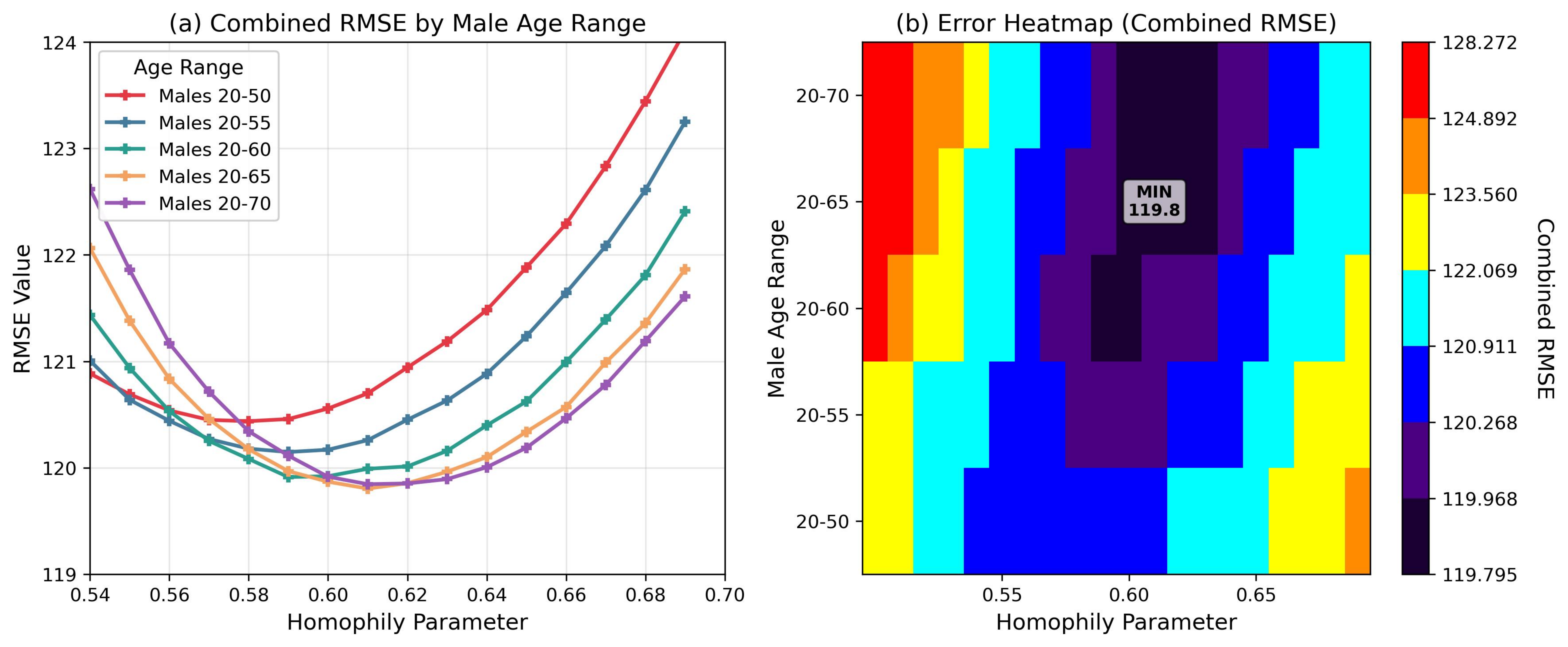}
    \caption{
        Grid-search results for the homophily parameter. The combined RMSE is 
        minimised at $h=0.61$ within the 20--65 male age range.
    }
    \label{fig:homophily_validation}

    \includegraphics[
        width=0.75\linewidth
    ]{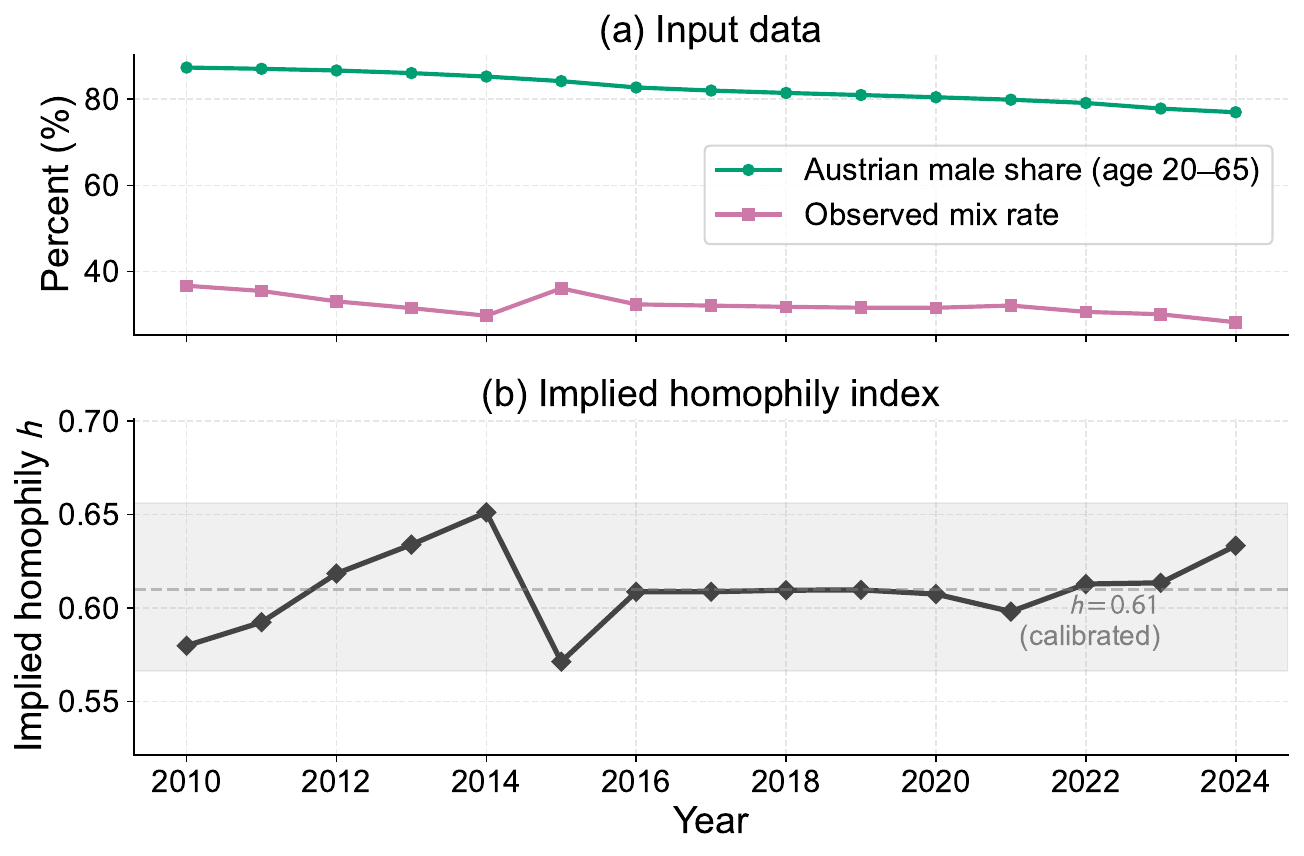}
    \caption{
        Reconstructed homophily index (2010--2024) from the annual birth registry 
        data, showing stability in the range 0.57--0.63 despite large changes 
        in population composition.
    }
    \label{fig:homophily_historical}
\end{figure}
}

{

\subsubsection{Austrian Population as Residual}\label{sec:residual}

The Austrian-national population at the cell level (region $\times$ sex $\times$ age $\times$ year) is obtained by subtracting the simulated foreign-national population from the total population produced by a separate total-population projection:
\begin{equation}
    P^{\text{Aut}}_{r,s,a,t} = P^{\text{Tot}}_{r,s,a,t} - P^{\text{For}}_{r,s,a,t}.
\end{equation}

The total-population projection uses the same cohort-component structure but does not distinguish nationality: it applies total (nationality-blind) mortality, fertility, and migration rates from the EUROPOP framework. This residual approach ensures that the Austrian and foreign populations sum exactly to the total in every cell.

Because the foreign and total projections are run independently with different input parameters, the residual can be negative in some cells, particularly at older ages in regions with high recent foreign inflows, where the simulated foreign stock may exceed the total. We handle negative residuals using the same Gaussian redistribution procedure described in the overview ($\sigma=5$~years), drawing the deficit from nearby positive-valued Austrian cells in the same region and sex. After redistribution, there are no remaining negative values that need to be set to zero. The resulting decomposition of total population into Austrian and foreign-national components is illustrated in Supplementary Figures~\ref{fig:population_projections} and \ref{fig:spatial_age_structure}.

\begin{figure}[htbp]
    \centering
    \includegraphics[width=0.9\textwidth]{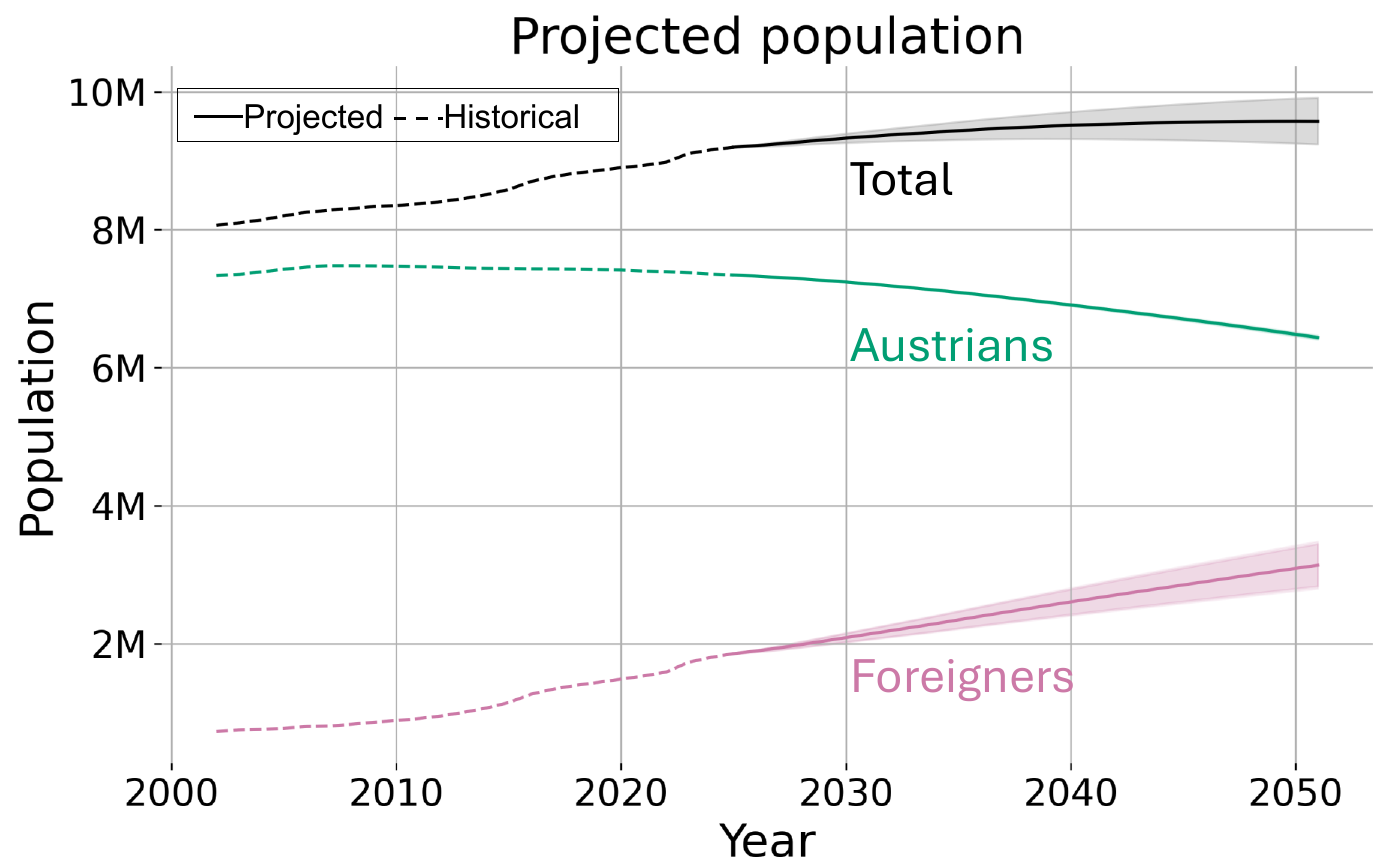}

    \vspace{0.8em}

    \includegraphics[width=\textwidth]{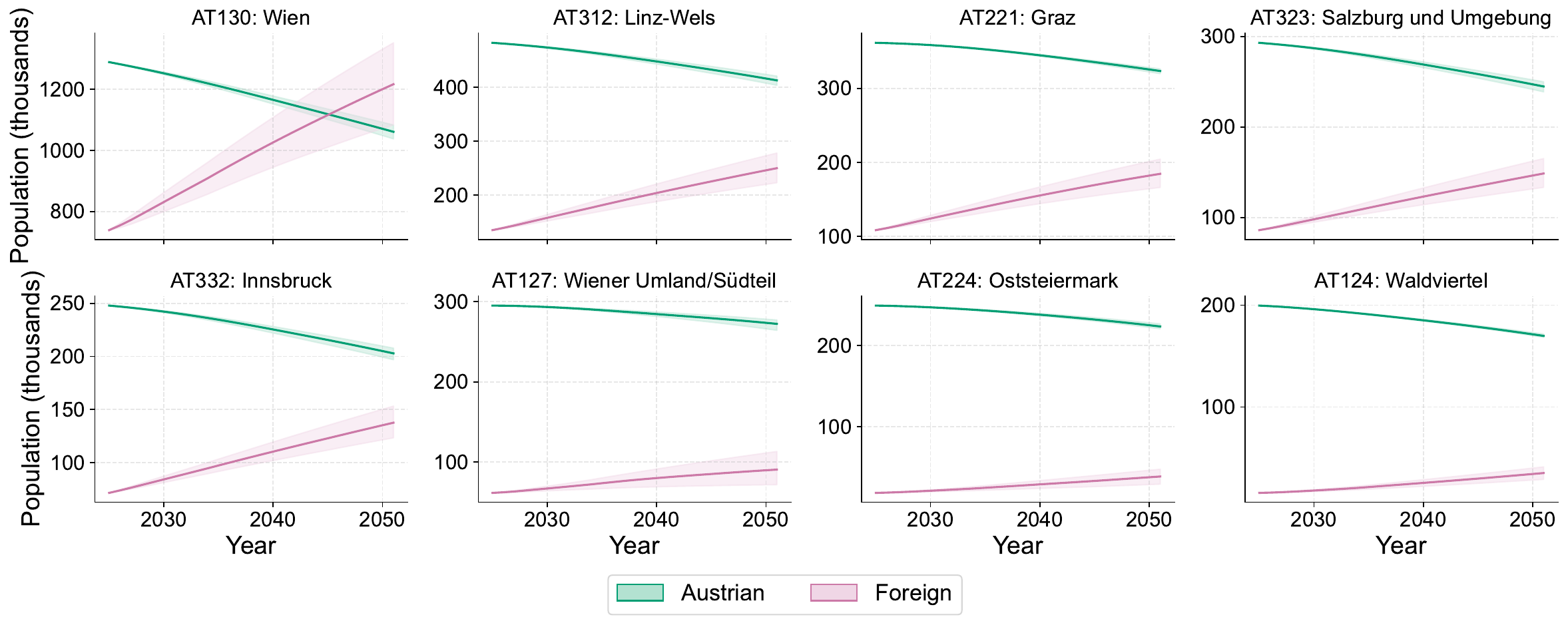}

    \caption{
    Population projections by nationality. 
    Top: National population projections (2025--2050) with historical values (2002--2024), showing Austrian (green) and foreign national (pink) populations. Shaded areas indicate the scenario range. 
    Bottom: Regional population projections for selected NUTS\,3 regions (2025--2050), decomposed by nationality.
    }
    \label{fig:population_projections}
\end{figure}

\begin{figure}[htbp]
    \centering
    \includegraphics[width=\textwidth]{figures/supp_methods/population_pct_change_map_2025_2050.pdf}

    \vspace{0.8em}

    \includegraphics[width=\textwidth]{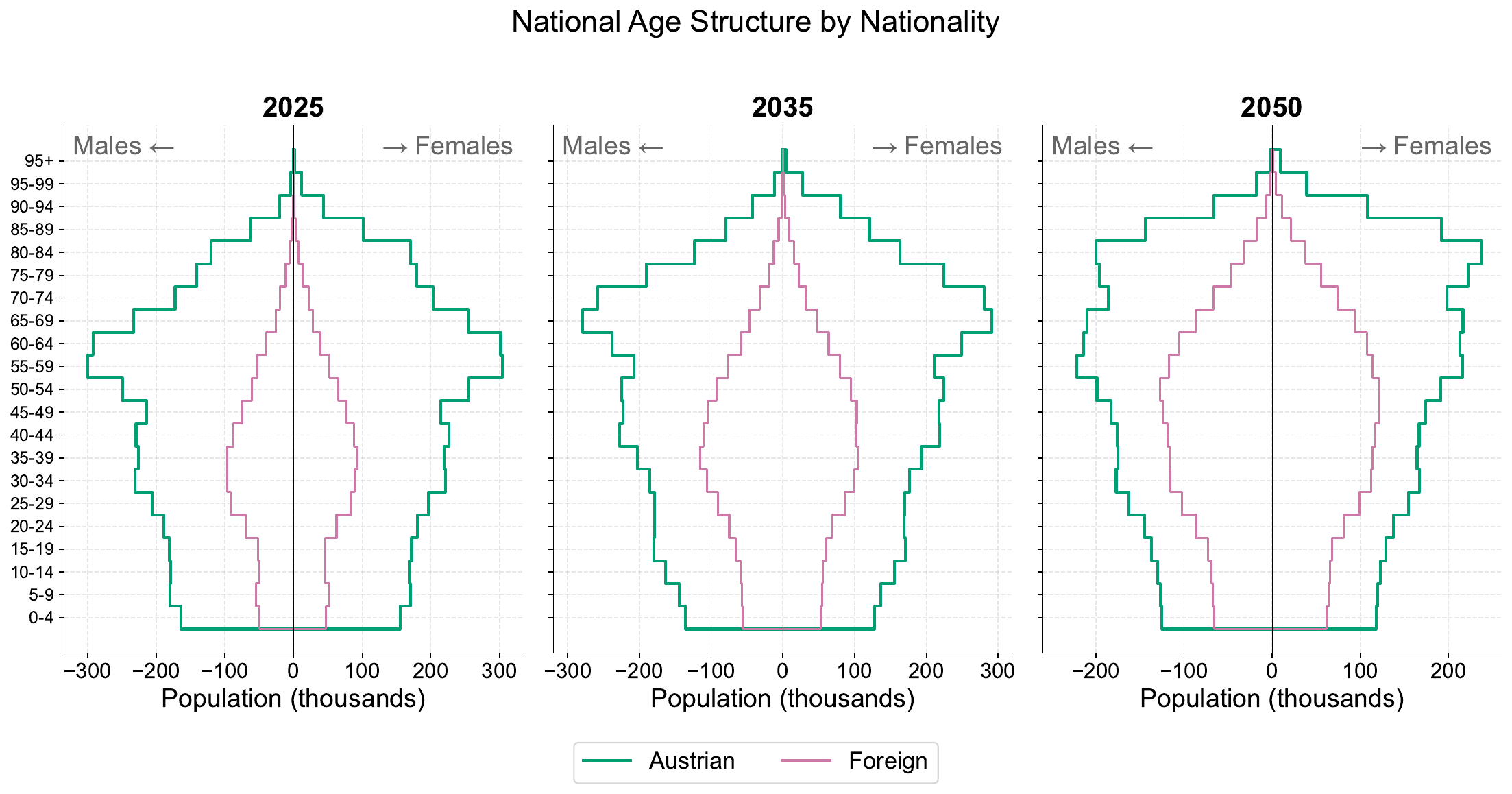}

    \caption{
    Projected spatial and demographic structure of the population.
    Top: Spatial distribution of projected population change between 2025 and 2050, disaggregated by Austrian and foreign nationals at the NUTS\,3 level. To ensure visual clarity and mitigate the influence of outliers, absolute change values are capped at the robust 95th percentile. Reported changes correspond to median projections across the migration-scenario ensemble.
    Bottom: Age structure of the projected population at selected years (e.g., 2025, 2035, 2050), decomposed by nationality, illustrating the younger age profile of the foreign-national population and its contribution to working-age cohorts.
    }
    \label{fig:spatial_age_structure}
\end{figure}
}

\subsection{Housing Demand: Technical Details}

\subsubsection{Data}

{
Three datasets from Statistik Austria serve as the empirical basis for the housing demand model. First, \emph{population data} provide annual counts of residents by NUTS\,3 region, single-year age (0--100), sex (male, female), and nationality (Austrian, foreign), and are available for the census years 2011, 2021, and 2022 \cite{statistikaustria2025population}. Second, the \emph{housing census} (Geb\"aude- und Wohnungsz\"ahlung) records the number of occupied dwellings classified by the demographic characteristics of the household reference person---specifically NUTS\,3 region, broad age group, sex, and nationality---for the same three years \cite{statistikaustria2025housing}. Third, \emph{institutional (non-private) household data} report the population residing in group quarters (e.g.\ nursing homes, student dormitories, correctional facilities) by NUTS\,3 region, five-year age group, sex, and nationality. The institutional data are available for 2012, 2021 \cite{statistikaustria2025registera}, and 2022 \cite{statistikaustria2025registerb}; the 2012 vintage is treated as corresponding to the 2011 census. In addition, household-size projections published by the Austrian Conference on Spatial Planning (\"OROK) provide region-level targets for average household size through 2050 and are used as external constraints during forecasting \cite{orok2024osterreichische}.
}

{
Dwellings recorded without a usual resident (age coded as ``Not applicable'') lack a household reference person and are excluded. Records with nationality flagged as ``Not applicable'' are dropped.
The resulting dwelling counts are aggregated by year, NUTS\,3 region, age group of the reference person, sex, and nationality. The institutional-household file is processed similarly, with one additional temporal adjustment: because the 2011 housing census used 2012 data to derive its institutional population figures, the year label is remapped from 2012 to 2011 to ensure temporal alignment with the census population. The age classification in this file uses five-year groups rather than single years, and these labels are retained for subsequent aggregation with the population data.
}

\subsubsection{Derived Variables}

{
The primary outcome of the housing demand model is the \emph{number of occupied dwellings} in each NUTS\,3 region and projection year. In this framework, one dwelling corresponds to one private household, so dwelling counts are equivalent to household counts. It is important to distinguish this quantity from the number of \emph{houses} (i.e.\ residential buildings), since a single building may contain multiple dwellings. The household count for a given demographic cell is obtained as:
\begin{equation}
    H_{r,a,s,n,t} \;=\; P_{r,a,s,n,t} \, \frac{\mathrm{HHFR}_{r,a,s,n,t}}{100},
    \label{eq:housing_dwellings}
\end{equation}
where $H$ denotes households (occupied dwellings), $P$ denotes population, $\mathrm{HHFR}$ is the Household Formation Rate (expressed as a percentage), and the subscripts index NUTS\,3 region ($r$), single-year age ($a$), sex ($s$), nationality ($n$), and year ($t$). Two intermediate quantities---the Group Quarters Rate and the headship rate---must be derived before $\mathrm{HHFR}$ can be computed.
}

{
The \emph{Group Quarters Rate} (GQR) measures the share of each demographic cohort living in institutional (non-private) households. At the age-group level, it is defined as:
\begin{equation}
    \mathrm{GQR}_{r,g,s,n,t} \;=\; \frac{P^{\mathrm{inst}}_{r,g,s,n,t}}{P^{\mathrm{total}}_{r,g,s,n,t}},
    \label{eq:gqr_agegroup}
\end{equation}
where $P^{\mathrm{inst}}$ is the institutional population and $P^{\mathrm{total}}$ is the total population within five-year age group $g$. Because population data are available at single-year resolution but institutional data use five-year groups, B-spline interpolation is used to produce smooth, age-specific GQR values across ages 0--100. Specifically, for each combination of year, NUTS\,3 region, sex, and nationality, the midpoints of the five-year age groups serve as knots, and the corresponding group-level GQR values serve as ordinates. A B-spline of degree $k = \min(3, m-1)$, where $m$ is the number of knots with positive population, is fitted. The resulting values are clipped to $[0,\,1]$. When the spline fit fails (e.g.\ due to insufficient knots), a linear interpolation fallback is employed; cohorts with only a single knot receive a constant GQR equal to that group's observed rate. The interpolated GQR is then applied to the single-year population to estimate the institutional and private-household populations at each age:
\begin{equation}
    P^{\mathrm{inst}}_{r,a,s,n,t} = P^{\mathrm{total}}_{r,a,s,n,t} \, \mathrm{GQR}_{r,a,s,n,t}, \qquad
    P^{\mathrm{priv}}_{r,a,s,n,t} = P^{\mathrm{total}}_{r,a,s,n,t} \, (1 - \mathrm{GQR}_{r,a,s,n,t}).
\end{equation}
}

{
The private-household population ($P^{\mathrm{priv}}$) is obtained by subtracting the institutional population from the total population, aggregated to the broad age groups used by the dwelling census. The population data (available at single-year ages) are first aggregated into the six age groups used in the census: under 15, 15--29, 30--49, 50--64, 65--84, and 85 and over. The institutional-household data, originally in five-year groups, are re-binned into the same six categories. Both aggregates are joined by year, NUTS\,3 region, age group, sex, and nationality, and the private-household population is computed as $P^{\mathrm{priv}} = P^{\mathrm{total}} - P^{\mathrm{inst}}$. For the 2011 census year, the total population already excludes institutional residents, so no subtraction is applied.
}

{
The \emph{headship rate} is defined as the ratio of occupied dwellings to the private-household population within each demographic cell:
\begin{equation}
    \mathrm{HR}_{r,g,s,n,t} \;=\; \frac{H_{r,g,s,n,t}}{P^{\mathrm{priv}}_{r,g,s,n,t}} \; 100,
    \label{eq:headship_rate}
\end{equation}
where the subscript $g$ denotes the broad census age group. Because the dwelling census reports only broad age groups, whereas the model requires rates at single-year resolution, a cubic-spline interpolation is applied. The midpoints of the six census age groups serve as knots (i.e.\ 7, 22, 39.5, 57, 74.5, 92.5), and the natural boundary condition is imposed. The spline is evaluated at each integer age from 15 to 100, and the resulting rates are clipped to $[0,\,100]$. Ages below 15 are assigned a headship rate of 0, since children do not form independent households in the Austrian data. Headship rates are computed at multiple levels of aggregation, from the fully disaggregated level (year $\times$ NUTS\,3 $\times$ age group $\times$ sex $\times$ nationality) down to single-variable margins, to support both the main forecast and diagnostic analyses.
}

{
The \emph{Household Formation Rate} (HHFR) combines the GQR and the headship rate into a single per-person probability of heading a private household:
\begin{equation}
    \mathrm{HHFR}_{r,a,s,n,t} \;=\; \bigl(1 - \mathrm{GQR}_{r,a,s,n,t}\bigr) \; \mathrm{HR}_{r,a,s,n,t},
    \label{eq:hhfr}
\end{equation}
where the GQR is at single-year age resolution (from the B-spline), and the headship rate $\mathrm{HR}$ is at single-year resolution (from the cubic spline). The HHFR is thus interpretable as the expected number of dwellings per person: an individual first survives the risk of being in an institutional household (factor $1 - \mathrm{GQR}$), and conditional on being in a private household, heads that household with probability $\mathrm{HR}/100$.
}

\subsubsection{Modelling Framework}

{
The housing demand model projects future dwelling counts by forecasting the HHFR forward in time and applying it to exogenous population projections. The forecasting proceeds in four stages: (i) estimation of cohort-specific HHFR trends in logit space, (ii) application of these trends to individual demographic cohorts, (iii) enforcement of external household-size constraints from \"OROK projections, and (iv) smoothing of age-profile discontinuities. Two demand scenarios are produced for each population projection: a \emph{trend} scenario, in which HHFR evolves according to the estimated historical trajectory, and a \emph{status quo} scenario, in which HHFR is held constant at its most recently observed level.
}

{
Because the HHFR is bounded between 0 and 100 (expressed as a percentage), trends are estimated in logit space to prevent extrapolation from producing infeasible values. The logit transform maps a rate $\mathrm{HHFR} \in (0,\,100)$ to the real line:
\begin{equation}
    \ell \;=\; \log\!\left(\frac{p}{1 - p}\right), \text{ \, with}\qquad p = \frac{\mathrm{HHFR}}{100},
    \label{eq:logit}
\end{equation}
with $p$ clipped to $[10^{-6},\, 1 - 10^{-6}]$ to avoid numerical singularities. The inverse (sigmoid) transform recovers a valid rate:
\begin{equation}
    \mathrm{HHFR} \;=\; \frac{100}{1 + \exp(-\ell)}.
    \label{eq:sigmoid}
\end{equation}
}

{
To obtain stable trends robust to small-sample noise at the fully disaggregated level, demographic cohorts are pooled across two dimensions before trend estimation. First, NUTS\,3 regions are grouped into five housing market clusters ($v$) based on the \"OROK (\textit{\"Osterreichische Raumordnungskonferenz}) classification \cite{orok2024osterreichische}: Vienna, urban agglomerations, regional centres, rural areas with good accessibility, and peripheral rural areas. This hierarchical structure enables robust trend estimation through regional pooling while preserving within-cluster heterogeneity. Second, single-year ages are grouped into eight five-year bands from 15--19 through 95--100, with ages 0--14 excluded (HHFR\,=\,0 according to Statistik Austria). Trends are then estimated for each combination of cluster ($v$), age group ($g$), sex ($s$), and nationality ($n$). For each such group, the trend slope in logit space is computed as:
\begin{equation}
    \beta_{v,g,s,n} \;=\; \frac{\ell_{v,g,s,n}^{(2021.5)} \;-\; \ell_{v,g,s,n}^{(2011)}}{2021.5 - 2011},
    \label{eq:trend_slope}
\end{equation}
where $\ell^{(2011)}$ is the population-weighted average logit-HHFR of the group in 2011, and $\ell^{(2021.5)}$ is the corresponding average over the pooled 2021--2022 data. Population-weighted averaging ensures that larger cohorts dominate the rate estimate within each group.
}

{
Each cohort (defined by NUTS\,3 region, single-year age, sex, and nationality) receives a base HHFR equal to its population-weighted average over 2021 and 2022. The forecasted HHFR for target year $t^{*}$ is then:
\begin{equation}
    \mathrm{HHFR}_{r,a,s,n}^{(t^{*})} \;=\; \sigma\!\left(\ell_{r,a,s,n}^{(\mathrm{base})} \;+\; \beta_{v(r),g(a),s,n} \; (t^{*} - 2021.5)\right),
    \label{eq:forecast_hhfr}
\end{equation}
where $\sigma(\,)$ denotes the sigmoid function (Eq.~\ref{eq:sigmoid}), $v(r)$ is the cluster to which region $r$ belongs, and $g(a)$ is the age group containing age $a$. The trend slope $\beta$ thus shifts the cohort's logit-HHFR linearly, while the sigmoid ensures that the result remains in $(0,\,100)$. Cohorts for which no trend can be estimated (e.g.\ due to missing 2011 data) retain their base HHFR unchanged.
}

{
After the initial trend-based forecast, the projected household sizes are compared against \"OROK regional targets \cite{orok2024osterreichische}. For each NUTS\,3 region, the implied average household size $\bar{Z}_r$ is computed as the ratio of the projected private-household population to the projected number of households. A multiplicative adjustment factor is derived as $\kappa_r = \bar{Z}_r \,/\, \bar{Z}_r^{(\mathrm{OEROK})}$, where $\bar{Z}_r^{(\mathrm{OEROK})}$ is the \"OROK target for the corresponding year, and is clipped to $[0.1,\, 5.0]$ to prevent extreme corrections.
Each cohort's HHFR within the region is then scaled by $\kappa_r$ and re-clipped to $[0,\,100]$:

\begin{equation}
    \mathrm{HHFR}_{r,a,s,n}^{(\mathrm{adj})} \;=\; \min\!\Bigl(100,\;\; \mathrm{HHFR}_{r,a,s,n}^{(t^{*})} \; \kappa_r\Bigr).
    \label{eq:oerok_adjustment}
\end{equation}

This procedure is applied at three key years (2030, 2040, 2050), and HHFR values for intermediate years are obtained by linear interpolation across these anchor points.
}

{
Because headship rates are estimated within broad age groups and then assigned to individual ages via spline interpolation, the resulting HHFR age profiles may exhibit residual discontinuities at group boundaries. A centred rolling-average smoother with a window of three ages is applied to each cohort's age profile (holding region, sex, nationality, and year fixed) to attenuate these artefacts. The smoothed values are clipped to $[0,\,100]$.
}

{
Final dwelling demand is computed by applying Eq.~\ref{eq:housing_dwellings} to each cell of the population projection. Two HHFR scenarios are crossed with the available population-projection scenarios. In the \emph{trend} scenario, HHFR evolves as described above (Eq.~\ref{eq:forecast_hhfr} with \"OROK constraints). In the \emph{status quo} scenario, HHFR is frozen at the population-weighted 2021--2022 average for each cohort, so that changes in dwelling demand arise exclusively from demographic shifts. Each combination of population and $\mathrm{HHFR}$ scenarios yields a distinct demand trajectory, enabling systematic exploration of future housing needs. The empirical fit of the base-period HHFR and the resulting projected age profiles under the trend scenario are illustrated in Supplementary Figure~\ref{fig:housing_fit} and Figure~\ref{fig:housing_trend}.
}

\begin{figure}[htbp]
    \centering
    \includegraphics[width=\textwidth]{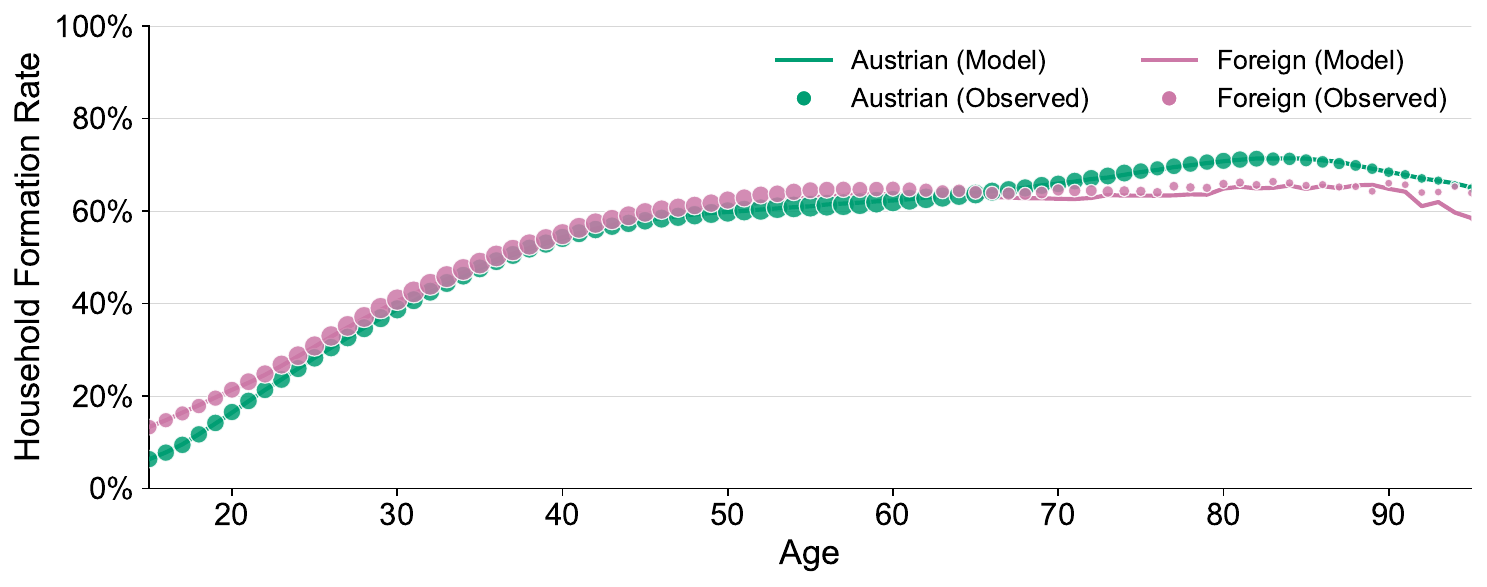}
    \caption{Comparison of observed and modelled household formation rates (HHFR) by 
    single-year age for Austrian nationals (green) and foreign nationals (pink). Scatter 
    points show the observed population-weighted HHFR from the 2022 base year, with 
    point size proportional to the underlying population. Solid lines show the status 
    quo HHFR, defined as the population-weighted average of the 2021 and 2022 observed 
    rates, which serves as the model's base-period anchor (t = 2021.5) from which 
    logit-space trend extrapolation departs. Divergence between the scatter and the line 
    at a given age indicates that the two-year averaging smooths year-to-year variation. 
    The divergence that grows systematically with age indicates cohort-specific dynamics 
    not captured by the pooled 2021–2022 base.}
    \label{fig:housing_fit}

    \includegraphics[width=\textwidth]{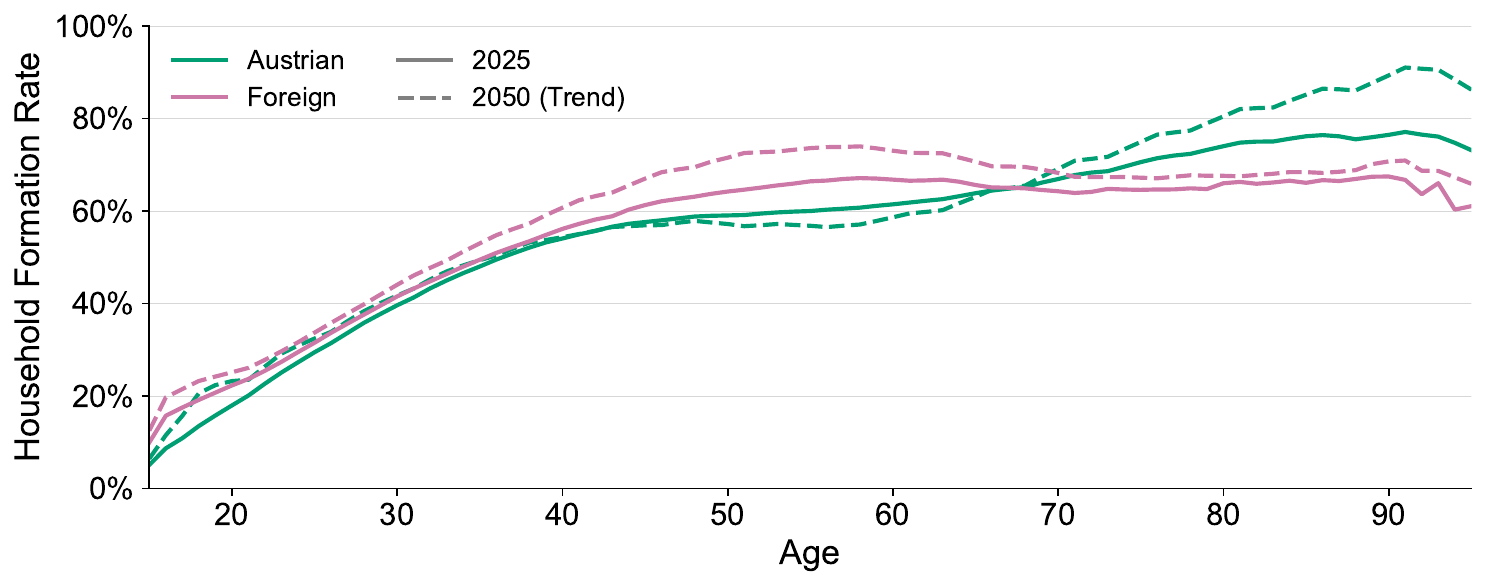}
    \caption{Projected household formation rate (HHFR) by single-year age under the 
    trend scenario, comparing the 2025 profile (solid lines) to the 2050 projection 
    (dashed lines), separately for Austrian nationals (green) and foreign 
    nationals (pink).}
    \label{fig:housing_trend}
\end{figure}

\subsubsection{Assumptions}

{
The model assumes that the two-level nationality classification (Austrian vs.\ foreign) remains a meaningful and stable predictor of household formation behaviour over the projection horizon.
In practice, the foreign-national population is heterogeneous, and compositional changes within this category (e.g.\ shifts in the origin-country mix of migration) may alter headship rates in ways that the model cannot capture.
}

{
The trend scenario assumes that the direction and approximate magnitude of change in logit-HHFR observed between 2011 and 2021/22 will persist through the projection horizon (up to 2050). Structural breaks, such as major housing-policy reforms, shifts in cultural norms around household formation, or economic shocks, could invalidate the extrapolation. For this reason, the \"OROK household-size projections are treated as authoritative external targets \cite{orok2024osterreichische}. The adjustment procedure assumes that discrepancies between the model's unconstrained forecast and the \"OROK targets should be resolved by uniformly scaling all cohorts' HHFR within a region, rather than by adjusting specific demographic subgroups. This uniform-scaling approach preserves within-region demographic differentials but may over- or under-correct particular subgroups.
}

{
Population projections are taken as exogenous and are not affected by the housing model's outputs. In reality, housing availability may influence internal migration patterns and, to some extent, fertility and immigration decisions. This one-directional coupling is a standard simplification in demographic demand modelling.
}

\subsubsection{Uncertainty and Robustness Tests}

{
We assess the sensitivity of the housing demand projections through eight tests targeting the trend extrapolation, external constraints, spatial aggregation, age-profile smoothing, nationality-specific household formation, and the overall variance structure.
}

{
\paragraph{Trend anchor sensitivity.}
Because the logit-space trend is estimated from only two temporal anchors (2011 and the 2021/2022 average), the projection is potentially sensitive to the choice of endpoint. Substituting 2021-only or 2022-only for the pooled endpoint shifts 2050 dwelling demand by less than $\pm 0.3\%$ ($-0.32\%$ and $+0.26\%$ respectively), confirming that the 2021--2022 average effectively averages out year-to-year noise. The slope distribution across 340 demographic groups is predominantly positive: $81.8\%$ of groups exhibit rising household formation rates, consistent with the secular trend toward smaller households observed across Europe. Foreign nationals show slightly steeper slopes ($\bar{\beta} = 0.0148$) than Austrians ($\bar{\beta} = 0.0124$), reflecting faster convergence toward Austrian national household formation patterns. In absolute terms, the weighted national HHFR increased by $+2.5$ percentage points for Austrians and $+3.4$ for foreign nationals between 2011 and the 2021--2022 average, indicating that both groups are forming independent households at higher rates, but with a narrowing gap.
}

{
\paragraph{External constraint impact.}
The \"OROK household-size constraints adjust the unconstrained trend projection by only $-0.67\%$ nationally, indicating close agreement between the model's endogenous extrapolation and the external targets. The regional adjustment factors range from $0.955$ (\"Ostliche Obersteiermark, where the model projects smaller households than \"OROK) to $1.058$ (Osttirol, where the model projects larger households). The distribution is near-symmetric around unity, with most factors within $\pm 3\%$. Urban regions tend to receive downward adjustments (Graz: $0.963$, Vienna: $0.971$), reflecting the model's tendency to slightly overproject household size reduction in dense areas, while peripheral regions receive upward adjustments (Tiroler Oberland: $1.047$), where the model underpredicts the household-size decline projected by \"OROK. These small corrections confirm that the logit-trend specification and the \"OROK targets produce broadly consistent trajectories, with the constraints serving as a calibration rather than a structural correction.
}

{
\paragraph{Spatial aggregation.}
National dwelling demand is remarkably insensitive to the level at which trends are pooled: 1-cluster (national), 5-cluster (\"OROK), and 35-region (NUTS\,3) specifications produce totals that differ by less than $0.2\%$. However, the underlying slope distributions diverge substantially. At the NUTS\,3 level, slopes range from $-1.31$ to $+2.63$, an order of magnitude wider than the cluster-level range ($-0.11$ to $+0.09$), and the correlation between cluster-level and region-level slopes is only $0.048$. This indicates that the 5-cluster pooling smooths region-specific noise, which cancels at the aggregate level but may matter for sub-national planning. We retain the 5-cluster specification as the preferred balance between robustness and spatial resolution: it eliminates the extreme slopes generated by small-sample NUTS\,3 cells while preserving systematic urban--rural differentials.
}

{
\paragraph{Spatial generalisability.}
Leave-one-cluster-out cross-validation yields effectively zero prediction error ($< 0.01\%$) for all five clusters. This means that predicting a cluster's dwelling demand using the average trend from the remaining four clusters produces results indistinguishable from using the cluster's own trend. The result is consistent with the cluster aggregation test: the five clusters have sufficiently similar mean slopes that cluster-specific trends contribute little additional information beyond the national average. This supports the model's robustness to the specific cluster assignment and suggests that the five-fold classification may be overly conservative, as pooling could be simplified to a national trend without materially affecting the results.
}

{
\paragraph{Age-profile smoothing.}
The choice of smoothing method has negligible impact on aggregate demand: all methods (rolling window of 3, 5, or 7; Gaussian with $\sigma = 1.0$ or $2.0$) produce totals within $0.04\%$ of the unsmoothed projection. The smoothing does, however, substantially reduce age-to-age discontinuities in the HHFR profile: the mean absolute age-to-age jump decreases from $2.44$ (unsmoothed) to $1.37$--$1.69$ depending on the method, and the maximum jump falls from $100$ (an artefact at age-group boundaries) to $20$--$34$. The rolling average with a window of $3$ is retained as the default because it achieves meaningful discontinuity reduction while introducing the least distortion of the underlying age profile.
}

{
\paragraph{Nationality-specific household formation.}
The HHFR profiles by nationality reveal a crossover pattern: foreign nationals exhibit higher headship rates than Austrians at young ages (ratio of $1.37$ at ages 15--24, $1.05$ at 25--34), near-parity at middle ages, and lower rates at older ages ($0.92$--$0.94$ at ages 65--84). The young-age overrepresentation reflects a higher propensity among young migrants to form independent households (including single-person households in shared housing), whereas the old-age underrepresentation reflects a greater prevalence of multi-generational households among the long-settled foreign population. Implied average household sizes at 2050 are nearly identical: $1.89$ persons per dwelling for Austrians and $1.94$ for foreign nationals.
}

{
\paragraph{Variance decomposition.}
To disentangle the sources of uncertainty in the 2050 projections, we decomposed the total variance of the projected bed demand into two components: (i) \emph{demographic uncertainty}, driven by the 54 different population projection scenarios (varying fertility, mortality, and migration), and (ii) \emph{parameter uncertainty}, driven by the standard errors of the estimated model coefficients. Results reveal that the behavioural model choice (trend vs.\ status quo) accounts for $83.4\%$ of total variance, while the demographic scenario accounts for only $16.6\%$. Although the volume effect of foreign nationals is the largest single contributor to gross demand ($+13.0$ pp), it largely offsets the decline in Austrian demand ($-10.0$ pp). Consequently, the behavioural trend toward smaller households becomes the primary driver of net system expansion and the dominant source of uncertainty: the behavioural span ($368{,}000$ dwellings) exceeds the demographic span ($284{,}000$ dwellings) across the ensemble of scenarios. This shows that the dominant source of uncertainty in the housing projections is the future path of the household composition rather than the population composition.
}


\subsection{Education Demand: Technical Details}

\subsubsection{Data}

{
The education demand model draws on three administrative data products from Statistik Austria. First, \emph{population data} provide annual counts of residents by NUTS\,3 region, single-year age, and nationality (Austrian vs.\ foreign) for the school years 2015--2023 \cite{statistikaustria2025population}. Second, \emph{pupil data} record enrolment counts by political district (Politischer Bezirk), school type, single-year age, and nationality for the same years \cite{statistikaustria2025school}. Third, \emph{teaching staff data} report headcounts of teachers by political district and school type for the corresponding years \cite{statistikaustria2025teaching}.
}

{
Pupil and teacher data are reported at the level of political districts, whereas the model's target spatial unit is the NUTS\,3 region. Most districts map one-to-one to a NUTS\,3 region via a static lookup table covering all 95 Austrian districts (including the 23 Viennese municipal districts, which are jointly mapped to \texttt{AT130}). However, seven districts straddle two NUTS\,3 regions (e.g.\ Baden spans \texttt{AT122} and \texttt{AT127}). For these split districts, pupil counts are allocated to each NUTS\,3 region in proportion to the age- and nationality-specific population of the respective region in the corresponding year. Teacher counts for split districts are allocated similarly, using total population weights at the district--year level. This population-based proportional allocation ensures that the spatial redistribution is demographically consistent.
}

{
The Austrian school system comprises a large number of specialised school types, which are reported separately in the administrative data. To obtain analytically tractable and policy-relevant categories, the original school types are consolidated into six groups through a two-stage mapping. In the first stage, fine-grained types are mapped to intermediate groups (e.g.\ ``Hauptschulen'', ``Modellversuch (Neue) Mittelschule an AHS'', and ``Mittelschule, Neue Mittelschule'' are all mapped to ``Mittelschulen''). In the second stage, these intermediate groups are mapped to six final categories: Primary School (Volksschulen), Lower Secondary (Mittelschulen), Polytechnic School (Polytechnische Schulen), Academic Secondary (allgemein bildende h\"ohere Schulen), Vocational Education (a consolidation of commercial, technical, agricultural, social, and pedagogical schools together with Berufsschulen), and Special Education (Sonderschulen). Four minor school types that lack an assigned teaching-staff counterpart in the administrative data (Akademien f\"ur Sozialarbeit, Bundessportakademien, Schulen im Gesundheitswesen, and Akademien im Gesundheitswesen) are excluded from the analysis. The same harmonised school-type mapping is applied to both the pupil and teacher datasets to ensure consistency.
}

{
In the pupil data, age is reported as age on 1 September of the school year. Children younger than six are recoded to age 5 (the modal pre-primary enrolment age), and individuals aged 50 or over are collapsed into a single category (age 50). Records with unknown age or unknown district are excluded. Nationality labels are retained at the two-level classification (Austrian, foreign) used throughout the model.
}

{
In the teacher data, only headcount records are retained; full-time-equivalent counts (Vollzeit\"aquivalente) are excluded. The school-year label is converted to the starting calendar year (e.g.\ ``2022/2023'' to year 2022). After school-type harmonisation and district-to-NUTS\,3 mapping, teacher counts are aggregated by year, NUTS\,3 region, and school type.
}

{
In a small number of region--school-type--year cells, pupils are enrolled, but no teachers are recorded (e.g.\ because a school type is served by teachers based in a neighbouring region). To avoid undefined student-to-teacher ratios, pupils in such zero-teacher cells are redistributed to geographically adjacent NUTS\,3 regions with positive teacher counts for the same school type and year. Adjacency is determined from the spatial geometry of the NUTS\,3 regions, and the redistribution weights are proportional to the number of teachers in each qualifying neighbour. This procedure is applied iteratively: if no first-order neighbours have teachers, second-order neighbours are considered.
}

\subsubsection{Derived Variables}

{
The primary outcome of the education demand model is the projected \emph{number of teachers} required in each NUTS\,3 region, school type, and year. Unlike the housing and healthcare sectors, where demand is expressed in terms of physical units (dwellings, beds), education demand is expressed in personnel terms, reflecting the policy-relevant question of teacher staffing. The projected number of teachers for a given cell is obtained by dividing projected student counts by the base-year student-to-teacher ratio:
\begin{equation}
    T_{r,k,t}^{(\mathrm{proj})} \;=\; \frac{S_{r,k,t}^{(\mathrm{proj})}}{\eta_{r,k,t_0}},
    \label{eq:teacher_demand}
\end{equation}
where $T$ is the number of teachers, $S$ is the projected number of students, $\eta$ is the student-to-teacher ratio in the base year $t_0$, and the subscripts index NUTS\,3 region ($r$), school type ($k$), and year ($t$). For historical years, actual teacher counts are used directly rather than the ratio-based estimate. The derivation of projected student counts and the student-to-teacher ratio is described below.
}

{
The \emph{enrolment ratio} measures the share of the age-eligible population that is enrolled in school.
For each combination of year, NUTS\,3 region, age, nationality, and school type, it is defined as:
\begin{equation}
    \mathrm{ER}_{r,a,n,k,t} \;=\; \frac{S_{r,a,n,k,t}}{P_{r,a,n,t}},
    \label{eq:enrollment_ratio}
\end{equation}
where $S$ is the observed student count and $P$ is the corresponding population. The enrolment ratio can exceed unity for certain age--nationality cells (e.g.\ when students commute across regional boundaries or when the student-age classification does not perfectly align with the population-age classification). This feature motivates the two-stage modelling approach (described in the next section).
}

{
The \emph{student-to-teacher ratio} ($\eta$) is computed at the NUTS\,3--school-type level by dividing the total number of students by the total number of teachers:
\begin{equation}
    \eta_{r,k,t} \;=\; \frac{\sum_{a,n} S_{r,a,n,k,t}}{T_{r,k,t}}.
    \label{eq:str}
\end{equation}

The base-year ratio $\eta_{r,k,t_0}$ (with $t_0 = 2023$) is carried forward as a fixed parameter in the projection, implying that staffing intensity per student is held constant at the most recently observed level.
}

{
Projected student counts are generated by the enrolment model (described in the modelling framework below) and represent the expected number of students in each NUTS\,3 region, school type, age, and nationality cell for each projection year. These projections are produced under two behavioural scenarios (status quo and trend) crossed with the available population-projection scenarios. The empirical fit of the enrolment model and the resulting projected age profiles under the trend specification are illustrated in Supplementary Figure~\ref{fig:education_fit} and Figure~\ref{fig:education_trend}.
}

\begin{figure}[htbp]
    \centering
    \includegraphics[width=\textwidth]{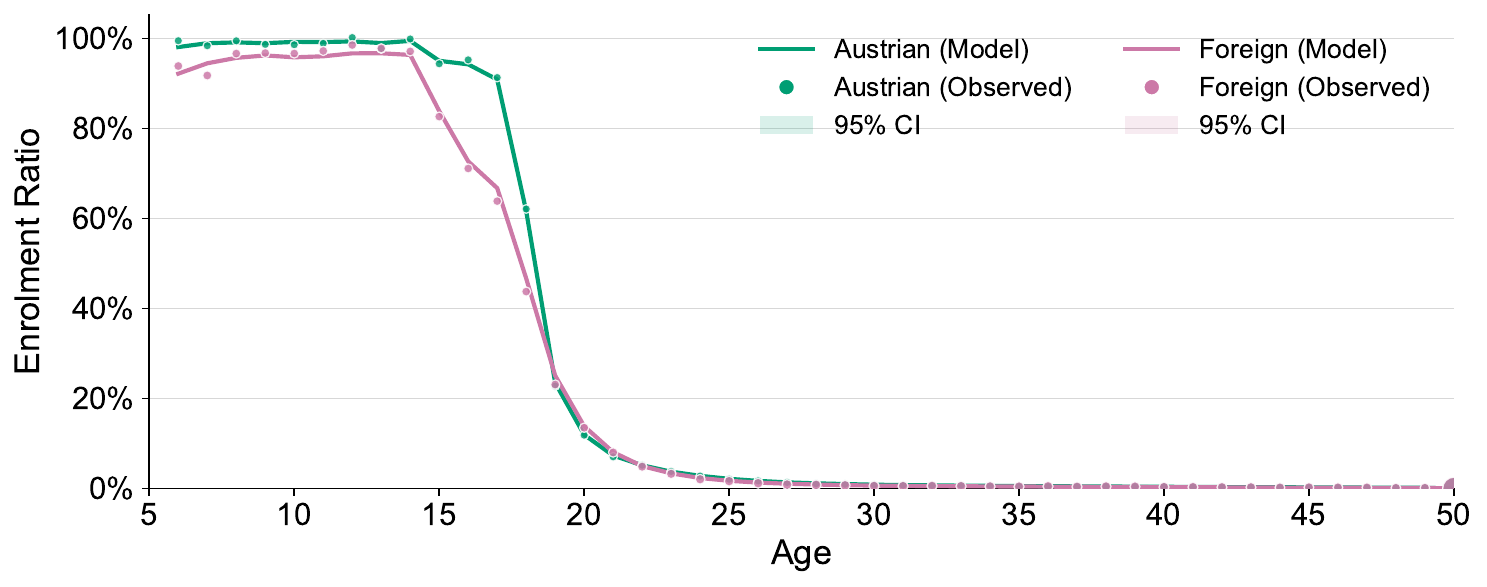}
    \caption{Observed versus fitted total enrolment ratio by single-year age (ages 6 and above) for Austrian nationals (green) and foreign nationals (pink). Scatter points show the observed enrolment ratio for 2023, with point size proportional to population. Solid lines show the fitted values from the enrolment model. Shaded bands show the 95\% credible interval from 1,000 Monte Carlo draws of the coefficient vector under the estimated covariance matrix.}
    \label{fig:education_fit}

    \includegraphics[width=\textwidth]{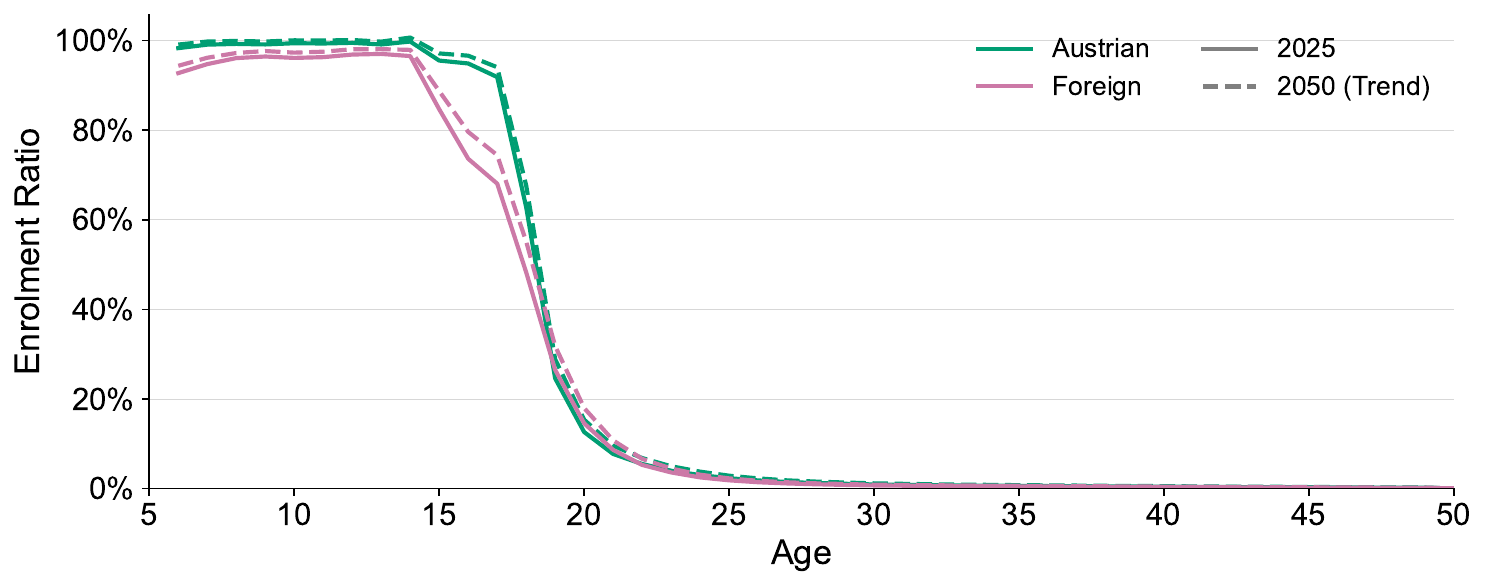}
    \caption{Projected total enrolment ratio by single-year age (ages 6 and above) under the trend specification of the enrolment model, comparing the 2025 profile (solid lines) to the 2050 projection (dashed lines), separately for Austrian nationals (green) and foreign nationals (pink).}
    \label{fig:education_trend}
\end{figure}

\subsubsection{Modelling Framework}

{
The education demand model employs a two-stage approach to project student enrolment. In the first stage, a binomial generalised linear model (GLM) estimates the \emph{base enrolment rate}, the probability that a person in a given demographic cell is enrolled in any school type. In the second stage, an empirical distribution allocates the predicted total enrolment across the six school types. A separate surplus component accounts for enrolment ratios that exceed unity. This structure accommodates the overdispersion observed in raw enrolment data (where, in some cells, more students are counted than residents) while preserving the demographic and spatial detail needed for teacher-demand projections.
}

{
Let $S_i^{\mathrm{total}}$ denote the total number of students (across all school types) in demographic cell $i$, and let $P_i$ denote the population.
The enrolled count is defined as $E_i = \min(S_i^{\mathrm{total}},\; P_i)$, i.e.\ the number of enrolments that can be accommodated within the population (capping at 100\% enrolment).
The binomial model specifies:

\begin{equation}
    E_i \;\sim\; \mathrm{Binomial}\!\left(P_i,\; \pi_i\right),
    \label{eq:binomial_enrollment}
\end{equation}
where the enrolment probability $\pi_i$ is linked to a linear predictor via the logit function:
\begin{equation}
    \mathrm{logit}(\pi_i) \;=\; \mathbf{x}_i^\top \boldsymbol{\beta}.
    \label{eq:binomial_logit}
\end{equation}

Two specifications of the linear predictor are fitted.
The \emph{status quo} specification includes categorical effects for nationality, age, and NUTS\,3 region:
\begin{equation}
   \mathrm{logit}(\pi_i) = \beta_0 + \beta_{\mathrm{nat}(i)} + \beta_{\mathrm{age}(i)} + \beta_{\mathrm{NUTS3}(i)} +  \beta_{\mathrm{age}(i)} \, \beta_{\mathrm{nat}(i)}.
    \label{eq:edu_sq}
\end{equation}

The \emph{trend} specification augments this with a standardised year covariate to capture secular trends in enrolment:

\begin{equation}
    \mathrm{logit}(\pi_i) = \beta_0 + \beta_{\mathrm{nat}(i)} + \beta_{\mathrm{age}(i)} + \beta_{\mathrm{NUTS3}(i)} + \beta_{\mathrm{age}(i)} \, \beta_{\mathrm{nat}(i)} +
    \beta_{\mathrm{year}} \, \widetilde{t}_i,
    \label{eq:edu_trend}
\end{equation}
where $\widetilde{t}_i = (t_i - \bar{t}) / \sigma_t$ is the year standardised to zero mean and unit variance over the training period.The interaction term $\beta_{\mathrm{age}(i)} \, \beta_{\mathrm{nat}(i)}$ allows the enrolment probability to vary differentially across age groups by nationality. This interaction is included to account for observed heterogeneity in enrolment patterns, such as a larger representation of Austrian nationals in post-primary education. Empirical robustness tests (detailed below) confirm that this term significantly improves model fit without introducing overfitting. Both models are estimated via maximum likelihood using the \texttt{statsmodels} GLM implementation with a binomial family and logit link.
}

{
For demographic cells where the observed enrolment ratio exceeds unity, the excess $S_i^{\mathrm{total}} / P_i - 1$ cannot be captured by the binomial model (which is bounded at $\pi = 1$). This surplus is modelled empirically: for each combination of age, NUTS\,3 region, and nationality, the historical average surplus rate $\bar{\varsigma}_{r,a,n}$ is computed and stored in a lookup table. During projection, the surplus rate for the matching cell is added to the binomial base rate to obtain the total enrolment ratio:
\begin{equation}
    \mathrm{ER}_i^{\mathrm{total}} \;=\; \hat{\pi}_i \;+\; \bar{\varsigma}_{r(i),a(i),n(i)}.
    \label{eq:surplus}
\end{equation}

}

{
The total projected student count is then $S_i^{\mathrm{proj}} = P_i \; \mathrm{ER}_i^{\mathrm{total}}$. The total projected enrolment is allocated across the six school types using empirical proportions. For each combination of NUTS\,3 region, age, and nationality, the historical (all-years-pooled) share of each school type in total enrolment is computed:
\begin{equation}
    \chi_{k \mid r,a,n} \;=\; \frac{\sum_t S_{r,a,n,k,t}}{\sum_t S_{r,a,n,\,,t}},
    \label{eq:school_shares}
\end{equation}
where $k$ indexes school type and the dot denotes summation over all types. The projected students in school type $k$ are then:
\begin{equation}
    S_{r,a,n,k,t}^{(\mathrm{proj})} \;=\; S_{r,a,n,\,,t}^{(\mathrm{proj})} \; \chi_{k \mid r,a,n}.
\end{equation}
}

{
Teacher demand is derived by dividing the projected student counts by the base-year student-to-teacher ratio (Eq.~\ref{eq:teacher_demand}). For historical years that fall within the training period, actual recorded teacher counts, distributed proportionally to school-type-level student counts within each NUTS\,3 region, are used instead of ratio-based estimates, ensuring a smooth transition from observed to projected values.
}

\subsubsection{Assumptions}

{
The trend scenario allows for a gradual change in the probability of enrolment for each demographic cell as captured by the year covariate, but assumes this linear (on the logit scale) trend continues through the projection horizon. Neither scenario can anticipate structural reforms to the education system (e.g.\ extension of compulsory schooling, introduction of new school types, or large-scale school closures).
}

{
The allocation of students across school types is based on historically pooled proportions and is held constant over the projection period. This implies that the relative attractiveness of, for example, vocational versus academic secondary education does not change, an assumption that may be violated if policy incentives, labour-market conditions, or cultural preferences shift.
}

{
The student-to-teacher ratio $\eta$ is fixed at its base-year (2023) value. This assumes no change in class-size policy, teaching-load regulations, or pedagogical staffing norms. In practice, policy-makers may respond to enrolment changes by adjusting staffing intensities, so the model's teacher projections should be interpreted as demand under current staffing standards rather than as forecasts of actual staffing levels.
}

{
The surplus enrolment component $\bar{\varsigma}$ is estimated from historical averages and held constant.
This component largely reflects cross-border commuting patterns and classification mismatches; changes in school-catchment areas or in the geographic distribution of the population could alter these patterns.
}

{
As in the housing model, population projections are taken as exogenous. Education outcomes (e.g.\ school availability) may in reality influence residential location choices, but this feedback is not modelled.
}

\subsubsection{Uncertainty and robustness tests}

{
We assess the sensitivity of the education demand projections to modelling choices, parameter assumptions, and data structure through eight complementary tests. These tests address three sources of uncertainty: model estimation error, parameter sensitivity, and the relative importance of scenario design choices.
}

{
\paragraph{Cross-validation.}
Leave-one-year-out cross-validation yields a mean absolute percentage error (MAPE) of $44.1\%$ ($\pm 3.0\%$) across nine folds, with Pearson correlations between observed and predicted enrolment consistently exceeding $r = 0.998$. The elevated MAPE reflects the granularity of the prediction task: errors concentrate in small demographic cells (low population, narrow age bins) where percentage deviations are mechanically amplified, while the near-perfect correlation confirms that the model captures the dominant structure of enrolment across regions, ages, and nationalities. Importantly, predictive accuracy is symmetric across citizenship groups (MAPE: $40.8\%$ for Austrians, $41.1\%$ for foreigners), confirming that the model does not systematically misrepresent either population. Across NUTS\,3 regions, Vienna exhibits the lowest error ($16.3\%$), consistent with the stabilising effect of large population denominators, while small rural regions such as AT112 (Waldviertel, $95.8\%$) show the highest errors.

Leave-one-region-out validation, which omits region fixed effects to test spatial generalisability, produces a higher mean MAPE ($81.6\%$), confirming that region-specific intercepts capture meaningful spatial heterogeneity. Vienna again achieves the lowest error ($19.1\%$), while the poorest performance occurs in small peripheral regions (AT127 Wiener Umland/S\"udteil, $210.8\%$). The systematic positive bias in several regions suggests that the national-level model overpredicts enrolment in areas with structurally lower participation, reinforcing the importance of the region fixed effects included in the main specification.
}

{
\paragraph{Model estimation uncertainty.}
Nonparametric bootstrap resampling of region--year blocks ($n = 200$ replications, 100\% convergence) yields extremely narrow uncertainty around aggregate predictions: the coefficient of variation of total predicted students is $0.12\%$, indicating that model estimation uncertainty is negligible relative to demographic scenario uncertainty. The nationality coefficient is robustly negative ($\hat{\beta} = -0.50$, 95\% bootstrap CI: $[-0.58, -0.42]$), confirming lower base enrolment probability for foreign nationals net of age and region effects. All age and region coefficients maintain consistent signs and magnitudes across bootstrap replications, with bootstrap standard errors typically below $0.10$ for age effects and below $0.05$ for region effects. The age–nationality interaction terms exhibit a non-linear pattern across the life cycle: foreign nationals face significantly lower enrolment probabilities at primary school ages (e.g. ages 6–11), a temporary convergence or even reversal around ages 14 and 19–23, and substantially higher relative enrolment at older adult ages (notably ages 49–50), with most interaction coefficients precisely estimated and bootstrap standard errors generally below 0.10.
}

{
\paragraph{Model specification.}
Comparison of alternative GLM specifications reveals that nationality--age interactions substantially improve fit (AIC reduction from $953{,}088$ to $909{,}499$), and the full interaction model incorporating both nationality--age and nationality--region terms achieves the lowest AIC ($870{,}413$). This indicates that enrolment probabilities differ across age groups by nationality (i.e., foreign nationals are underrepresented in primary and lower secondary education and overrepresented in vocational tracks at older ages). While other high-dimensional terms mechanically improve in-sample fit, they substantially increase parameter dimensionality, amplify variance, and reduce spatial generalizability in cross-validation exercises. We therefore retain the nationality–age interaction specification in the main analysis, as it captures demographically meaningful heterogeneity in educational trajectories while preserving parsimony and external validity.

A critical finding concerns the trend specification. The year coefficient reverses sign between the pre-COVID window (2015--2019: $\hat{\beta}_{\text{year}} = +0.036$, $p < 0.001$) and the post-COVID window (2019--2023: $\hat{\beta}_{\text{year}} = -0.026$, $p < 0.001$; 2020--2023: $-0.032$). This instability implies that linear trend extrapolation is sensitive to the training window: pre-pandemic data suggest rising enrolment intensity, whereas post-pandemic data suggest contraction. The full-period estimate ($+0.027$) reflects a weighted average that masks this structural break.
}

{
\paragraph{School-type distribution.}
The empirical allocation of students across school types, held constant in the projection, exhibits low but non-negligible temporal variation. Over 2015--2023, primary school shares increase by approximately $+0.23$ percentage points per year ($p < 0.05$), while vocational education shares decline almost symmetrically ($-0.21$ pp/year, $p < 0.05$). If these trends continued linearly to 2050, the primary share would rise by roughly $6$ pp at the expense of vocational education. However, overall coefficients of variation remain small (below $5\%$ for all school types), and the difference between pooled and most-recent-year (2023) shares does not exceed $1.2$ pp for any type. Nationality-disaggregated shares reveal structural differences: foreign students are overrepresented in primary ($35.6\%$ vs.\ $30.0\%$) and lower secondary ($22.5\%$ vs.\ $18.4\%$) and underrepresented in vocational education ($21.9\%$ vs.\ $29.0\%$), which are preserved through the nationality-specific empirical distributions used in the projection.
}

{
\paragraph{Student-to-teacher ratio.}
The base-year student-to-teacher ratio ($\bar{\eta} = 7.4$) has remained stable over the observation period (range: $7.3$--$7.5$ from 2015 to 2023), lending empirical support to the fixed-ratio assumption. Because teacher demand is inversely proportional to $\eta$, perturbations translate directly into demand: a $\pm 10\%$ shift in $\eta$ produces a $\mp 9$--$12\%$ change in projected teacher requirements. Variation across school types is substantial, ranging from $3.5$ in special education to $10.0$ in vocational education, implying that compositional shifts in the student body across school types can alter aggregate staffing needs even if the type-specific ratios remain constant.
}

{
\paragraph{Surplus enrolment.}
The surplus component---capturing enrolment ratios exceeding unity due to cross-regional commuting and classification mismatches---affects only $7.7\%$ of demographic cells and exerts minimal influence on aggregate projections. Bounding the surplus at its cell-specific historical maximum increases total predicted students by $1.8\%$ relative to the mean-surplus baseline; eliminating the surplus reduces them by a negligible amount. The surplus is spatially concentrated in regions with cross-border educational catchments (e.g., AT321 Lungau, AT122 Niederösterreich-Süd), and its aggregate impact is minimal next to the demographic and behavioural uncertainty captured by the scenario ensemble.
}

{
\paragraph{Variance decomposition.}
To disentangle the sources of uncertainty in the 2050 projections, we decomposed the total variance of the projected bed demand into two components: (i) \emph{demographic uncertainty}, driven by the 54 different population projection scenarios (varying fertility, mortality, and migration), and (ii) \emph{parameter uncertainty}, driven by the standard errors of the estimated model coefficients. Results reveal that demographic assumptions account for $63.0\%$ of total variance, while the choice between status quo and trend behavioural specifications accounts for $37.0\%$, with no detectable interaction. The demographic scenario ensemble spans approximately $95{,}400$ students (from $971{,}000$ under low migration to $1{,}067{,}000$ under high migration), while the behavioural model gap spans $42{,}000$ students (status quo: $997{,}300$; trend: $1{,}039{,}500$). This confirms that the dominant source of uncertainty in the education projections is the future path of the population composition rather than the internal modelling of enrolment behaviour. The absence of interaction indicates that the behavioural model choice shifts the level of projected demand uniformly across demographic scenarios without altering their relative ordering.
}


\subsection{Healthcare Demand: Technical Details}

\subsubsection{Data}

{
The healthcare demand model draws on two main data sources. First, \emph{individual-level patient records} are provided by the Austrian Federal Ministry of Health (\textit{Bundesministerium f\"ur Arbeit, Soziales, Gesundheit, Pflege und Konsumentenschutz}) and contain, for each inpatient episode, the patient identifier, year, number of hospital stays, number of hospital bed-days, sex, age group, nationality, and region of residence at NUTS\,2 level. The healthcare model thus operates at NUTS\,2 resolution, one level coarser than the housing and education models. These records are for 2019 and cover public and private inpatient hospital admissions in Austria. Second, \emph{population data} from Statistik Austria provides the total resident population by NUTS\,2 region, five-year age group, sex, and nationality (Austrian vs.\ foreign) for 2019 \cite{statistikaustria2025population}.
}

{
In the observed data, a small number of patients exhibit region changes within a year (e.g.\ due to relocation). When a patient appears multiple times within a year, the first recorded age group and region are retained. Additionally, patients with an unknown region are excluded (0.26\% of all patient records).
}

{
The Poisson model requires a dataset in which every demographic cell, defined by the cross-classification of year, sex, age group, nationality, and NUTS\,2 region, is represented, including cells with zero hospitalisations. This is achieved by constructing a full factorial for the base year (2019). The procedure proceeds as follows: (i)~the number of distinct patients is counted within each demographic cell; (ii)~this count is subtracted from the cell's population to obtain the number of ``healthy'' (non-hospitalised) individuals; (iii)~zero-outcome records are generated for each of these individuals; (iv)~the patient and zero-outcome records are concatenated.
Cells where the patient count exceeds the population (due to small-sample noise or cross-border treatment) are clipped to zero surplus. The resulting dataset contains one row per individual in the 2019 population, with exact hospitalisation outcomes for patients and zeros for non-patients.
}

\subsubsection{Derived Variables}

{
The primary outcome of the healthcare demand model is the projected \emph{average number of hospital beds occupied per day}, computed for each NUTS\,2 region and projection year. This quantity is obtained by dividing the total annual bed-days by 365:
\begin{equation}
    \bar{BD}_{r,t} \;=\; \frac{1}{365} \sum_{a,s,n} BD_{r,a,s,n,t},
    \label{eq:avg_beds}
\end{equation}
where $BD_{r,a,s,n,t}$ denotes the total bed-days consumed by the cohort defined by NUTS\,2 region ($r$), age group ($a$), sex ($s$), nationality ($n$), and year ($t$). The bed-day total for a projected cohort is, in turn, derived from the per-capita bed-day rate and the cohort's population, as described below.
}

{
Two hospitalisation outcomes are recorded for each patient: the \emph{number of stays}, counting distinct inpatient episodes within the year, and the \emph{number of bed-days}, counting the total days spent in hospital. For the full-factorial dataset, non-hospitalised individuals carry values of zero for both variables.
}

{
The individual-level data are aggregated into demographic cohorts for model fitting. For each cohort (defined by age group, sex, nationality, and NUTS\,2 region), the following quantities are computed: total bed-days, total stays, mean bed-days per person, mean stays per person, and cohort size (population count). These aggregates serve as the sufficient statistics for the Poisson model.
}

{
The Poisson model (described in the next section) produces a predicted \emph{bed-day rate per capita} for each cohort:
\begin{equation}
    \hat{\lambda}_{r,a,s,n} \;=\; \exp\!\left(\mathbf{x}_{r,a,s,n}^\top \hat{\boldsymbol{\beta}}\right),
    \label{eq:predicted_rate}
\end{equation}
where $\mathbf{x}$ is the vector of demographic indicators and $\hat{\boldsymbol{\beta}}$ is the vector of estimated coefficients. The projected total bed-days for a future cohort are then:
\begin{equation}
    BD_{r,a,s,n,t} \;=\; \hat{\lambda}_{r,a,s,n} \; P_{r,a,s,n,t},
\end{equation}
and the projected average daily beds follow from Eq.~\ref{eq:avg_beds}.
}

\subsubsection{Modelling Framework}

{
A Poisson generalised linear model is used to estimate the relationship between demographic characteristics and hospital bed-day utilisation.
Let $Y_i$ denote the total bed-days consumed by cohort $i$ and $P_i$ its population.
The model specifies:
\begin{equation}
    Y_i \;\sim\; \mathrm{Poisson}\!\left(\nu_i\right), \text{\, with}\qquad \nu_i = P_i  \lambda_i,
    \label{eq:poisson_model}
\end{equation}
where $\lambda_i$ is the per-capita bed-day rate linked to the covariates via the log function:
\begin{equation}
    \log(\lambda_i) \;=\; \mathbf{x}_i^\top \boldsymbol{\beta}.
    \label{eq:poisson_link}
\end{equation}
Equivalently, $\log(P_i)$ enters as an offset in the log-linear model for total bed-days:
\begin{equation}
    \log(\nu_i) \;=\; \log(P_i) \;+\; \mathbf{x}_i^\top \boldsymbol{\beta}.
    \label{eq:poisson_offset}
\end{equation}
The covariate vector includes categorical indicators for age group (19 levels), sex (2 levels), nationality (2 levels), and NUTS\,2 region (9 levels):
\begin{equation}
    \mathbf{x}_i^\top \boldsymbol{\beta} \;=\; \beta_0 \;+\; \beta_{\mathrm{age}(i)} \;+\; \beta_{\mathrm{sex}(i)} \;+\; \beta_{\mathrm{nat}(i)} \;+\; \beta_{\mathrm{NUTS2}(i)} \;+\; \beta_{\mathrm{age}(i)} \, \beta_{\mathrm{nat}(i)} \;+\; \beta_{\mathrm{age}(i)} \, \beta_{\mathrm{sex}(i)}.
    \label{eq:poisson_formula}
\end{equation}

The age--nationality interaction accounts for differential utilisation patterns across age cohorts by citizenship status, such as a widening utilisation gap at older ages. The age--sex interaction captures divergences in hospitalisation rates between males and females at advanced ages, reflecting sex-specific morbidity patterns. These interactions are justified by robustness tests (detailed below), which demonstrate substantial improvements in model fit and reduced residual bias compared to additive models.
The model is fitted by maximum likelihood using the \texttt{statsmodels} GLM implementation with a Poisson family and log link. Because the Poisson variance assumption ($\mathrm{Var}(Y_i) = \nu_i$) is typically violated in health-utilisation data due to overdispersion, heteroskedasticity-consistent (HC3) robust standard errors are used for inference, yielding a quasi-Poisson interpretation of the model.
}

{
For projection, the model's estimated coefficients are applied to the population-projection data.
Since the model is cross-sectional (fitted only to the base year 2019) and does not include a time covariate, the predicted per-capita rates are constant across years. Future demand thus changes only through the compositional effect of the projected population:
\begin{equation}
    BD_{r,a,s,n,t} \;=\; P_{r,a,s,n,t}^{(\mathrm{proj})} \; \hat{\lambda}_{r,a,s,n}.
\end{equation}
This corresponds to a status quo assumption in which utilisation rates remain at 2019 levels, and all demand changes are driven by demographic shifts (population growth, ageing, and changes in nationality composition).
}

\begin{figure}[htbp]
    \centering
    \includegraphics[width=\textwidth]{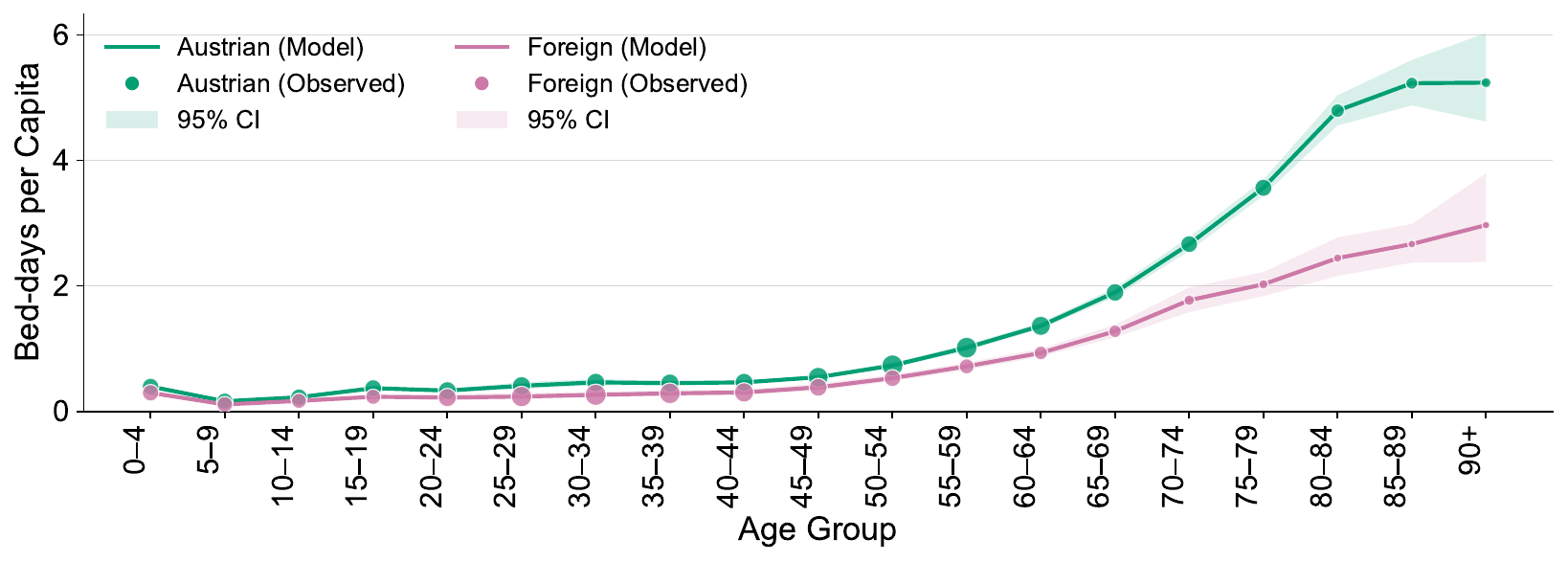}
    \caption{Observed versus fitted hospital bed-day utilisation rate (bed-days per capita) by five-year age group for Austrian nationals (green) and foreign nationals (pink). Scatter points show the observed rate from 2019, with point size proportional to the cell population. Solid lines show the fitted per-capita rate from the fitted model. Shaded bands show the 95\% interval from 1,000 Monte Carlo draws of the coefficient vector under the robust covariance matrix.}
    \label{fig:healthcare_fit}
\end{figure}

\subsubsection{Assumptions}

{
The model is fitted exclusively on 2019 data, chosen because it is the most recent pre-COVID year in the dataset. This assumes that 2019 utilisation patterns are representative of ``normal'' healthcare demand. To the extent that post-pandemic care-seeking behaviour, hospital capacity, or treatment protocols have shifted permanently, the 2019 baseline may not reflect the steady state to which the healthcare system converges.
}

{
The cross-sectional Poisson model yields a single set of per-capita rates held constant over the projection horizon. This is analogous to the status quo scenario in the housing and education models: all projected demand changes arise from demographic composition effects, not from changes in care intensity. Factors such as medical-technology adoption, preventive-care improvements, policy changes in hospital admission criteria, or shifts in the burden of disease are not captured.
}

{
Healthcare demand is modelled at the NUTS\,2 level (federal states), whereas the other two sectors use NUTS\,3 regions. This coarser resolution reflects the administrative level at which patient data are available. However, it implies that within-state heterogeneity in healthcare access and utilisation is averaged out.
}

{
As with the other two sectors, population projections are exogenous. Healthcare outcomes may influence mortality and, indirectly, population projections, but this feedback loop is not modelled.
}

{
The model assumes that the variance of bed-days equals the mean, which is typically violated in healthcare data. The use of HC3 robust standard errors addresses this for inference (confidence intervals and hypothesis tests remain valid even under overdispersion), but the point estimates of the coefficients are unchanged. The Monte Carlo simulation interval thus reflects parameter uncertainty under the robust covariance but does not account for residual overdispersion in individual outcomes.
}

\subsubsection{Uncertainty and robustness tests}

{
We assess the sensitivity of healthcare demand projections through eight tests that target distributional assumptions, model specification, estimation uncertainty, and the relative importance of uncertainty sources.}

{
\paragraph{Overdispersion.}
The model assumes that the variance of the outcome equals its mean (equidispersion). In healthcare data, unobserved patient heterogeneity typically leads to variance exceeding the mean (overdispersion). Our model exhibits significant overdispersion, with a Pearson dispersion statistic of $145.5$. To test whether this violation affects the accuracy of our projections, we fitted a Negative Binomial model, which relaxes the variance assumption by introducing a dispersion parameter $\alpha$ (where $\mathrm{Var}(Y) = \mu + \alpha \mu^2$). We fixed $\alpha=1.0$ to test the model's sensitivity to a standard geometric dispersion structure. The Negative Binomial specification yields a lower AIC ($12{,}804$ vs.\ $96{,}260$), confirming that it better describes the variance in the data. However, the mean estimates, which drive the projections, remain robust. The coefficient for foreign nationality shifts from $-0.31$ (Poisson) to $-0.44$ (Negative Binomial), indicating that the Poisson model is, if anything, conservative regarding the lower utilisation of foreign nationals. Since the Poisson estimator is consistent for the mean even under overdispersion, and we employ HC3 robust standard errors to correct the inference, we retain the Poisson specification for the central projection to maintain consistency with the demographic accounting framework.
}

{
\paragraph{Model specification.}
We compared five model specifications to justify the inclusion of interaction terms. The purely additive model yields an AIC of $251{,}155$. Adding age--nationality interactions reduces the AIC to $245{,}557$, while adding age--sex interactions yields a dramatic reduction to $101{,}849$. The chosen parsimonious model, which includes both interaction sets, yields an AIC of $96{,}260$.

While a ``full interaction'' model (including region--nationality terms) achieves a marginally lower AIC ($88{,}384$), we retain the parsimonious specification. The region--nationality interactions would require estimating distinct parameters for small migrant populations in rural federal states, increasing the risk of overfitting. The parsimonious model captures the essential demographic dynamics while remaining statistically robust. Crucially, the interaction terms reveal a widening gap in utilisation between nationals and foreign nationals among advanced-age individuals. The rate ratio (foreign vs.\ Austrian) is approximately $0.73$ for infants and children but declines to $\approx 0.49$ for the 85+ age group. This confirms that the ``healthy migrant'' effect \cite{dervic2024healthcare}, or return-migration selection, is strongest at ages when healthcare demand is highest.
}

{
\paragraph{Spatial robustness.}
Leave-one-region-out cross-validation yields a mean MAPE of $32.9\%$ across the nine NUTS\,2 regions, an improvement over the $40.6\%$ error of the additive model. Performance varies, with Upper Austria (AT31, $12.9\%$) and Vienna (AT13, $19.0\%$) predicted well, while Lower Austria (AT12, $55.2\%$) remains difficult to predict out-of-sample due to specific hospital utilisation patterns not explained by demographics alone. The in-sample MAPE with region fixed effects is $29.9\%$, and aggregate totals are reproduced exactly (pseudo $R^2 = 0.986$, correlation $= 0.993$). The out-of-sample errors confirm that region fixed effects capture genuine inter-state differences in healthcare infrastructure and utilisation intensity, particularly the low-utilisation profile of Burgenland and the high-utilisation profile of Styria (AT21). These regional intercepts are retained in the projection, ensuring that the spatial distribution of demand reflects observed federal-state heterogeneity rather than a national average.
}

{
\paragraph{Estimation uncertainty.}
Nonparametric bootstrap resampling of cohort-level data ($n = 200$, 100\% convergence) yields a coefficient of variation of $0.82\%$, corresponding to a 95\% confidence interval of $[34{,}960, 36{,}121]$ beds. The width of this interval ($1{,}162$ beds) represents approximately $\pm 1.6\%$ of the central estimate ($35{,}487$), indicating that parameter uncertainty contributes only marginally to the overall range of potential outcomes. Furthermore, the median of the bootstrap distribution ($35{,}464$) is virtually identical to the point estimate from the reference model, confirming that the estimator is unbiased.
}

{
\paragraph{Variance decomposition.}
To disentangle the sources of uncertainty in the 2050 projections, we decomposed the total variance of the projected bed demand into two components: (i) \emph{demographic uncertainty}, driven by the 54 different population projection scenarios (varying fertility, mortality, and migration), and (ii) \emph{parameter uncertainty}, driven by the standard errors of the estimated model coefficients. 
The analysis reveals a relatively balanced split: demographic scenarios account for $60.1\%$ of the total variance, while parameter uncertainty accounts for the remaining $39.9\%$. This contrasts with the education sector, where demographic uncertainty is the overwhelming driver ($>60\%$). The heightened relevance of parameter uncertainty in healthcare stems from the model's cross-sectional design. Because the model is fitted to a single year (2019), the estimated shape of the age--morbidity curve, particularly the steep gradient at older ages, carries statistical uncertainty that propagates strongly when projected onto a rapidly ageing population.

However, while the relative share of parameter uncertainty is notable, the absolute magnitude of total uncertainty is small. The span of outcomes across all demographic scenarios is only $810$ beds ($2.3\%$ of the mean), and the 95\% confidence interval from the Monte Carlo parameter simulation is similarly narrow ($1{,}005$ beds, or $2.8\%$ of the mean). This indicates that the central finding, a demand increase of approximately 30\%, is structurally robust. Neither extreme demographic shifts nor statistical noise in the utilisation coefficients is sufficient to substantially alter the projection's trajectory.
}

{
\paragraph{Residual diagnostics.}
The model demonstrates a good fit to the base-year data, achieving a pseudo $R^2$ of $0.994$. Aggregate bed-day totals are reproduced with near-exact precision (bias $< 0.01\%$) for both the Austrian and foreign populations. The analysis of residuals across demographic subgroups confirms that the model captures the complex relationship between age, nationality, and healthcare demand without systematic error. The mean bias within specific age--nationality cells—most notably for elderly foreign nationals—is $0.00\%$. This absence of bias indicates that the interaction terms in the model specification adequately capture the non-linearities in utilisation rates. Consequently, the estimated decline in the foreign-to-national rate ratio at older ages can be interpreted as a demographic feature reflected in the model coefficients, rather than as an artifact of residual lack of fit.
}

\subsection{Analytical Framework}

{
The analytical framework translates raw demand projections into interpretable metrics that address the paper's central question: to what extent do foreign nationals drive future demand of services of general interest, and how does this vary across sectors and regions? Raw demand trajectories (expressed in sector-specific units, such as teachers, dwellings, or hospital bed-days) are not directly comparable across sectors, and they do not reveal whether a group's service consumption is proportionate to its demographic weight. We therefore construct three complementary layers of analysis. The first layer decomposes the change in total demand into its constituent drivers, separating the contributions of population size, age structure, and behavioural shifts. The second layer defines standardised indices that enable cross-sectoral and cross-group comparison of demand evolution. The third layer formalises the spatial relationship between migration patterns and demand geography, distinguishing between the cumulative effect of the settled foreign population and the marginal effect of new arrivals. 

This framework is applied independently to each sector $k \in \{\text{Education, Housing, Healthcare}\}$. In the following equations, $D$ denotes the sector-specific demand unit (e.g., students, households, or bed-days) and $P$ denotes the relevant population denominator. All metrics are computed separately for Austrian ($n = \text{Aut}$) and foreign ($n = \text{For}$) nationality groups, as well as for the total population.
}

\subsubsection{Decomposition of growth drivers}
{
Aggregate demand growth in any service sector conflates three distinct mechanisms: changes in the number of potential users (population volume), shifts in their age composition (which alter per-capita consumption intensity even at constant behaviour), and changes in utilisation behaviour itself (e.g., evolving household formation norms or school enrolment patterns). Attributing demand pressure to migration without disentangling these mechanisms risks confounding a volume effect (more people) with a structural effect (an ageing population consuming more per capita). The decomposition below isolates each channel, enabling statements such as ``of the projected $X\%$ increase in healthcare demand, $Y$ percentage points are attributable to the ageing of the Austrian nationals and $Z$ percentage points to foreign population growth.''
}
{
The decomposition proceeds in two stages. First, the \emph{behavioural effect} ($E_{\text{Beh}}$) is extracted as the difference between the trend scenario projection ($D^{\text{Trend}}_{t}$) and the status quo counterfactual ($D^{\text{SQ}}_{t}$), both evaluated at the same projected population:
\begin{equation}
    E_{\text{Beh}} = D^{\text{Trend}}_{t} - D^{\text{SQ}}_{t}.
\end{equation}
This quantity captures the change in demand attributable solely to evolving per-capita utilisation intensity, for instance, the trend toward smaller household sizes in housing or shifting enrolment patterns in education. For healthcare, where only a status quo projection is available, the behavioural effect is zero by construction, and total demand change is fully allocated to the remaining two components.

Second, the residual change within the status quo scenario is decomposed into a \emph{volume effect} and an \emph{ageing effect} using a symmetric (Fisher-type) decomposition that avoids the index-number problem inherent in asymmetric decompositions. Let $I_{t} = D^{\text{SQ}}_{t} / P^{\text{SQ}}_{t}$ denote the per-capita demand intensity under the status quo at time $t$, and let $t_0 = 2025$ denote the base year. The volume effect isolates the change attributable to population growth, evaluated at the inter-period average intensity:
\begin{equation}
    E_{\text{Vol}} = \left(P^{\text{SQ}}_{t} - P_{t_0}\right) \frac{I_{t_0} + I^{\text{SQ}}_{t}}{2}.
\end{equation}
The ageing effect captures the change attributable to shifts in age composition---and hence in per-capita intensity---evaluated at the inter-period average population:
\begin{equation}
    E_{\text{Age}} = \left(I^{\text{SQ}}_{t} - I_{t_0}\right) \frac{P_{t_0} + P^{\text{SQ}}_{t}}{2}.
\end{equation}

By construction, the three effects sum to the total demand change: $\Delta D_{t} = E_{\text{Vol}} + E_{\text{Age}} + E_{\text{Beh}}$. The symmetric weighting ensures that neither the base-year nor the target-year structure is privileged, yielding a unique decomposition that is invariant to the direction of comparison.
}

{
To enable cross-sectoral comparison---for instance, to state that ageing contributes more percentage points to healthcare growth than to housing growth---the absolute effects are standardised as Contribution variables ($C$), expressed in percentage points relative to the base-year demand level ($D_{t_0}$):
\begin{equation}
    C_{\text{Vol}} = \frac{E_{\text{Vol}}}{D_{t_0}} \; 100, \qquad
    C_{\text{Age}} = \frac{E_{\text{Age}}}{D_{t_0}} \; 100, \qquad
    C_{\text{Beh}} = \frac{E_{\text{Beh}}}{D_{t_0}} \; 100.
\end{equation}

These contributions are computed separately for the Austrian and foreign sub-populations. The sum of all six sub-components (three effects $\times$ two nationality groups) equals the total percentage change in demand from 2025 to the target year. This nationality disaggregation is essential for the paper's research question: it allows us to determine, for each sector, whether the dominant pressure originates from the volume of foreign arrivals, the ageing of the Austrian population, or behavioural shifts that operate independently of citizenship status.
}

\subsubsection{Standardised comparative indices}
{
Three indices are tracked annually from 2025 to 2050, each answering a distinct policy-relevant question: Is demand growing or shrinking? Which group generates it? And is that group's consumption proportionate to its demographic size?
}

{
\paragraph{Cumulative demand growth.}
The cumulative demand growth ($CDG$) index tracks the aggregate trajectory of service requirements relative to the base year, expressed as a percentage deviation:
\begin{equation}
    CDG_{t} = \left(\frac{D_{t}}{D_{t_0}} - 1\right) \; 100.
\end{equation}

Positive values indicate expansion (the system requires more capacity than in 2025); negative values indicate contraction. Because cumulative demand growth is unit-free, it permits direct comparison of expansion pressure across sectors measured in fundamentally different units. For example, stating that healthcare cumulative demand growth reaches $+30\%$ while education cumulative demand growth reaches $-11\%$ immediately reveals that the two sectors face opposite pressures, a contrast that would be obscured by comparing absolute numbers of beds and teachers.
}

{
\paragraph{Demand Share.}
The Demand Share ($DS$) quantifies each nationality group's proportion of total sectoral demand at a given point in time:
\begin{equation}
    DS_{n,t} = \frac{D_{n,t}}{D^{\text{tot}}_{t}},
\end{equation}
where $D_{n,t}$ is the demand generated by group $n$ and $D^{\text{tot}}_{t}$ is the aggregate demand. Tracking $DS_{n,t}$ over time reveals compositional shifts in the user base, for instance, whether the share of foreign nationals in education demand is rising because their absolute demand grows or because nationals' demand contracts more rapidly. The demand share is a descriptive measure that does not adjust for demographic weight; a group that constitutes 30\% of the population and generates 30\% of demand has a demand share of $0.30$, as does a group that constitutes 10\% of the population and generates 30\% of demand. The representation index below provides this adjustment.
}

{
\paragraph{Representation Index.}
The \emph{representation index} ($RI_{n,t}$) normalises each group's Demand Share by its share of the total population, yielding a measure of proportionality between demographic weight and service consumption:
\begin{equation}
    RI_{n,t} = \frac{DS_{n,t}}{P_{n,t} / P^{\text{tot}}_{t}}.
\end{equation}
A value of $RI = 1.0$ indicates exact parity: the group consumes services in proportion to its population size. Values above $1.0$ indicate over-representation (the group's demand share exceeds its population share), while values below $1.0$ indicate under-representation. The representation index is central to evaluating claims of disproportionate service consumption. For instance, if foreign nationals have an $RI_{n,t}$ of $0.50$ in healthcare, they consume services at half the rate implied by their population share. Conversely, an $RI_{n,t}$ near $1.0$ in housing indicates proportionate consumption, implying that the growth in foreign housing demand is a mechanical consequence of population growth rather than excessive per-capita use. The temporal evolution of the $RI_{n,t}$ further reveals whether groups are converging toward or diverging from parity as the population ages and grows.
}
\subsubsection{Spatial heterogeneity metrics}
{
National-level results may mask substantial regional variation. A sector in which demand appears migration-driven at the aggregate level may in fact exhibit growth that is spatially uncorrelated with migrant settlement patterns, indicating that the growth stems from country-wide demographic processes (e.g., ageing) rather than from the localised presence of foreign nationals. Conversely, low national demand may concentrate in specific regions, creating localised infrastructure pressures. To distinguish between these possibilities, we estimate two regression models for each sector and each projection year $t \in [2026, 2050]$, yielding time-varying coefficients that reveal whether and when demand ``decouples'' from migration geography.
}

{
\paragraph{Spatial Concentration.}
The \emph{spatial concentration} tests whether the cumulative settlement pattern of foreign nationals determines the spatial distribution of demand growth. For each year $t$, we regress the cumulative demand growth in region $r$ on the regional foreign population share:
\begin{equation}
    g_{r,t} = \alpha^{\text{conc}}_t + \beta^{\text{conc}}_t \, \psi_{r,t} + \epsilon_{r,t},
\end{equation}
where $g_{r,t} = (D_{r,t} - D_{r,t_0}) / D_{r,t_0}$ is the percentage change in demand relative to 2025, and $\psi_{r,t} = P_{r,t}^{\text{For}} / P_{r,t}^{\text{Total}}$ is the share of foreign nationals in the regional population. The slope coefficient $\beta^{\text{conc}}_t$ is the primary quantity of interest. A positive and statistically significant $\beta^{\text{conc}}_t$ indicates spatial concentration: regions with higher foreign population shares experience systematically larger demand growth, suggesting that the accumulated presence of foreign nationals shapes the geography of service needs. A coefficient indistinguishable from zero indicates spatial independence, i.e. demand growth is unrelated to the foreign share, pointing instead to universally operating drivers such as population ageing. The temporal trajectory of $\beta^{\text{conc}}_t$ reveals whether concentration is intensifying (rising slope) or dissipating (declining slope) over the projection horizon.
}

{
\paragraph{Foreign Population Growth Elasticity.}
While the spatial concentration captures the spatial footprint of the settled foreign population, the \emph{foreign population growth elasticity} measures the marginal sensitivity of demand to foreign population growth:
\begin{equation}
    g_{r,t} = \alpha^{\text{growth}}_t + \beta^{\text{growth}}_t \, f_{r,t} + \nu_{r,t},
\end{equation}
where $f_{r,t} = (P_{r,t}^{\text{For}} - P_{r,t_0}^{\text{For}}) / P_{r,t_0}^{\text{For}}$ is the cumulative percentage growth of the foreign population in region $r$ since 2025. The two coefficients have distinct interpretations. The intercept $\alpha^{\text{growth}}_t$ captures the baseline demand trajectory in the hypothetical absence of foreign population growth; a negative intercept indicates that the Austrian population alone would generate a structural contraction in demand, as is the case in education. The slope $\beta^{\text{growth}}_t$ is the elasticity of demand with respect to foreign population growth.  A value near $1.0$ implies a roughly one-to-one relationship: each percentage point increase in foreign population growth corresponds to a comparable percentage increase in demand growth. Values below $1.0$ indicate dampening, in which foreign-driven demand growth is partially offset by the simultaneous decline in Austrian nationals or by ageing effects. A coefficient statistically indistinguishable from zero implies full decoupling: demand is independent of foreign population growth and is instead governed by the demographic momentum of the broader resident population.

The combination of the two models enables a precise diagnosis of each sector's relationship to migration. A sector with a rising $\beta^{\text{conc}}$ but a declining $\beta^{\text{growth}}$, as observed in education, exhibits increasing spatial concentration of demand in regions with high foreign population shares, yet this concentration is no longer sensitive to how fast the foreign population grows. This pattern arises because the spatial footprint of the foreign population is already established; the demand it generates derives from the age structure and reproductive momentum of that settled population rather than from continued inflows. This distinction is critical for policy: it implies that restricting new arrivals would not neutralise the spatial redistribution of demand, because the driver has already shifted from border flows to domestic demographic processes.
}

{
Both models are estimated via ordinary least squares for each projection year independently, with parameter uncertainty quantified by OLS standard errors and 95\% confidence intervals constructed as $CI = \hat{\beta} \pm 1.96 \times SE$. Housing and education regressions use $N = 35$ NUTS\,3 regions as observations, while healthcare uses $N = 9$ NUTS\,2 regions, reflecting the coarser spatial resolution of the healthcare data. All coefficients are reported as medians across the migration scenario ensemble; shaded bands in figures represent the scenario range rather than parameter uncertainty. A coefficient is interpreted as statistically significant when its confidence interval does not include zero.
}

\renewcommand{\figurename}{Supplementary Figure}
\renewcommand{\tablename}{Supplementary Table}
\setcounter{section}{2}
\renewcommand{\thefigure}{B.\arabic{figure}}
\renewcommand{\thetable}{B.\arabic{table}}
\renewcommand{\thesection}{B}

\section{Supplementary Tables}\label{sec:supp_tables}
{
Aggregate service demand trajectories from 2025 to 2050 are reported for each sector, disaggregated by citizenship status, providing the headline projection figures underlying the main results (Supplementary Table \ref{tab:summary}).

\begin{table}[htbp]
\centering
\caption{Aggregate service demand trajectories across primary welfare sectors (2025--2050). Projections detail required capacities disaggregated by citizenship status. Values represent the median across the scenario ensemble, with brackets [Min, Max] denoting the full uncertainty envelope. Baseline and projected values for education and healthcare are in thousands; housing is in millions.}
\label{tab:summary}
\resizebox{\textwidth}{!}{%
\begin{tabular}{llcccc}
\toprule
\textbf{Sector} & \textbf{Nationality} & \textbf{Baseline 2025} & \textbf{Projected 2050} & \textbf{Cumulative growth (\%)} & \textbf{Demand share (\%)} \\
\midrule
\textit{Education} & Austrian & $108.3$ & $80.8$ [$79.8$, $81.7$] & $-25.4$ [$-26.4$, $-24.6$] & $66.0$ [$63.1$, $69.0$] \\
(thousand teachers) & Foreign & $29.5$ & $41.7$ [$35.8$, $47.8$] & $+41.2$ [$+21.2$, $+61.8$] & $34.0$ [$31.0$, $36.9$] \\
 & \textbf{Total} & $\mathbf{137.9}$ & $\mathbf{122.5}$ [$\mathbf{115.6}$, $\mathbf{129.5}$] & $\mathbf{-11.1}$ [$\mathbf{-16.2}$, $\mathbf{-6.1}$] & $\mathbf{100.0}$ \\
\addlinespace
\textit{Housing} & Austrian & $3.43$ & $3.42$ [$3.41$, $3.44$] & $-0.1$ [$-0.5$, $+0.3$] & $68.2$ [$66.2$, $70.3$] \\
(million dwellings) & Foreign & $0.79$ & $1.59$ [$1.44$, $1.75$] & $+102.3$ [$+82.7$, $+122.3$] & $31.8$ [$29.7$, $33.8$] \\
 & \textbf{Total} & $\mathbf{4.21}$ & $\mathbf{5.02}$ [$\mathbf{4.85}$, $\mathbf{5.19}$] & $\mathbf{+19.1}$ [$\mathbf{+15.1}$, $\mathbf{+23.1}$] & $\mathbf{100.0}$ \\
\addlinespace
\textit{Healthcare} & Austrian & $24.9$ & $30.0$ [$30.0$, $30.1$] & $+20.8$ [$+20.5$, $+21.2$] & $84.5$ [$83.6$, $85.5$] \\
(thousand beds per day) & Foreign & $2.3$ & $5.5$ [$5.1$, $5.9$] & $+143.9$ [$+126.1$, $+162.0$] & $15.5$ [$14.5$, $16.4$] \\
 & \textbf{Total} & $\mathbf{27.1}$ & $\mathbf{35.5}$ [$\mathbf{35.1}$, $\mathbf{36.0}$] & $\mathbf{+31.1}$ [$\mathbf{+29.2}$, $\mathbf{+32.9}$] & $\mathbf{100.0}$ \\
\bottomrule
\end{tabular}
}
\end{table}
}

{
The total change in national system demand is decomposed into its three constituent drivers (i.e. volume, ageing, and behavioural effects) expressed in percentage points relative to the 2025 baseline, separately for Austrian and foreign nationals in each sector (Supplementary Table \ref{tab:waterfall_national}).

\begin{table}[htbp]
\centering
\small
\caption{Decomposition of demand drivers at the national level (2025--2050). Values represent the contribution of each driver to the total change in national system demand, expressed in percentage points relative to the 2025 baseline. The volume effect captures changes attributable to population size; the ageing effect captures changes attributable to shifts in age composition; and the behavioural effect captures changes in per-capita utilisation intensity. The healthcare model assumes constant utilisation rates and therefore has no behavioural component. Results are reported as the median across the scenario ensemble, with the full uncertainty range in brackets.}
\label{tab:waterfall_national}
\begin{tabular}{llrrr}
\toprule
\textbf{Sector} & \textbf{Nationality} & \textbf{Volume (pp)} & \textbf{Ageing (pp)} & \textbf{Behavioural (pp)} \\
\midrule
\textbf{Education} & Austrian & $-8.3$ [$-8.6$, $-8.0$] & $-13.6$ [$-14.0$, $-13.2$] & $+1.9$ [$+1.9$, $+1.9$] \\
 & Foreign & $+13.0$ [$+9.4$, $+16.7$] & $-5.5$ [$-6.0$, $-5.0$] & $+1.3$ [$+1.1$, $+1.5$] \\
\addlinespace
\textbf{Healthcare} & Austrian & $-13.1$ [$-13.6$, $-12.6$] & $+32.2$ [$+31.9$, $+32.5$] & --- \\
 & Foreign & $+6.8$ [$+5.1$, $+8.4$] & $+5.2$ [$+4.9$, $+5.5$] & --- \\
\addlinespace
\textbf{Housing} & Austrian & $-10.0$ [$-10.4$, $-9.6$] & $+5.3$ [$+5.2$, $+5.4$] & $+4.6$ [$+4.6$, $+4.6$] \\
 & Foreign & $+13.0$ [$+9.6$, $+16.4$] & $+2.1$ [$+1.9$, $+2.2$] & $+4.1$ [$+3.7$, $+4.6$] \\
\bottomrule
\end{tabular}
\end{table}
}

{
The same three-component decomposition is expressed relative to each citizenship group's own 2025 baseline demand, enabling direct comparison of the relative magnitude of drivers within each group rather than their contribution to the national total (Supplementary Table \ref{tab:decomposition}).

\begin{table}[htbp]
\centering
\caption{Decomposition of demand drivers by citizenship group (2025--2050). Percentage points are expressed relative to the 2025 baseline demand for each citizenship group separately. The volume effect captures changes driven by population size; the ageing effect captures changes driven by shifts in age composition; and the behavioural effect captures shifts in per-capita utilisation intensity. The healthcare model assumes constant utilisation rates and therefore has no behavioural component. Total growth indicates the net percentage change in demand for each group by 2050. Values represent the median across the scenario ensemble, with the full uncertainty range in brackets.}
\label{tab:decomposition}
\resizebox{\textwidth}{!}{%
\footnotesize
\begin{tabular}{llrrrr}
\toprule
\textbf{Sector} & \textbf{Nationality} & \textbf{Volume (pp)} & \textbf{Ageing (pp)} & \textbf{Behavioural (pp)} & \textbf{Total growth (\%)} \\
\midrule
Education & Austrian & $-10.6$ [$-11.0$, $-10.2$] & $-17.3$ [$-17.7$, $-16.8$] & $+2.4$ [$+2.4$, $+2.4$] & $-25.4$ [$-26.3$, $-24.6$] \\
 & Foreign & $+60.7$ [$+43.9$, $+78.3$] & $-25.1$ [$-27.7$, $-22.8$] & $+4.8$ [$+4.1$, $+5.5$] & $+40.4$ [$+20.4$, $+61.0$] \\
\addlinespace
Healthcare & Austrian & $-14.3$ [$-14.8$, $-13.7$] & $+34.5$ [$+34.2$, $+34.9$] & --- & $+20.2$ [$+19.9$, $+20.6$] \\
 & Foreign & $+82.9$ [$+62.2$, $+103.4$] & $+69.0$ [$+65.7$, $+72.6$] & --- & $+151.9$ [$+134.1$, $+169.9$] \\
\addlinespace
Housing & Austrian & $-12.3$ [$-12.8$, $-11.8$] & $+6.5$ [$+6.4$, $+6.6$] & $+5.7$ [$+5.7$, $+5.7$] & $-0.1$ [$-0.5$, $+0.3$] \\
 & Foreign & $+69.3$ [$+51.1$, $+87.7$] & $+11.0$ [$+10.2$, $+11.9$] & $+22.0$ [$+19.6$, $+24.5$] & $+102.3$ [$+82.7$, $+122.3$] \\
\bottomrule
\end{tabular}
}
\end{table}
}

{
The national demand decomposition is disaggregated to the level of federal states, showing each region's contribution to national system growth or contraction by driver, and supporting the spatial analysis presented in the main text (Supplementary Table \ref{tab:waterfall_regional}).

\begin{table}[htbp]
\centering
\caption{Regional contribution to national system demand growth by driver (2025--2050). Percentage points are expressed relative to total national system demand in 2025 for each sector. The volume effect captures growth driven by regional population size changes; the ageing effect captures regional shifts in age composition; and the behavioural effect captures changes in regional per-capita utilisation intensity. The healthcare model assumes constant utilisation rates and therefore has no behavioural component. The net impact is the sum of all drivers, representing each region's total contribution to national demand growth or contraction. Values represent the median across the scenario ensemble, with the full uncertainty range in brackets.}
\label{tab:waterfall_regional}
\scriptsize
\begin{tabular}{llrrrr}
\toprule
\textbf{Sector} & \textbf{Region} & \textbf{Volume (pp)} & \textbf{Ageing (pp)} & \textbf{Behavioural (pp)} & \textbf{Net impact (pp)} \\
\midrule
\textbf{Education}
 & Vienna        & $+2.9$ [$+1.3$, $+4.6$] & $-3.9$ [$-4.1$, $-3.7$] & $+0.7$ [$+0.7$, $+0.8$] & $-0.3$ [$-2.0$, $+1.5$] \\
 & Salzburg      & $+0.1$ [$-0.2$, $+0.5$] & $-1.1$ [$-1.2$, $-1.0$] & $+0.2$ [$+0.2$, $+0.2$] & $-0.8$ [$-1.2$, $-0.3$] \\
 & Tyrol         & $+0.3$ [$-0.1$, $+0.7$] & $-1.4$ [$-1.6$, $-1.2$] & $+0.3$ [$+0.3$, $+0.3$] & $-0.9$ [$-1.5$, $-0.2$] \\
 & Vorarlberg    & $+0.4$ [$+0.1$, $+0.6$] & $-0.9$ [$-1.0$, $-0.8$] & $+0.1$ [$+0.1$, $+0.1$] & $-0.4$ [$-0.7$, $-0.0$] \\
 & Upper Austria & $+0.7$ [$-0.1$, $+1.5$] & $-3.5$ [$-3.7$, $-3.3$] & $+0.5$ [$+0.4$, $+0.5$] & $-2.3$ [$-3.3$, $-1.3$] \\
 & Styria        & $+0.2$ [$-0.3$, $+0.8$] & $-2.5$ [$-2.6$, $-2.3$] & $+0.3$ [$+0.3$, $+0.4$] & $-1.9$ [$-2.5$, $-1.2$] \\
 & Carinthia     & $-0.2$ [$-0.4$, $-0.0$] & $-1.2$ [$-1.3$, $-1.1$] & $+0.1$ [$+0.1$, $+0.2$] & $-1.3$ [$-1.5$, $-1.0$] \\
 & Lower Austria & $+0.3$ [$-0.6$, $+1.3$] & $-3.8$ [$-4.3$, $-3.3$] & $+0.5$ [$+0.5$, $+0.6$] & $-2.9$ [$-4.4$, $-1.4$] \\
 & Burgenland    & $+0.1$ [$-0.1$, $+0.3$] & $-0.8$ [$-0.9$, $-0.8$] & $+0.1$ [$+0.1$, $+0.1$] & $-0.7$ [$-0.9$, $-0.4$] \\
\addlinespace
\textbf{Healthcare}
 & Vienna        & $-0.6$ [$-1.6$, $+0.5$] & $+6.6$ [$+6.2$, $+7.0$] & --- & $+6.1$ [$+4.9$, $+7.2$] \\
 & Salzburg      & $-0.5$ [$-0.7$, $-0.3$] & $+2.5$ [$+2.4$, $+2.6$] & --- & $+2.0$ [$+1.9$, $+2.1$] \\
 & Tyrol         & $-0.5$ [$-0.8$, $-0.2$] & $+3.5$ [$+3.4$, $+3.6$] & --- & $+3.0$ [$+2.8$, $+3.2$] \\
 & Vorarlberg    & $0.0$ [$-0.1$, $+0.1$] & $+1.6$ [$+1.5$, $+1.6$] & --- & $+1.6$ [$+1.5$, $+1.7$] \\
 & Upper Austria & $-0.9$ [$-1.3$, $-0.6$] & $+6.9$ [$+6.8$, $+7.0$] & --- & $+6.0$ [$+5.7$, $+6.2$] \\
 & Styria        & $-1.5$ [$-1.8$, $-1.3$] & $+5.7$ [$+5.6$, $+5.9$] & --- & $+4.2$ [$+4.0$, $+4.4$] \\
 & Carinthia     & $-1.3$ [$-1.6$, $-1.1$] & $+4.1$ [$+4.0$, $+4.2$] & --- & $+2.8$ [$+2.5$, $+3.0$] \\
 & Lower Austria & $-0.6$ [$-1.1$, $-0.2$] & $+5.2$ [$+5.0$, $+5.4$] & --- & $+4.6$ [$+4.1$, $+5.1$] \\
 & Burgenland    & $-0.2$ [$-0.3$, $-0.1$] & $+1.3$ [$+1.2$, $+1.3$] & --- & $+1.1$ [$+0.9$, $+1.2$] \\
\addlinespace
\textbf{Housing}
 & Vienna        & $+2.1$ [$+0.5$, $+3.8$] & $+2.1$ [$+2.0$, $+2.2$] & $+1.1$ [$+1.0$, $+1.2$] & $+5.3$ [$+3.5$, $+7.1$] \\
 & Salzburg      & $+0.1$ [$-0.2$, $+0.4$] & $+0.4$ [$+0.3$, $+0.4$] & $+0.6$ [$+0.6$, $+0.6$] & $+1.1$ [$+0.8$, $+1.4$] \\
 & Tyrol         & $+0.4$ [$-0.0$, $+0.8$] & $+0.5$ [$+0.5$, $+0.6$] & $+0.9$ [$+0.9$, $+1.0$] & $+1.8$ [$+1.4$, $+2.3$] \\
 & Vorarlberg    & $+0.4$ [$+0.2$, $+0.6$] & $+0.3$ [$+0.3$, $+0.3$] & $+0.5$ [$+0.5$, $+0.5$] & $+1.1$ [$+0.9$, $+1.3$] \\
 & Upper Austria & $+0.4$ [$-0.3$, $+1.2$] & $+1.2$ [$+1.1$, $+1.3$] & $+1.6$ [$+1.5$, $+1.7$] & $+3.2$ [$+2.4$, $+4.1$] \\
 & Styria        & $-0.2$ [$-0.7$, $+0.4$] & $+0.9$ [$+0.8$, $+1.0$] & $+1.2$ [$+1.2$, $+1.3$] & $+2.0$ [$+1.4$, $+2.6$] \\
 & Carinthia     & $-0.4$ [$-0.6$, $-0.1$] & $+0.3$ [$+0.3$, $+0.4$] & $+0.6$ [$+0.6$, $+0.6$] & $+0.6$ [$+0.4$, $+0.8$] \\
 & Lower Austria & $+0.1$ [$-0.9$, $+1.1$] & $+1.3$ [$+1.2$, $+1.5$] & $+1.9$ [$+1.7$, $+2.0$] & $+3.3$ [$+2.2$, $+4.4$] \\
 & Burgenland    & $-0.0$ [$-0.2$, $+0.2$] & $+0.2$ [$+0.2$, $+0.3$] & $+0.4$ [$+0.3$, $+0.4$] & $+0.6$ [$+0.3$, $+0.8$] \\
\bottomrule
\end{tabular}
\end{table}
}

\renewcommand{\figurename}{Supplementary Figure}
\renewcommand{\tablename}{Supplementary Table}

\setcounter{section}{3} 

\renewcommand{\thefigure}{C.\arabic{figure}}
\renewcommand{\thetable}{C.\arabic{table}}
\renewcommand{\thesection}{C}

\section{Supplementary Notes}\label{sec:supp_notes}
This supplementary section provides contextual background that motivates and situates the main analysis. The first part documents the Austrian political context, describing how narratives linking migration to service overburden have translated into concrete policy outcomes and how public attitudes toward the social rights of foreign nationals compare across European countries. This context establishes why citizenship, rather than migration background or country of birth, is the analytically relevant category for evaluating overburden claims. The remaining three parts present sector-specific background for each of the three welfare domains examined in the paper. For each sector, we describe the data sources and spatial coverage used in the analysis, document historical patterns of service utilisation disaggregated by citizenship status, and highlight the structural features of demand that shape the projection results discussed in the main text.

\subsection{Austrian political contex}
{
Austria offers a particularly instructive case for examining the dynamics between migration narratives and service demand. The narrative linking migration to SGI overburden has produced concrete policy outcomes. Austria's withdrawal from the Global Compact for Migration exemplifies how claims of an impending ``fight for resources'' gained sufficient political traction to shape national policy, despite the agreement's legally non-binding nature \cite{conrad2021posttruth}. Most recently, in April 2025, the Austrian parliament enacted legislation authorising the government to suspend applications for family reunification by refugees. The Austrian chancellor justified this measure by stating that reception capacities, particularly in education and integration, were exhausted and systems ``overburdened'' by these challenges \cite{parlamentoesterreich2025pause}. Yet Austria also demonstrates that this narrative is contested. Campaigns such as \textit{Wir sind Oberösterreich} (We are Upper Austria) challenge the image of migrants as pure resource consumers by depicting them in professional uniforms as firefighters, medics, and nurses, thereby emphasising their contributions to staffing essential services \cite{dennison2020basic}.
}

{
The political focus on migration as a driver of service overburden resonates with public attitudes toward the eligibility of foreign nationals for social support. Data from the European Social Survey (ESS Round 8) quantify the extent of welfare chauvinism in Austria relative to its European peers \cite{europeansocialsurveyeuropeanresearchinfrastructureesseric2016ess}. Respondents were asked: \textit{``Thinking of people coming to live in [country] from other countries, when do you think they should obtain the same rights to social benefits and services as citizens already living here?''} The results reveal a stark preference for restrictiveness among the Austrian public (Supplementary~Table~\ref{tab:ess_survey}). Nearly 42\% of Austrian respondents believe social rights should be either conditional on citizenship ($26.4\%$) or withheld entirely ($15.4\%$). The proportion of respondents believing foreign nationals should \textit{never} receive the same rights ($15.4\%$) is notably high for a Western European welfare state, contrasting sharply with Germany ($2.3\%$), Switzerland ($3.4\%$), and Sweden ($0.8\%$), and clustering closer to rates observed in Lithuania ($16.5\%$) and Russia ($18.8\%$). Conversely, only $19.2\%$ of Austrians support granting rights upon arrival or after one year of residence. This attitudinal landscape, in which legal status (citizenship) serves as the primary gatekeeper to social legitimacy, underscores the importance of decomposing service demand by citizenship rather than by migration background.
}

\begin{table}[htbp]
\centering
\small
\caption{Public attitudes toward immigrants' rights to social benefits (European Social Survey). Percentage of respondents selecting each condition for when immigrants should obtain the same rights as citizens. Austria (highlighted) exhibits one of the highest rates of unconditional exclusion (``Never'') in Western Europe. Source: ESS Round 8 Data (2016). Values represent percentages of valid responses ($N_{AT}=1923$).}
\label{tab:ess_survey}
\begin{tabular}{lrrrr}
\toprule
\textbf{Country} & \textbf{Immediately/1 Year} & \textbf{After Working/Taxes} & \textbf{Once Citizen} & \textbf{Never} \\
\midrule
Sweden (SE)       & 38.8 & 32.9 & 27.5 & 0.8 \\
Germany (DE)      & 24.3 & 50.2 & 23.3 & 2.3 \\
Netherlands (NL)  & 16.3 & 33.9 & 47.4 & 2.3 \\
Switzerland (CH)  & 23.4 & 54.3 & 19.0 & 3.4 \\
France (FR)       & 23.3 & 50.1 & 19.7 & 6.9 \\
\textbf{Austria (AT)} & \textbf{19.2} & \textbf{39.0} & \textbf{26.4} & \textbf{15.4} \\
Hungary (HU)      & 5.9  & 36.2 & 27.5 & 30.4 \\
\bottomrule
\end{tabular}
\end{table}

\subsection{Housing in Austria}\label{sec:supp_data_housing}

{
Housing demand data are derived from the Austrian population and housing censuses administered by Statistics Austria. The analysis draws on three reference years: 2011, 2021, and 2022, providing a decade-long baseline for trend estimation. Data are disaggregated by single-year age (0–100), sex, citizenship status (Austrian vs. foreign national), and NUTS\,3 region (35 regions). The housing census captures occupied dwelling counts classified by household reference person characteristics, enabling the construction of cohort-specific household formation rates.
}

\begin{figure}[htbp]
\centering
\includegraphics[width=0.7\textwidth]{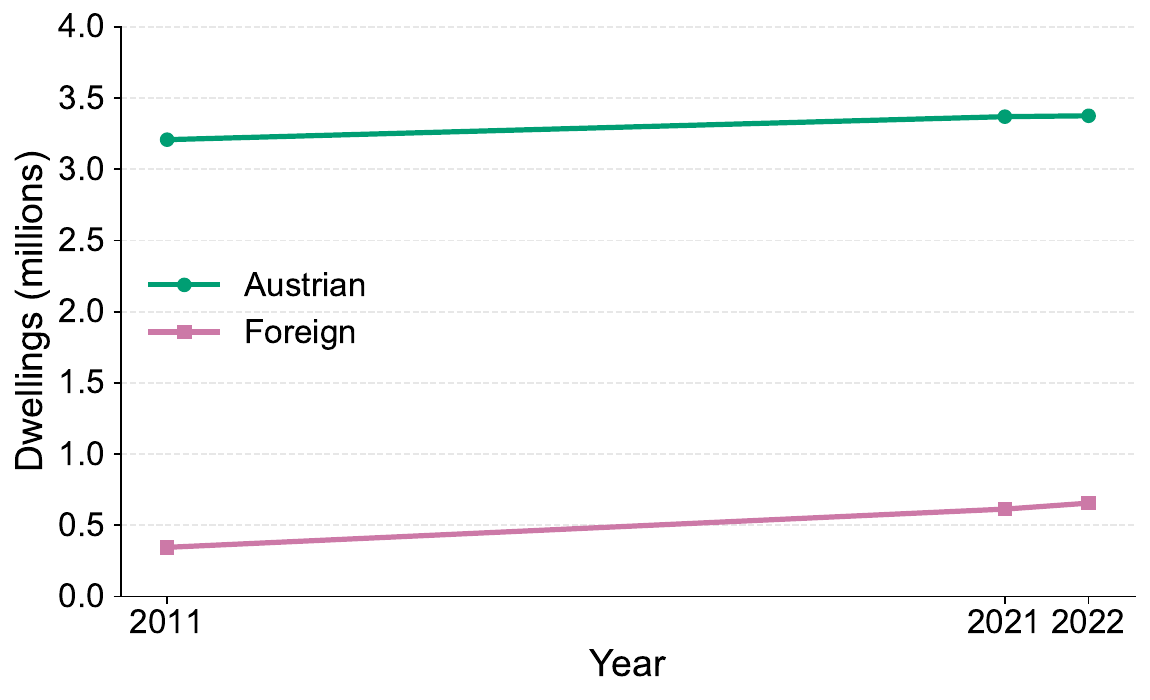}
\caption{Historical dwelling demand by citizenship status (2011–2022). Total dwellings are disaggregated by the household reference person's citizenship status. Austrian dwelling demand grew by 5.2\% over the decade, while foreign dwelling demand increased by 89.9\%, raising the foreign share from 9.7\% to 16.3\% of total stock.}
\label{fig:housing_historical}
\end{figure}

{
Austria's total housing stock grew from 3.55 million dwellings in 2011 to 4.03 million in 2022, an increase of 13.5\% over the decade (Supplementary Figure \ref{fig:housing_historical}) \cite{statistikaustria2025housing}. However, this aggregate growth masks a pronounced divergence by citizenship status. Dwellings attributable to Austrian nationals increased modestly from 3.21 million to 3.38 million (+5.2\%), while dwellings attributable to foreign nationals nearly doubled from 0.35 million to 0.66 million (+89.9\%). As a result, the foreign share of total dwelling demand rose from 9.7\% to 16.3\%. This shift occurred despite a slight decline in the Austrian population (7.46 million to 7.39 million) and a substantial increase in the foreign population (0.91 million to 1.59 million), indicating that the divergent growth in dwelling demand reflects population dynamics rather than changes in household formation behaviour.
}

\begin{figure}[htbp]
\centering
\includegraphics[width=\textwidth]{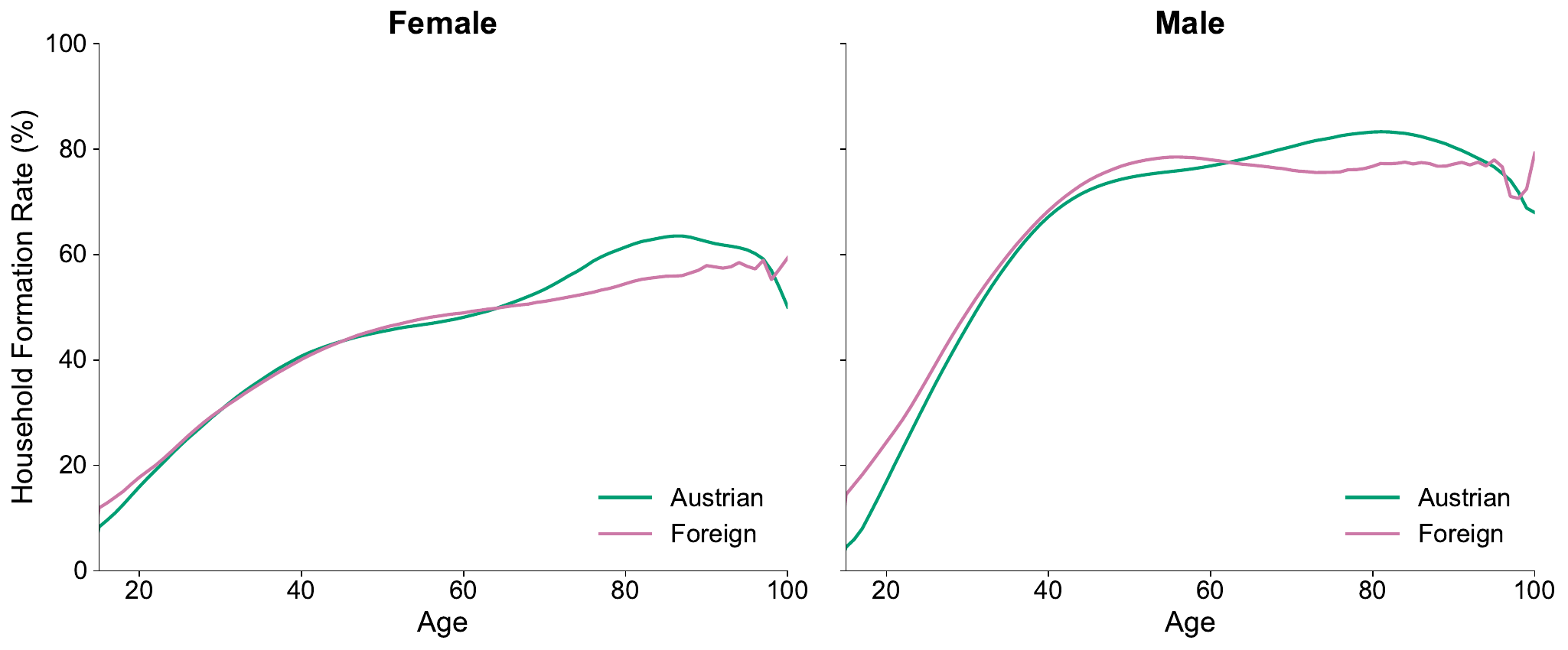}
\caption{Age-specific household formation rates by citizenship and sex (2021–2022). The household formation rate ($\mathrm{HHFR}$) is the probability that an individual serves as the household reference person, given residence in the private household population. Rates are population-weighted averages across all NUTS\,3 regions. Male rates (right panel) are consistently higher than female rates (left panel), reflecting household reference person conventions. Citizenship groups exhibit similar lifecycle trajectories, indicating that dwelling demand differences are driven by population structure rather than behavioural divergence.}
\label{fig:housing_hhfr}

\centering
\includegraphics[width=0.8\textwidth]{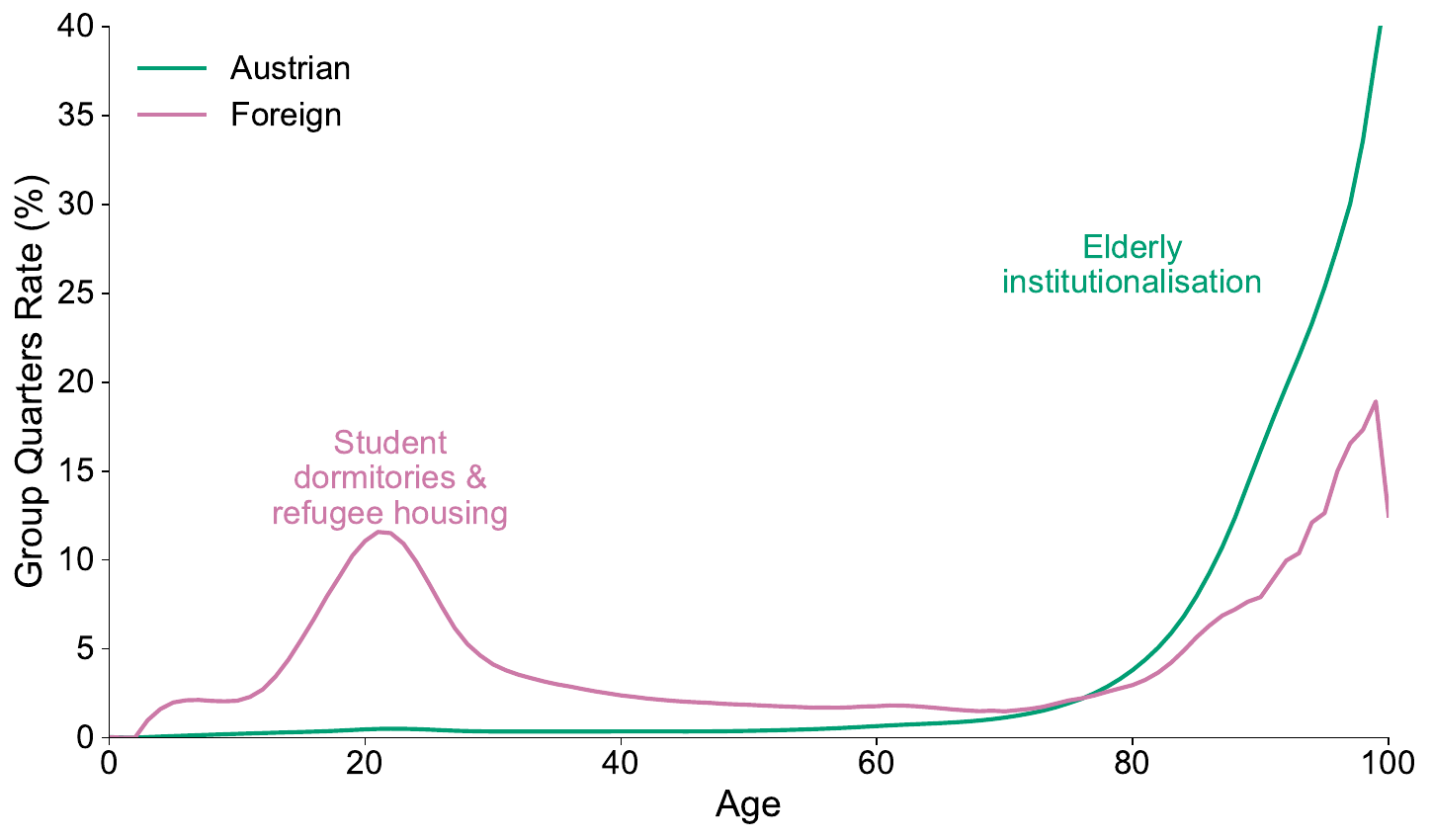}
\caption{Age-specific group quarters rates by citizenship (2021–2022). The group quarters rate ($\mathrm{HHFR}$) represents the proportion of the population residing in institutional settings rather than private households. Austrian nationals (blue) exhibit elevated institutionalisation only at advanced ages, reflecting nursing home residence. Foreign nationals (red) show a distinctive peak in early adulthood attributable to student dormitories and refugee accommodation, but lower institutionalisation at elderly ages. Rates are population-weighted averages across all NUTS\,3 regions and both sexes.}
\label{fig:housing_gqr}
\end{figure}

{
The household formation rate ($\mathrm{HHFR}$), the probability that an individual serves as a household reference person, conditional on not residing in institutional settings (see Supplementary Methods A), exhibits a characteristic lifecycle pattern that varies by sex but is largely convergent across citizenship groups (Supplementary Figure \ref{fig:housing_hhfr}). For both Austrian and foreign nationals, $\mathrm{HHFR}$ rises steeply from near zero at age 18 through early adulthood, reflecting the transition to independent living \cite{statistikaustria2025housing, statistikaustria2025registera, statistikaustria2025registerb}. By age 30, rates are similar across groups: 38.7\% for Austrians and 40.2\% for foreign nationals. The curves diverge by sex rather than citizenship: male $\mathrm{HHFR}$ peaks at approximately 80\% in the 60s and 70s, while female $\mathrm{HHFR}$ peaks at approximately 63\% in the 80s, reflecting traditional household reference person conventions whereby men are more frequently designated as the household head in partnered households. At advanced ages, $\mathrm{HHFR}$ declines as institutionalisation increases, particularly for women. The convergence of $\mathrm{HHFR}$ across citizenship groups implies that differences in dwelling demand are driven primarily by population composition (age structure and total size) rather than behavioural differences in household formation.
}

{
The group quarters rate ($\mathrm{GQR}$)---the proportion of the population residing in institutional settings such as nursing homes, student dormitories, refugee accommodation, and correctional facilities---reveals distinct patterns across citizenship and age groups (Supplementary Figure \ref{fig:housing_gqr}). For Austrian nationals, $\mathrm{HHFR}$ remains below 1\% throughout working ages before rising sharply after age 75, reaching 21.3\% among those aged 90 and over \cite{statistikaustria2025registera, statistikaustria2025registerb,statistikaustria2025population}. This pattern reflects the transition to residential care among the oldest-old. Foreign nationals exhibit a qualitatively different profile: $\mathrm{HHFR}$ peaks at approximately 10--12\% in the late teens and early twenties, driven by residence in student dormitories and, particularly following the 2015 refugee crisis, organised refugee accommodation. Among elderly foreign nationals, $\mathrm{HHFR}$ is substantially lower than among Austrians (7.8\% for ages 85+ compared to 14.6\%), reflecting both the younger age structure of the foreign population and potentially differential access to institutional care. These divergent $\mathrm{HHFR}$ profiles have implications for the household formation rate: the elevated young-adult $\mathrm{HHFR}$ among foreign nationals temporarily removes this population from the private housing market, while the elevated elderly $\mathrm{HHFR}$ among Austrians reduces late-life housing demand.
}

\begin{figure}[htbp]
\centering
\includegraphics[width=\textwidth]{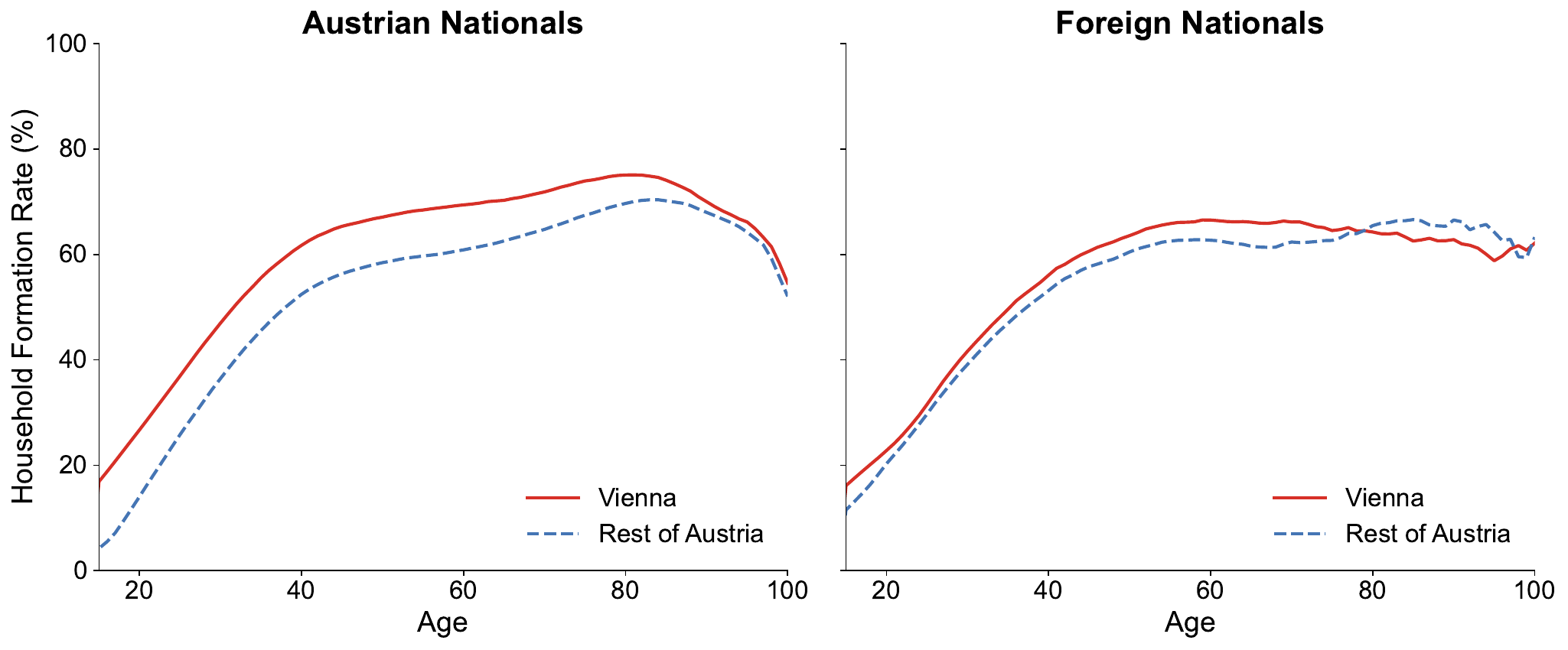}
\caption{Regional variation in household formation rates (2021–2022). Household formation rates are compared between Vienna (solid red) and the rest of Austria (dashed blue) for Austrian nationals (left) and foreign nationals (right). Vienna exhibits consistently higher rates across all ages, reflecting smaller average household sizes in the capital. The regional gap is more pronounced for Austrian nationals than for foreign nationals.}
\label{fig:housing_hhfr_regional}

\centering
\includegraphics[width=0.7\textwidth]{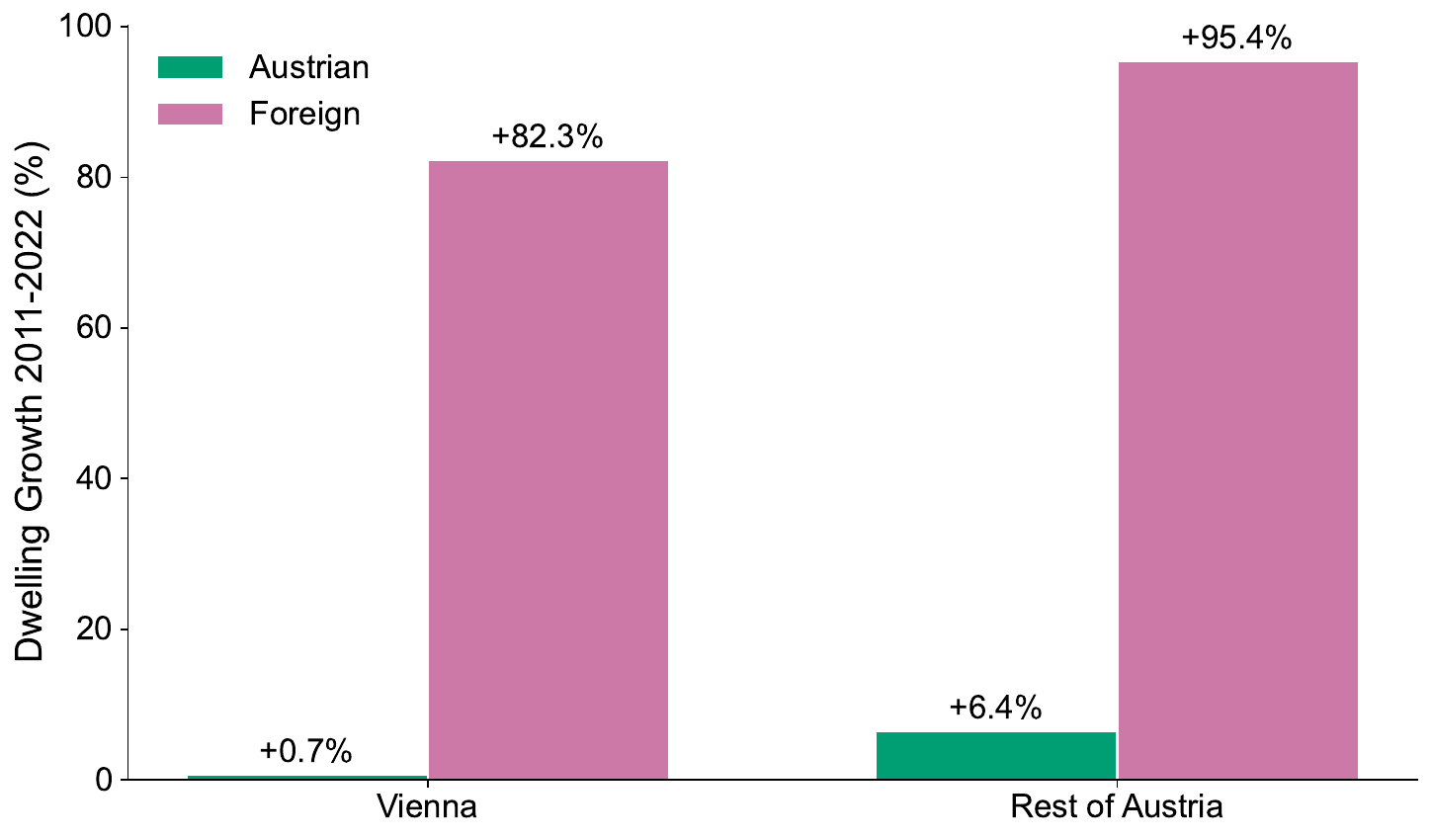}
\caption{Dwelling demand growth by region and citizenship (2011–2022). Growth rates are shown for Vienna and the rest of Austria, disaggregated by citizenship status. Austrian dwelling demand grew minimally in both regions (+0.7\% in Vienna, +6.4\% elsewhere), while foreign dwelling demand approximately doubled in both contexts (+82.3\% in Vienna, +95.4\% elsewhere).}
\label{fig:housing_regional_growth}
\end{figure}

{
Household formation rates vary substantially across Austria's housing markets (Supplementary Figure \ref{fig:housing_hhfr_regional}). Vienna exhibits consistently higher $\mathrm{HHFR}$ than the rest of Austria across all ages and both citizenship groups, reflecting the capital's smaller average household sizes and higher prevalence of single-person households. At age 30, $\mathrm{HHFR}$ in Vienna reaches 47.0\% for Austrians compared to 36.5\% elsewhere, a gap of over 10 percentage points \cite{statistikaustria2025housing, statistikaustria2025registera, statistikaustria2025registerb,statistikaustria2025population}. This Vienna premium persists into middle age (67.1\% vs. 58.5\% at age 50 for Austrians) and then narrows somewhat in later life. The regional differential is less pronounced among foreign nationals (41.7\% vs. 39.1\% at age 30), suggesting that foreign household structures are more similar across regions than Austrian household structures. These regional differences have important implications for dwelling demand projections: a given level of population growth in Vienna generates more dwelling demand than an equivalent level of growth in rural Austria.
}

{
The regional dimension of dwelling demand growth reveals that the citizenship divergence is a nationwide phenomenon (Supplementary Figure \ref{fig:housing_regional_growth}). In Vienna, Austrian dwelling demand grew by only 0.7\% over the decade, while foreign demand increased by 82.3\% \cite{statistikaustria2025housing}. Outside Vienna, Austrian dwelling demand grew by 6.4\%, while foreign demand surged by 95.4\%. By 2022, foreign nationals accounted for 28.0\% of dwelling demand in Vienna (where they constitute 32.2\% of the population) and 12.7\% in the rest of Austria (where they constitute 13.7\% of the population). The near-parity between dwelling and population shares confirms that the divergent demand growth is driven by population dynamics rather than by differential housing consumption behaviour.
}

{
The data imply an average household size of 2.23 persons per dwelling in 2022, with foreign households slightly larger (2.42 persons) than Austrian households (2.19 persons) \cite{statistikaustria2025housing, statistikaustria2025registera, statistikaustria2025registerb,statistikaustria2025population}. This differential reflects the younger age structure of the foreign population and has implications for housing adequacy: given that new construction in Vienna disproportionately yields smaller units \cite{statistikaustria2025wohnen}, the mismatch between unit size and household composition contributes to overcrowding conditions among foreign nationals.
}

\subsection{Education in Austria}\label{sec:supp_data_education}

{
Education demand data are derived from administrative records maintained by Statistics Austria, including school enrolment registers and teaching staff statistics. The analysis covers academic years 2015 through 2023, providing an eight-year baseline for trend estimation. Data are disaggregated by single-year age (5–50), sex, citizenship status (Austrian vs. foreign national), NUTS\,3 region (35 regions), and school type. The Austrian school system is consolidated into six analytical categories aligned with the International Standard Classification of Education: primary school (\textit{Volksschulen}), lower secondary (\textit{Mittelschule}), academic secondary (AHS), polytechnic school, vocational education, and special education.
}

\begin{figure}[htbp]
\centering
\includegraphics[width=\textwidth]{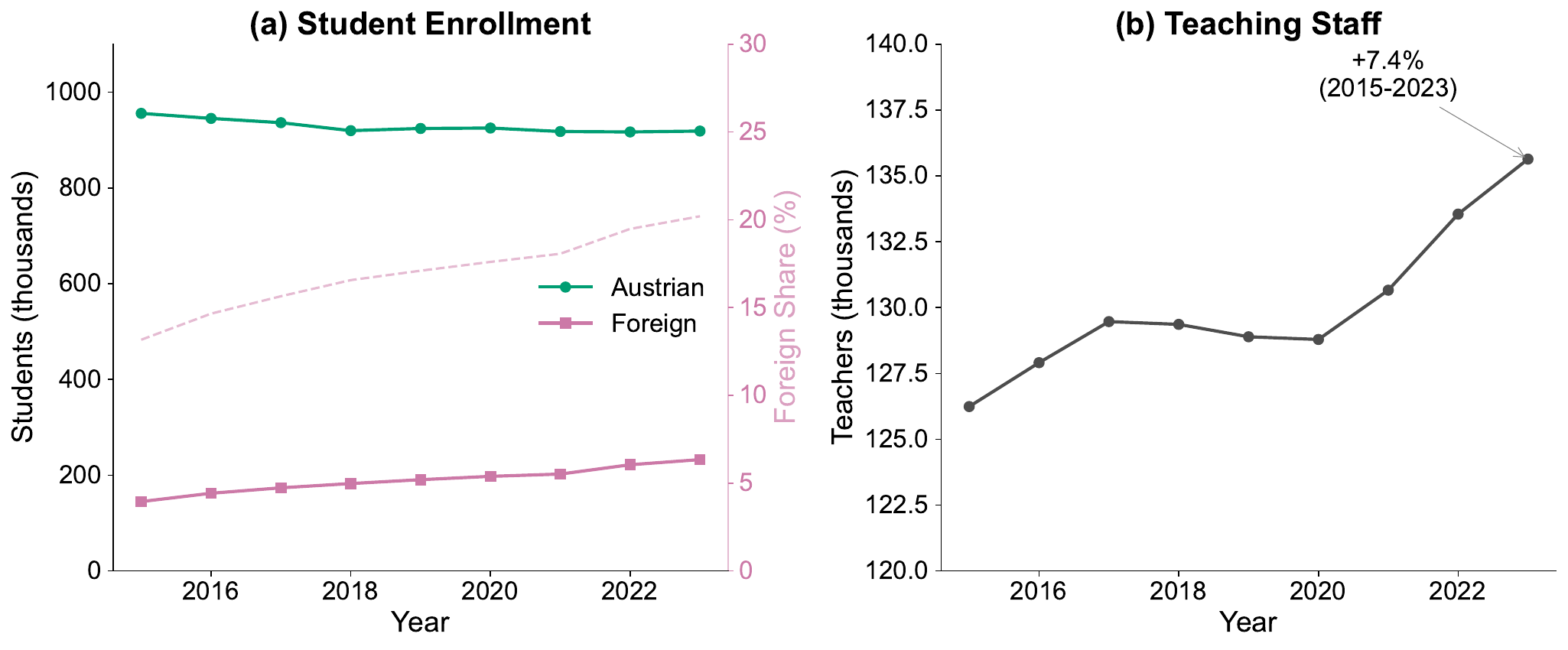}
\caption{Historical trends in student enrolment and teaching staff (2015–2023). (a) Student enrolment disaggregated by citizenship status. Austrian enrolment declined by 3.9\% while foreign enrolment increased by 60.5\%, raising the foreign share from 13.2\% to 20.2\% (dashed line, right axis). (b) Total teaching staff in full-time equivalents. The teaching workforce grew by 7.4\% over the period, resulting in a declining student-to-teacher ratio.}
\label{fig:education_historical}
\end{figure}

{
Austria's education system experienced divergent enrolment trajectories by citizenship status over the observation period (Supplementary Figure \ref{fig:education_historical}a). Austrian student enrolment declined modestly from 955,600 in 2015 to 918,700 in 2023 ($-3.9\%$), while foreign student enrolment increased substantially from 144,900 to 232,600 ($+60.5\%$) \cite{statistikaustria2025school}. As a result, the foreign share of total enrolment rose from 13.2\% to 20.2\%, a 7 percentage-point increase over eight years. This divergence reflects both the declining school-age population in Austria and the growing foreign population, rather than differential enrolment behaviour. The teaching workforce expanded by 7.4\% over the same period, from 126,200 to 135,600 full-time equivalent positions (Supplementary Figure \ref{fig:education_historical}b). The national student-to-teacher ratio consequently declined from 8.72 to 8.49, indicating modest improvements in staffing intensity despite the compositional shift in the student body.
}

\begin{figure}[htbp]
\centering
\includegraphics[width=0.75\textwidth]{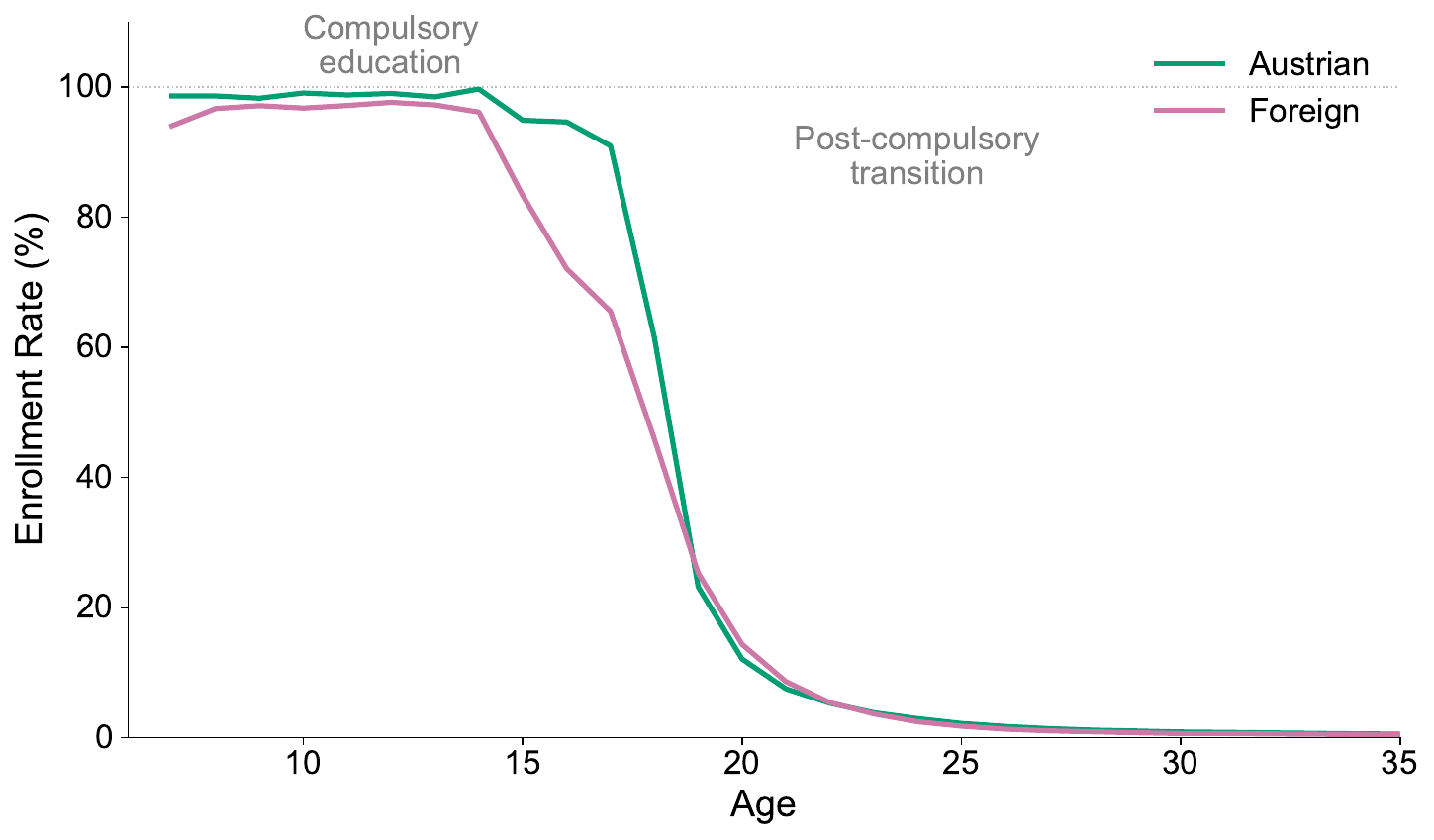}
\caption{Age-specific school enrolment rates by citizenship (2021–2023 average). Enrolment rates represent the proportion of the age-specific population enrolled in any educational institution. Both citizenship groups exhibit near-universal enrolment during compulsory education (ages 6–14), but diverge substantially at post-compulsory ages (15–18), with foreign nationals exiting the education system at higher rates. The gap reaches 25 percentage points at age 17.}
\label{fig:education_enrollment}
\end{figure}

{
The enrolment rate---the proportion of the age-specific population enrolled in any educational institution---exhibits a characteristic lifecycle pattern with near-universal participation during compulsory schooling followed by divergence at post-compulsory ages (Supplementary Figure \ref{fig:education_enrollment}). During the compulsory education period (ages 6–14), enrolment rates are high and largely convergent across citizenship groups: 98.8\% for Austrians and 96.1\% for foreign nationals on average \cite{statistikaustria2025school,statistikaustria2025population}. The gap during the compulsory years may reflect recent arrivals who have not yet been integrated into the school system or administrative delays in registration.

The critical divergence emerges at the post-compulsory transition (ages 15–18). At age 15, the enrolment gap is 11.6 percentage points (94.9\% Austrian vs 83.3\% foreign) \cite{statistikaustria2025school,statistikaustria2025population}. This gap widens to 25.4 percentage points by age 17 (91.0\% vs. 65.5\%) and then narrows slightly at age 18, as Austrian students complete upper secondary education. On average across post-compulsory ages, Austrian enrolment stands at 85.5\% compared to 66.7\% for foreign nationals, a gap of nearly 19 percentage points. This pattern indicates that, while compulsory education achieves near-universal coverage regardless of citizenship, the transition to upper secondary and vocational tracks marks a significant divergence, with foreign nationals substantially more likely to exit the education system at the end of compulsory schooling.
}

\begin{figure}[htbp]
\centering
\includegraphics[width=0.85\textwidth]{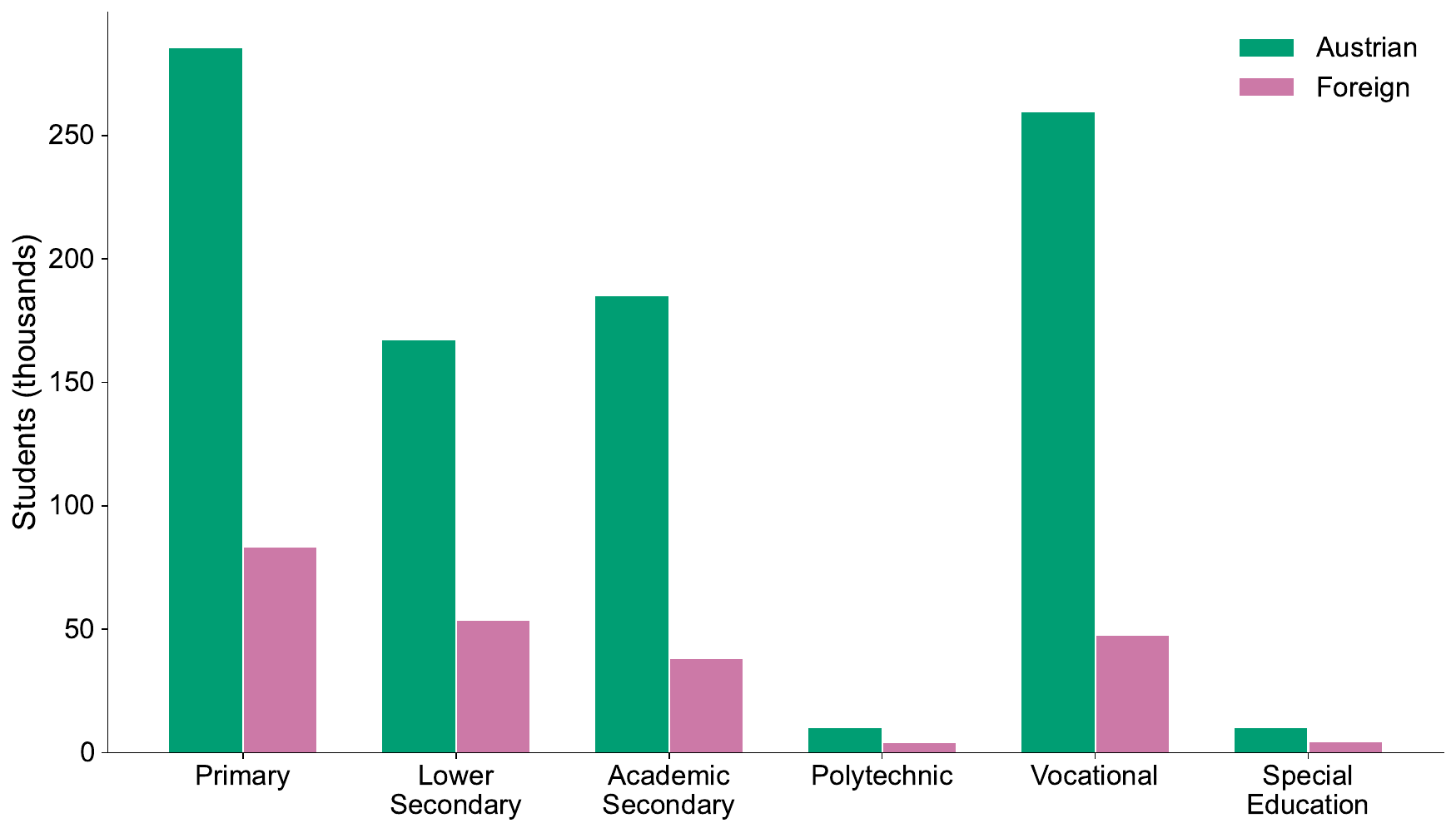}
\caption{Student distribution by school type and citizenship (2023). Absolute student counts (thousands) across six school categories. Foreign nationals are over-represented in lower secondary and special education relative to their population share (20.2\%), and under-represented in academic secondary and vocational education.}
\label{fig:education_school_type}
\end{figure}

{
The distribution of students across school types reveals differential tracking patterns by citizenship (Supplementary Figure \ref{fig:education_school_type}). Primary education accounts for the largest share of both groups, with foreign nationals comprising 22.6\% of primary enrolment in 2023 \cite{statistikaustria2025school,statistikaustria2025population}. At the secondary level, foreign students are over-represented in lower secondary schools (24.4\% foreign share) and under-represented in academic secondary schools (17.1\% foreign share), suggesting differential sorting into academic versus general tracks. The pattern is more pronounced in vocational education, where foreign nationals constitute only 15.5\% of enrolment despite comprising 20.2\% of the overall student population. Special education exhibits the highest foreign share at 31.3\%, which may reflect both genuine special needs and the channelling of students with language barriers into special education settings. These tracking patterns have implications for educational outcomes and subsequent labour market integration.
}

\begin{figure}[htbp]
\centering
\includegraphics[width=0.75\textwidth]{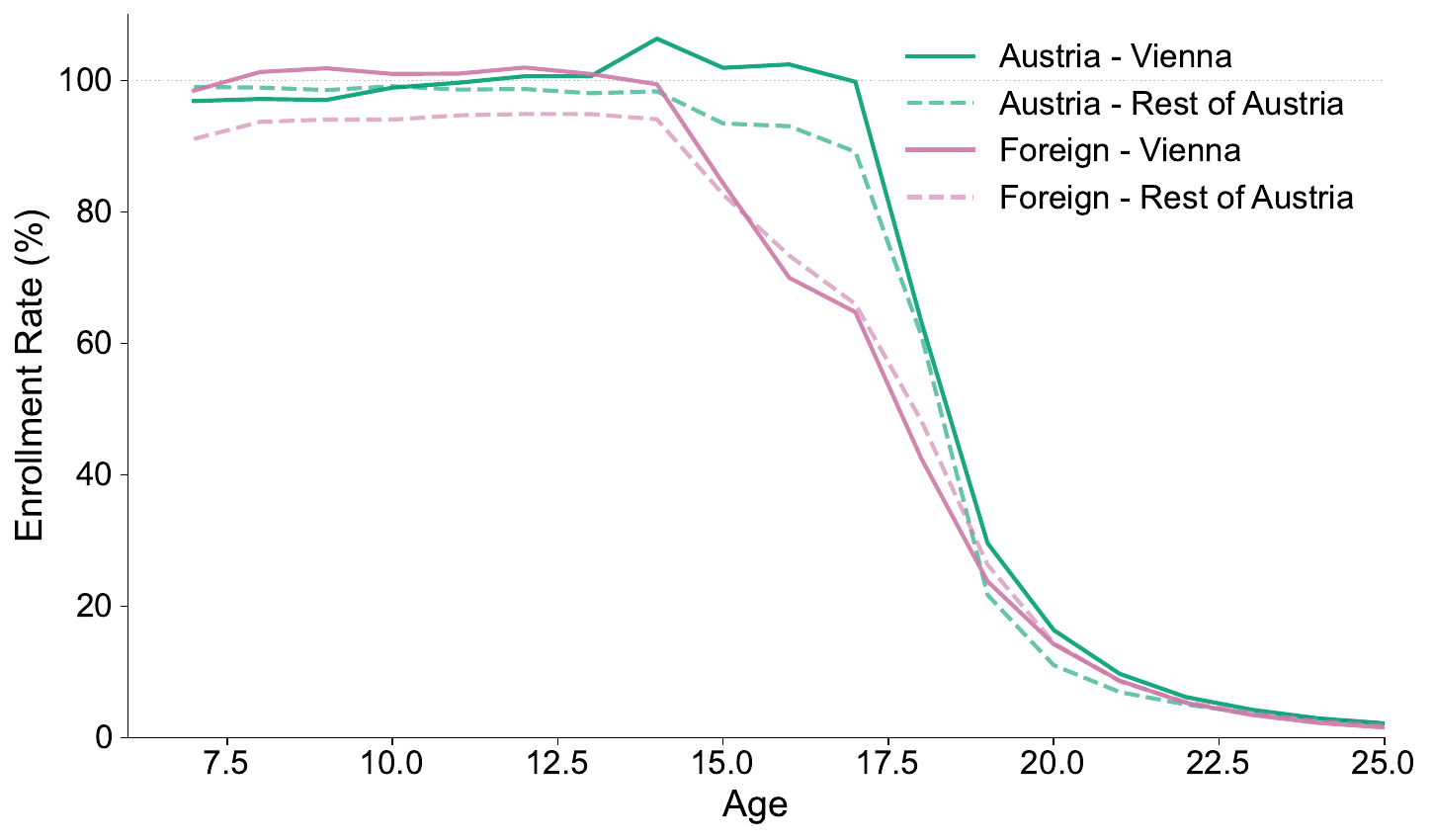}
\caption{Regional variation in enrolment rates (2021–2023 average). Enrolment rates are compared between Vienna and the rest of Austria for both citizenship groups. Vienna exhibits higher post-compulsory enrolment for Austrian students but a steeper drop-off for foreign students at the compulsory-to-post-compulsory transition.}
\label{fig:education_regional}
\end{figure}

{
Regional variation in enrolment rates reflects both the spatial concentration of the foreign population and differences in educational infrastructure (Supplementary Figure \ref{fig:education_regional}). Vienna, where foreign nationals constitute 36.9\% of student enrolment compared to 15.4\% in the rest of Austria, exhibits distinctive patterns \cite{statistikaustria2025school,statistikaustria2025population}. Austrian students in Vienna show higher enrolment rates at post-compulsory ages than their counterparts elsewhere, likely reflecting the concentration of academic secondary schools and higher education institutions in the capital. For foreign nationals, the regional pattern is more complex: Vienna shows higher enrolment during compulsory ages but a steeper drop-off at the post-compulsory transition. The convergence of all groups toward similar rates by age 20 reflects the transition to tertiary education and labour market entry, which follows similar patterns across citizenship and regions.
}

{
These patterns have several implications for education demand modelling. First, the near-universal enrolment during compulsory ages means that primary and lower secondary demand is driven almost entirely by demographic structure rather than behavioural variation. Second, the substantial citizenship gap at post-compulsory ages introduces compositional effects: a population with a higher foreign share will, all else equal, generate lower per capita upper secondary demand. Third, the spatial concentration of foreign students in Vienna (36.9\% of the capital's students) implies that capacity planning in the capital must account for a fundamentally different student composition than in rural regions. Finally, the rising foreign share, combined with stable student-to-teacher ratios, suggests that the challenge facing the education system is less about quantitative capacity than about qualitative adaptation, particularly in language support and integration services for an increasingly diverse student body.
}

\subsection{Healthcare in Austria}\label{sec:supp_data_healthcare}

{
Healthcare utilisation data derive from patient-level hospital admission records obtained from the Austrian Federal Ministry of Health for the reference year 2019. This year was selected as the most recent complete observation period before the COVID-19 pandemic, which distorted hospitalisation patterns. Data are disaggregated by five-year age groups (0–4 through 90+), sex, citizenship status (Austrian vs. foreign national), and NUTS\,2 region (nine federal states). The primary outcome measures are total hospital stays (admissions) and total bed-days (length of stay), from which per-capita utilisation rates are derived.
}

\begin{figure}[htbp]
\centering
\includegraphics[width=0.6\textwidth]{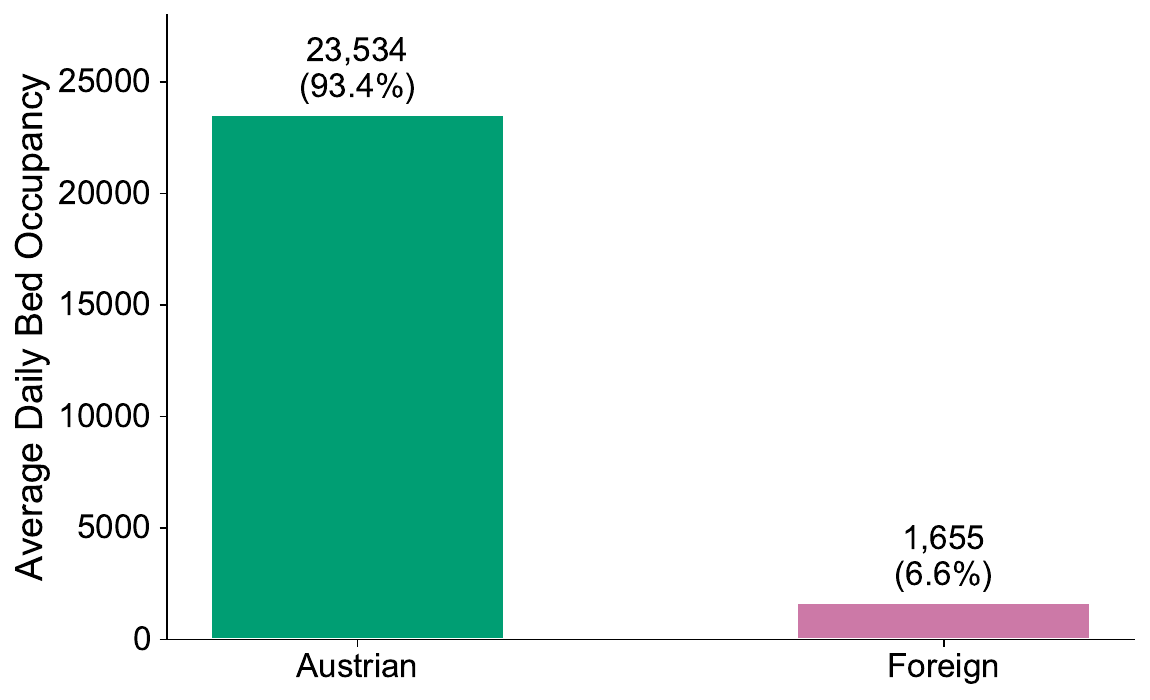}
\caption{Hospital bed utilisation by citizenship (2019). Average daily bed occupancy disaggregated by citizenship status. Austrian nationals account for 93.4\% of total hospital bed utilisation despite comprising 91.4\% of the population, reflecting higher per-capita utilisation rates.}
\label{fig:healthcare_total}
\end{figure}

{
In 2019, Austria's hospital system recorded approximately 9.2 million bed-days, equivalent to an average daily occupancy of 25,188 beds. Austrian nationals accounted for 93.4\% of total bed-days (23,534 beds daily), while foreign nationals contributed 6.6\% (1,655 beds daily). This distribution reflects substantial under-utilisation by foreign nationals relative to their population share: in the sample, foreign nationals constituted 8.6\% of the population but consumed only 6.6\% of hospital bed-days (Supplementary Figure \ref{fig:healthcare_total}). On a per-capita basis, Austrian nationals averaged 7.17 bed-days annually compared to 5.34 for foreign nationals, a differential of 34\% \cite{statistikaustria2025population}. This aggregate pattern is consistent with the ``healthy migrant effect'' documented in European literature \cite{dervic2024healthcare}, whereby recent migrants exhibit lower morbidity than the national population due to selection effects and the health requirements of migration itself.
}

\begin{figure}[htbp]
\centering
\includegraphics[width=0.85\textwidth]{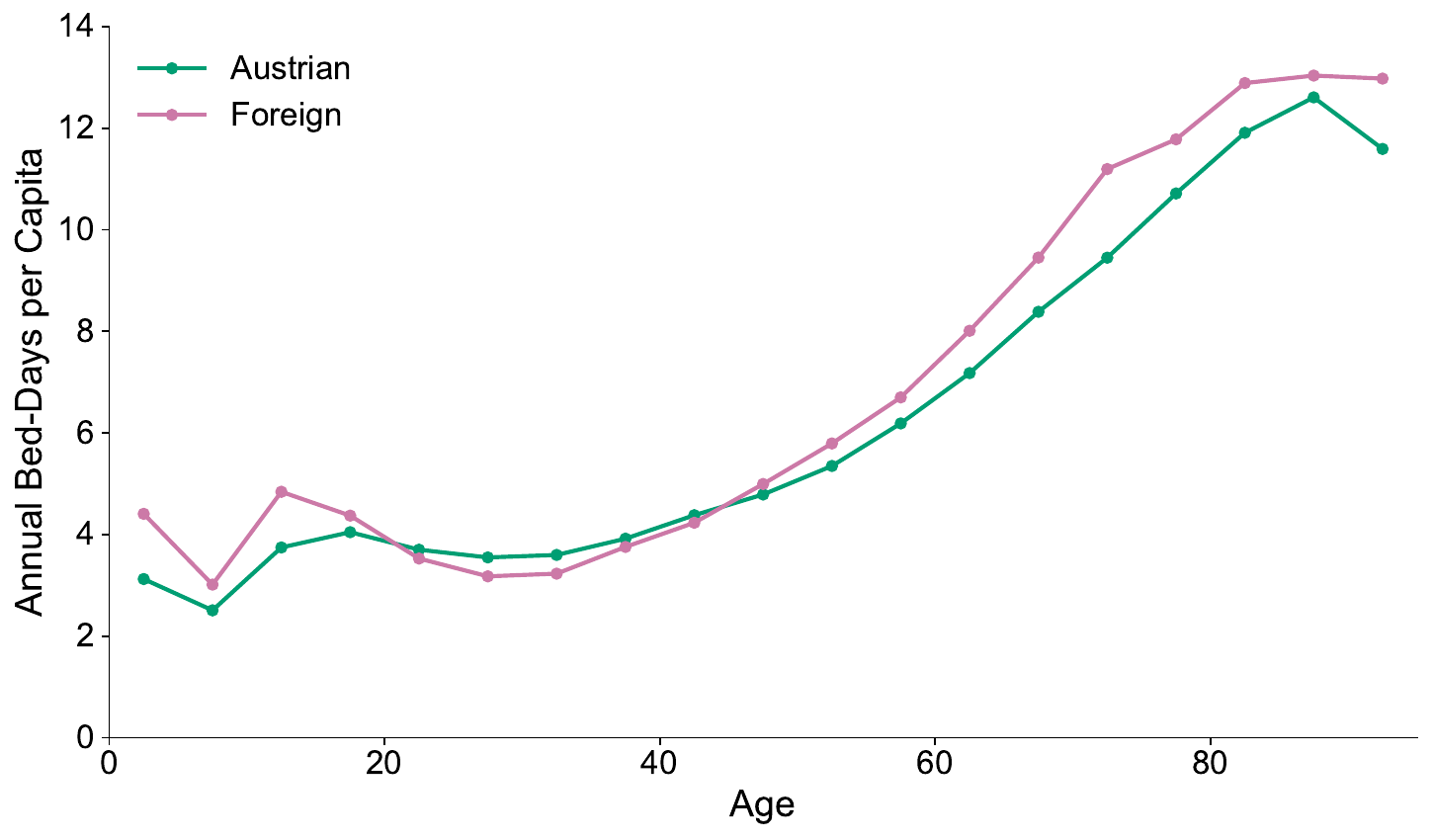}
\caption{Age-specific hospital utilisation by citizenship (2019). Annual bed-days per capita across five-year age groups. Both groups exhibit steep age gradients, with utilisation increasing approximately fourfold between working-age and advanced-age groups. The citizenship differential varies across the lifecycle: foreign nationals exhibit higher rates in childhood and at ages 70+, whereas Austrians exhibit higher rates during working ages.}
\label{fig:healthcare_age}

\centering
\includegraphics[width=\textwidth]{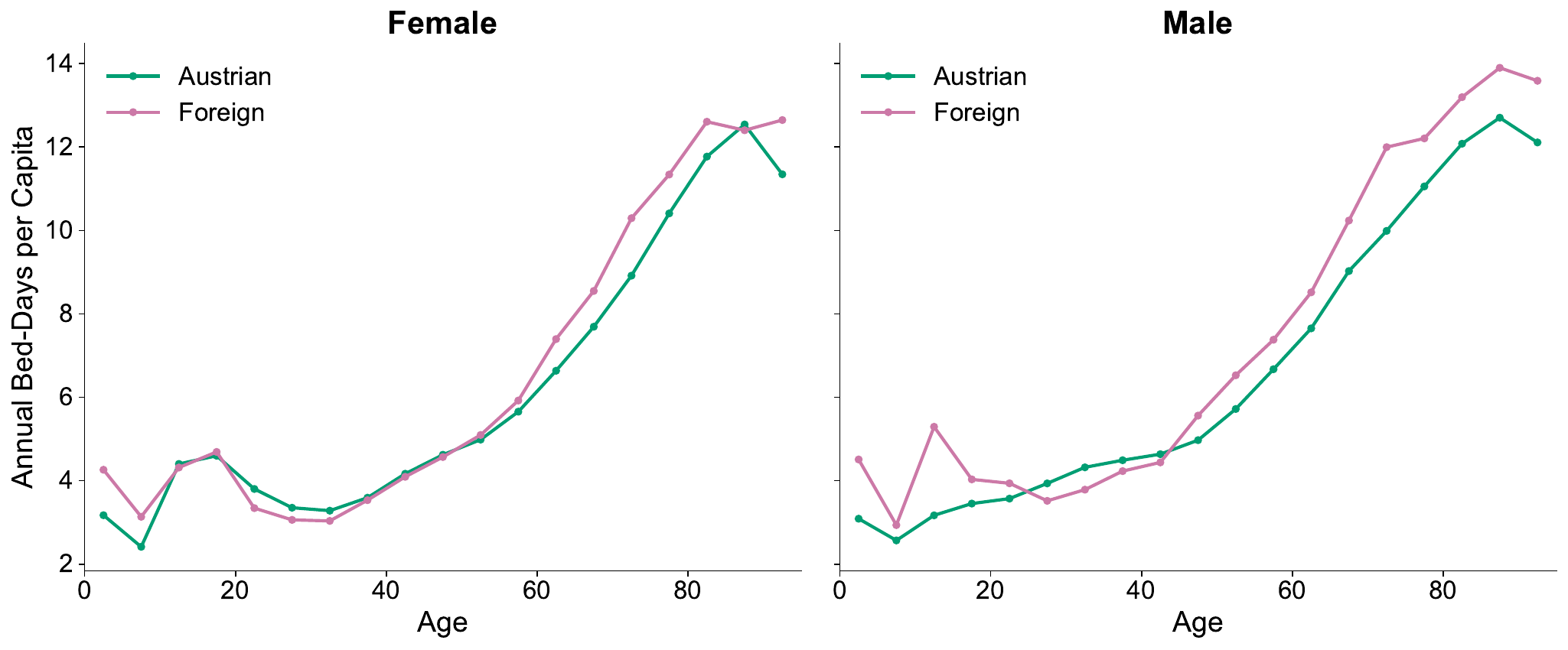}
\caption{Hospital utilisation by age, sex, and citizenship (2019). Annual bed-days per capita are shown separately for females (left) and males (right). Males exhibit higher utilisation across most ages for both citizenship groups. The citizenship differential, with foreign nationals showing lower rates during working ages but higher rates at advanced ages, is consistent across sexes.}
\label{fig:healthcare_age_sex}
\end{figure}

{
The age profile of hospitalisation reveals a steep gradient that dominates the demographic structure of healthcare demand (Supplementary Figure \ref{fig:healthcare_age}). For both citizenship groups, per-capita bed-days increase approximately fourfold between ages 40 and 85, from approximately 4 days annually to over 12 days. This exponential relationship between age and healthcare consumption is the fundamental driver of the ``structure effect'' identified in the main analysis: as populations age, per-capita demand intensity rises mechanically even without changes in age-specific utilisation behaviour.

The citizenship differential varies across the age distribution in a nuanced pattern. At young ages (0–20), foreign nationals exhibit higher hospitalisation rates than Austrians, with rates approximately 30\% higher in childhood. This pattern may reflect delayed access to preventive care among migrant families, resulting in more acute presentations requiring hospitalisation, or differences in maternal and infant health outcomes. During working ages (25–50), the pattern reverses: Austrian rates exceed foreign rates, consistent with the healthy migrant effect observed in aggregate statistics. At advanced ages (70+), the pattern reverses again: foreign nationals exhibit higher hospitalisation rates than Austrians. Among those aged 70 and over, foreign nationals average 11.85 bed-days annually compared to 10.99 for Austrians. This elderly crossover may reflect selection effects (survivors exhibit higher morbidity), reduced access to community-based care alternatives, or compositional differences in the elderly foreign population.
}

{
Hospitalisation patterns differ by sex in ways that interact with citizenship (Supplementary Figure \ref{fig:healthcare_age_sex}). Males exhibit higher hospitalisation rates than females across most ages for both citizenship groups, except for reproductive ages, where female rates include maternity-related admissions. Among Austrian nationals, males average 7.47 bed-days annually compared to 6.90 for females. Among foreign nationals, the sex differential is more pronounced: males average 6.22 bed-days compared to 4.72 for females, a gap of 32\%. This larger sex differential among foreign nationals may reflect occupational differences (with foreign males concentrated in higher-risk manual occupations) or more pronounced sex differences in healthcare-seeking behaviour among migrant populations.
}

\begin{figure}[htbp]
\centering
\includegraphics[width=0.75\textwidth]{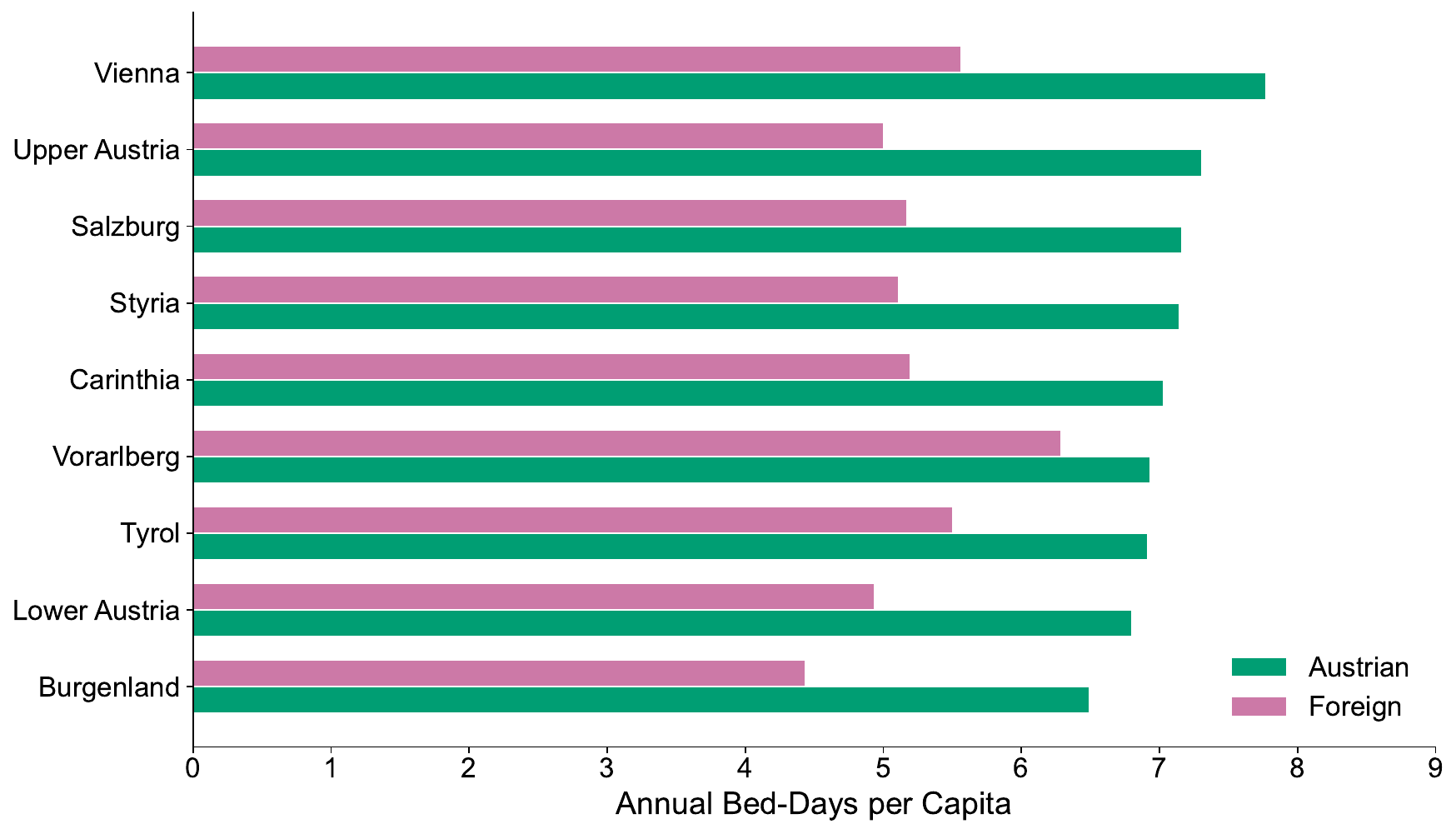}
\caption{Regional variation in hospital utilisation (2019). Annual bed-days per capita by federal state (NUTS\,2) and citizenship status. Foreign nationals exhibit lower utilisation than Austrians in all regions, with the gap ranging from 1.4 to 2.4 bed-days per capita. Vienna shows the highest Austrian rates but only moderate foreign rates.}
\label{fig:healthcare_regional}
\end{figure}

{
Hospital utilisation varies across Austria's federal states, with Austrian nationals exhibiting rates ranging from 6.49 bed-days per capita in Burgenland to 7.77 in Vienna (Supplementary Figure \ref{fig:healthcare_regional}). Foreign national rates range from 4.44 in Burgenland to 6.29 in Vorarlberg. Notably, the citizenship differential is consistent across all regions: foreign nationals consume fewer bed-days per capita than Austrians in every federal state, with the gap ranging from 1.4 to 2.4 bed-days. Vienna exhibits the highest Austrian utilisation rate but only moderate foreign utilisation, resulting in one of the largest citizenship gaps. This regional consistency suggests that the citizenship differential reflects demographic and health characteristics rather than region-specific barriers to healthcare access.
}

{
These patterns have several implications for healthcare demand modelling. First, the steep age gradient in utilisation, with rates increasing fourfold between working-age and advanced-age groups, confirms that demographic ageing is the dominant driver of future healthcare demand, irrespective of citizenship composition. Second, the lower aggregate utilisation by foreign nationals implies that population growth through migration generates less healthcare demand per capita than equivalent Austrian population growth, at least in the short to medium term. Third, the elderly crossover, whereby foreign nationals at ages 70+ exhibit higher utilisation than Austrians, suggests that as the foreign population ages, the healthy migrant advantage may attenuate or reverse. Fourth, the single-year baseline (2019) precludes trend estimation, motivating the status quo assumption in the projection model; actual future utilisation may diverge due to technological advances, policy reforms, or changes in care delivery patterns. The healthcare demand model, therefore, isolates the demographic contribution to future demand, providing a baseline against which policy scenarios can be evaluated.
}

\end{document}